\documentclass[pdflatex,sn-mathphys-num]{sn-jnl}

\usepackage{graphicx}%
\usepackage{multirow}%
\usepackage{amsmath,amssymb,amsfonts}%
\usepackage{amsthm}%
\usepackage{mathrsfs}%
\usepackage[title]{appendix}%
\usepackage{xcolor}%
\usepackage{textcomp}%
\usepackage{manyfoot}%
\usepackage{booktabs}%
\usepackage{algorithm}%
\usepackage{algorithmicx}%
\usepackage{algpseudocode}%
\usepackage{listings}%
\usepackage{mathtools}

\usepackage{color,xcolor,ucs}
\usepackage{mathtools}   \usepackage{tikz} 
\usepackage{ amssymb }
\usepackage{extarrows} 
\usepackage{pgf,tikz}
\usepackage{float}
\usetikzlibrary{positioning}
\usetikzlibrary{shapes.geometric}
\usetikzlibrary{shapes.misc}
\usetikzlibrary{arrows}
\usepackage{caption}
\usepackage{mathrsfs}
\usetikzlibrary{arrows,shapes,automata,backgrounds,petri,positioning}
\usetikzlibrary{decorations.pathmorphing}
\usetikzlibrary{decorations.shapes}
\usetikzlibrary{decorations.text}
\usetikzlibrary{decorations.fractals}
\usetikzlibrary{decorations.footprints}
\usetikzlibrary{shadows}
\usetikzlibrary{calc}
\usetikzlibrary{spy}
\usepackage{amsmath}
\usepackage{array}
\usepackage{ amssymb }
\usepackage{braket}
\usepackage{qcircuit}
\usepackage{soul}
\usepackage{braket} 
\usepackage{relsize}

\usepackage{amsmath}
\usepackage{ amssymb }
\usepackage{braket}
\usepackage{qcircuit}
\usepackage{soul}
\usepackage{braket}

\theoremstyle{thmstyleone}%
\theoremstyle{thmstyletwo}%

\theoremstyle{thmstylethree}%

\begin{document}

\title[Article Title]{Multiplayer parallel repetition without dependency-breaking and anchoring variables: monotonic, concave amplification }


\author*{\fnm{Pete} \sur{Rigas}}\email{pbr43@cornell.edu}

\affil*{Newport Beach, CA 92625}

\abstract{We obtain quantitative estimates on the decay of the multiplayer optimal value under parallel repetition. In comparison to a previous work of the author in 2025 (arXiv: 2508.09380) which sought to generalize dependency-breaking and anchoring variables from two-player Quantum games, being able to establish quantitative estimates on the decay of the optimal value of a multiplayer game under parallel repetition is of interest to establish under different assumptions. Specifically, independently of the dependency-breaking and anchoring variables that have previously been employed to remove correlations from entangled information shared between Alice and Bob (hence removing dependencies), monotonic concave functions can be used in place of such variables to obtain rates of decay on the optimal value. The game-theoretic setting with two players was first analyzed with monotonic concave functions by Lanzenberger and Maurer. For $q_i , x_i > 0$ $\forall 1 \leq i \leq N$ where $N > 0 $ is the total number of players we adddress an open question raised in their work regarding potential generalizations of two-player monotonic concave functions, through amplification functions of the form  $\Psi_{\textit{Mult}} \equiv  \Psi =   N  - \underset{1 \leq i \leq N}{\prod} \mathrm{exp} \big[ - q_i x_i \big]$, which in the multiplayer game-theoretic setting have more intricate combinatorial structures. \footnote{\textbf{MSC Class}: 81P02; 81Q02}}

\maketitle


\section{Introduction}

\subsection{Overview}

\noindent Amplification results are central to several areas of research interest within Quantum information theory, with related progress providing quantitative estimates on anchoring and fortification, [2], nonlocality, [3,8,10,12,15,18,19], entanglement, [4], numerical approximations of nonlinear problems, [5,6,7,9,13,14,16], as well as security [17,20,22,23,24]. Given developments in the theory of hardness amplification in [10], it is of great interest to further explore one direction of future research interest mentioned by the authors, particularly relating to classes of novel concave functions that can be used for more general game-theoretic settings. To provide several quantitative estimates on amplification of the winning probability, Lanzenberger and Maurer make use of several properties of concave functions, for demonstrating: (1) how expected values supported over the probability distributions of questions for Alice and Bob can be related to expected values over the probability distributions for questions of one player at a time; (2) how the specific form of the concave function, namely $\psi \big( x \big) = 1 - \big( 1 - x \big)^{q}$, for $q \in \textbf{N}$, impacts the upper bound for Alice and Bob's winning probability; (3) whether any other forms of concave functions can be considered in future work [10].

The approach relying upon concave functions in [10] is not only of great interest to further explore in its own right, but also through connections of previous works of the author which further developed dependency-breaking variables first put forth in [2,3,4]. In the multiplayer setting, previous work of the author in [21] developed generalizations beyond the two player game-theoretic setting, particularly in: (1) quantifying how complexity theoretic parameters appearing in $\mathrm{O} \big( \cdot \big)$ notation are related to computing expected values of dependency-breaking variables; (2) determining how the exponential rate of decay of the winning probability is dependent upon the number of rounds of parallel repetition; (3) underlying connections with the nonlocality of Quantum information [2,3,4,11,12,19]. While Quantum strategies previously formulated by the author in the multiplayer setting, [19], consist of error bounds and deviations from optimal Quantum strategies through duality gaps and primal feasible solutions of well posed semidefinite programs (SDPs), quantifying how optimality arises in amplification related settings is of interest to formalize. 

Albeit the fact that Alice and Bob in the most simple two-player setting can amplify their winning probabilities under parallel repetition through a suitably defined concave function, [10], several additional questions of interest remain. First and foremost, if there were additional participants in a game that can make use of entanglement in Quantum strategies, what properties of concave functions would one expect to hold? Straightforwardly, while concave functions considered in the amplification framework, [10], rely on the dynamic weakly verifiable puzzle (DWVP), hint queries, verification queries and the forgery probability, we demonstrate how more general classes of concave functions can be obtained for amplification results in multiplayer game-theoretic settings. For the multiplayer game-theoretic setting, in comparison to the simplest two player game-theoretic setting, one encounters: (1) additional sources of entanglement from other participants rather than Alice and Bob; (2) more general families of concave functions, say $\psi, \psi^{\prime}, \psi^{\prime\prime}, \cdots , \psi^{\prime\prime\cdots\prime}$, where,

\begin{align*}
   \# \big\{  1 \leq i \leq C  :  \big( \psi ,  \psi^{\prime}, \psi^{\prime\prime}, \cdots , \psi^{\prime\prime\cdots\prime} \big) \mapsto \big( \psi_1 , \cdots , \psi_i , \psi_{i+1} , \cdots \psi_C \big)   \big\} = N ,
\end{align*}

\noindent as opposed to $\psi,\psi^{\prime}$, would satisfy the properties: each concave function within the family $\psi, \psi^{\prime}, \psi^{\prime\prime}, \cdots , \psi^{\prime\prime\cdots\prime}$ for upper bounding the expected value,

\begin{align*}
       \textbf{E}_{ Q_1 \times \cdots \times Q_N}   \big[   \mu \big( Q_1, Q_2 , \cdots , Q_N \big) \big]   
        \sim  \textbf{E}_{\mathcal{Q}_1 \cdots \mathcal{Q}_{i-1} \mathcal{Q}_{i+1} \cdots  \mathcal{Q}_N} \big[        \psi^{\prime {\cdots} \prime}_{i} \big(   \textbf{E}_{\mathcal{Q}_i}     \big)     \big]         \leq \epsilon_i \psi^{\prime {\cdots} \prime}_{i} \big[  \underset{j \in [ n ]  , j \neq i}{\prod} \delta_j      \big]
\end{align*}    

\noindent for $\big\{ \delta_k \big\}_{1 \leq k \leq C} \in \textbf{R} \cap \big[ 0 , 1 \big]$, where $\textbf{E} \big[ \cdot \big]$ denotes the expectation with respect to the probability measure $\textbf{P} \big[ \cdot \big]$,

\begin{align*}
    \textbf{P} \big[ \cdot \big] \equiv \textit{Probability measure over questions distributed uniformly at random from the referee's } \\ \textit{probability distribution $\pi$} ,
\end{align*}

\noindent $[ n ] \equiv \big( 0, \cdots , n \big)$ for some $n>0$, and $\big\{ \epsilon_k \big\}_{1 \leq k \leq n}$ is a sequence of nonnegative constants. In addition to the inequality stated above between the expectation and $\psi^{\prime\cdots\prime}_{i-1}$, additional inequalities from the concave functions take the following form. Fix $N^{\prime} < N$, for which:

{\small \[   \left\{\!\begin{array}{ll@{}>{{}}l} 
\textit{Amplification across two players}.\text{ }  \text{The inequality corresponding to amplification for two players takes the form,}\\ \\  \\  \textbf{E}_{\mathcal{Q}_1 \cdots \mathcal{Q}_{i-1} \mathcal{Q}_{i+1} \cdots \mathcal{Q}_{j-1} \mathcal{Q}_{j+1} \cdots \mathcal{Q}_N} \big[ \psi^{\prime\cdots\prime}_{i} \big( \textbf{E}_{\mathcal{Q}_i} \big)   \psi^{\prime\cdots\prime}_{j} \big( \textbf{E}_{\mathcal{Q}_j} \big)         \big] \leq  \frac{1}{\sqrt{2} } \bigg[  \epsilon_i   \psi^{\prime\cdots\prime}_{i} \big[  \underset{k \in \{ n \}  ,  k \neq i  }{\prod}  \delta_k   \big]               +  \epsilon_j  \psi^{\prime\cdots\prime}_{j}  \big[  \underset{k \in \{ n \}  ,  k \neq i \neq j }{\prod} \delta_k  \big]    \bigg] \\ \text{, where } i \neq j \text{ satisfy } 1 \leq i \neq j \leq N

,  \\ \\ \\  \textit{Amplification across three players}.\text{ }  \text{The inequality corresponding to amplification for three players takes the form,} \\  \\  \\ \textbf{E}_{\mathcal{Q}_1 \cdots \mathcal{Q}_{i-1} \mathcal{Q}_{i+1} \cdots \mathcal{Q}_{j-1} \mathcal{Q}_{j+1} \cdots \mathcal{Q}_{k-1} \mathcal{Q}_{k+1}  \cdots \mathcal{Q}_N} \big[ \psi^{\prime\cdots\prime}_{i} \big( \textbf{E}_{\mathcal{Q}_i} \big)   \psi^{\prime\cdots\prime}_{j} \big( \textbf{E}_{\mathcal{Q}_j} \big)   \psi^{\prime\cdots\prime}_{k-1} \big( \textbf{E}_{\mathcal{Q}_k} \big)         \big] \leq  \frac{1}{\sqrt{3} } \bigg[  \epsilon_i   \psi^{\prime\cdots\prime}_{i} \big[  \underset{l \in \{ n \}  ,  l \neq i  }{\prod}  \delta_l   \big]                +  \epsilon_j \\  \\ \times \psi^{\prime\cdots\prime}_{j}  \big[  \underset{l \in \{ n \}  ,  l  \neq i \neq j }{\prod}   \delta_l \big]  +     \epsilon_j  \psi^{\prime\cdots\prime}_{k}  \big[  \underset{l \in \{ n \}  ,  l  \neq i \neq j \neq k }{\prod}   \delta_l \big]         \bigg]    \text{, where } i \neq j \neq k \text{ satisfy } 1 \leq i \neq j \neq k \leq N ,    \\ \vdots 
\\ 
\textit{Amplification across N players}.\text{ }  \text{The inequality corresponding to amplification for N players takes the form,} \\ \\  \\     \textbf{E}_{\mathcal{Q}_1 \cdots \mathcal{Q}_{i-1} \mathcal{Q}_{i+1} \cdots \mathcal{Q}_{j-1} \mathcal{Q}_{j+1} \cdots \mathcal{Q}_{k-1} \mathcal{Q}_{k+1}  \cdots \mathcal{Q}_N} \big[ \psi^{\prime\cdots\prime}_{i} \big( \textbf{E}_{\mathcal{Q}_i} \big)   \psi^{\prime\cdots\prime}_{j} \big( \textbf{E}_{\mathcal{Q}_j} \big)   \psi^{\prime\cdots\prime}_{k} \big( \textbf{E}_{\mathcal{Q}_k} \big)   \times \cdots \times  \psi^{\prime\cdots\prime}_{N^{\prime}} \big( \textbf{E}_{\mathcal{Q}_{N^{\prime}}} \big)       \big] \end{array}\right. 
\]  }

{\small \[   \left\{\!\begin{array}{ll@{}>{{}}l}  \leq  \frac{1}{\sqrt{N} } \bigg[  \epsilon_i   \psi^{\prime\cdots\prime}_{i} \big[  \underset{l \in \{ n \}  ,  l \neq i  }{\prod}  \delta_l   \big]                +  \epsilon_j \psi^{\prime\cdots\prime}_{j}  \big[  \underset{l \in \{ n \}  ,  l  \neq i \neq j }{\prod}   \delta_l \big]  +     \epsilon_j  \psi^{\prime\cdots\prime}_{k-1}  \big[  \underset{l \in \{ n \}  ,  l  \neq i \neq j \neq k }{\prod}   \delta_l \big]       + \cdots +    \epsilon_{N^{\prime}}  \psi^{\prime\cdots\prime}_{N^{\prime} }  \\ \\ \times  \big[  \underset{l \in \{ n \}  ,  l  \neq i \neq j \neq k \neq \cdots \neq N^{\prime} }{\prod}   \delta_l \big]                   \bigg] ,    \text{ where } i \neq j \neq k \neq \cdots \neq N^{\prime} \text{ satisfy } 1 \leq i \neq j \neq k \neq \cdots \neq N^{\prime} \leq N .        \\  \\  \tag{$\textbf{1}$} \\  
\end{array}\right. 
\]  } 

\bigskip

\noindent The normalizations, either $\frac{1}{\sqrt{2}}$, $\frac{1}{\sqrt{3}}$ or $\frac{1}{\sqrt{N}}$ appearing in the above inequalities resemble several factors appearing in error bounds previously characterized by the author in [19,21]; it is typical for one to encounter normalizations proportional to the square root of the number of observations that parties of two, or more, players can make while responding to questions drawn uniformly at random from the referee's probability distribution of questions. However, besides dynamics of the above form corresponding to amplification through products of concave functions, straightforwardly one could also be interested in how game-theoretic objects can be related with amplification. To this end one crucial observation provided in [10] surrounds the optimal value which can be defined for direct product of games. Specifically, under a suitable class of restrictions, the optimal values satisfy,

{\small \begin{align*}
  \omega \big( G \big) \lesssim \omega \big( G , \textit{restrictions} \big)  .
\end{align*} }

\noindent where,

{\small \begin{align*}
   \omega \big( G \big) \propto \# \big\{ \textit{strategy : strategy has no restrictions} \big\}   , \end{align*}

   \begin{align*} \omega \big( G , \textit{restrictions} \big) \propto \# \big\{ \textit{strategy : strategy satisfies any of the constraints} \\ \textit{in restrictions} \big\}   . \\ 
\end{align*} } 

\noindent Intuitively, the collection of restrictions that one can place on the optimal value $\omega \big( G \big)$,

\begin{align*}
 \textit{restrictions}^{\prime} \equiv \textit{Restrictions placed on the strategies used by Alice and Bob at the second} \\ \textit{ round of parallel repetition}   , \\ \vdots  \\   \textit{restrictions}^{\prime\cdots \prime} \equiv    \textit{Restrictions placed on the strategies used by Alice and Bob at the N th} \\ \textit{ round of parallel repetition}      , 
\end{align*}

\noindent for a two-player game with Alice and Bob could stipulate,

{\small \begin{align*}
 \textit{restrictions}^{\prime} \equiv \textit{Restriction $H \big( X, Y \big) \geq \delta_2$ at the second round of parallel repetition}   , \\ \vdots  \\   \textit{restrictions}^{\prime\cdots \prime} \equiv    \textit{Restrictions $H \big( X, Y \big) \geq \delta_N$ at the N th round of parallel repetition}     ,  \\ 
\end{align*} }

\noindent given a sequence of nonnegative constants $\big\{ \delta_j \big\}_{1 \leq j \leq C} \in \textbf{R} \cap \big[ 0 , 1 \big]$. The min-entropy, $H \big( \cdot , \cdot \big)$, between the two random variables which Alice and Bob use for responding to questions drawn uniformly at random from the referee's probability distribution,

\begin{align*}
  H_{\mathrm{min}} \big(  X  , Y    \big)  \equiv H \big( X , Y  \big)  = \underset{Y \in \mathcal{Y}}{\underset{X \in \mathcal{X}}{\sum}}   \textbf{P} \big[ X , Y \big]      \mathrm{ln} \big[  \textbf{P} \big[ X , Y \big]      \big] =  \underset{Y \in \mathcal{Y}}{\underset{X \in \mathcal{X}}{\sum}}  \textbf{P} \big[ X , Y \big]     \mathrm{log}_2  \big[  \textbf{P} \big[ X , Y \big]      \big]  .
\end{align*}

\noindent Intuitively, to characterize aspects of the joint min-entropy on countably many questions that players respond to for amplification across all coordinates $C$ in the multiplayer setting, one considers the entropy,

{\small \begin{align*}
 {\tiny H_{\mathrm{min}} \big(  Q_1 , Q_2, \cdots , Q_N    \big)  \equiv H \big(  Q_1 , Q_2, \cdots , Q_N    \big)  = \underset{Q_N \in \mathcal{Q}_N}{\underset{\vdots}{\underset{Q_2 \in \mathcal{Q}_2}{\underset{Q_1 \in \mathcal{Q}_1 }{\sum}}}}   \textbf{P} \big[  Q_1 , Q_2, \cdots , Q_N  \big]      \mathrm{ln} \big[  \textbf{P} \big[  Q_1 , Q_2, \cdots , Q_N  \big]      \big]}  \end{align*}

  \begin{align*} =  \underset{Q_N \in \mathcal{Q}_N}{\underset{\vdots}{\underset{Q_2 \in \mathcal{Q}_2}{\underset{Q_1 \in \mathcal{Q}_1 }{\sum}}}}   \textbf{P} \big[  Q_1 , Q_2, \cdots , Q_N  \big]      \mathrm{log}_2  \big[ \textbf{P} \big[  Q_1 , Q_2, \cdots , Q_N  \big]      \big]  , \\ 
\end{align*}  }

\noindent for the joint probability measure,

\begin{align*}
   \textbf{P} \big[ Q_1 , Q_2 , \cdots , Q_N \big] \equiv \underset{1 \leq i \leq N}{\prod}  \textbf{P} \big[ Q_i \big]  \sim \underset{1 \leq i \leq N}{\prod}  \textbf{P} \big[ \mathcal{Q}_i \big]    . 
\end{align*}

\noindent In the two-player setting, if one does not make use of dependency-breaking, and anchoring, variables, [2,3,4], arguments presented in [10] using concave functions are dependent upon:

\begin{itemize}
    \item[$\bullet$] \textit{The asymmetric role between Alice and Bob in the two-player setting}. In comparison to multiplayer settings analyzed with dependency-breaking, and anchoring, variables, [21], amplification in the two-player game theoretic setting consists of applying Jensen's inequality. Given a concave function $f$ and expectation $\textbf{E} \big[ \cdot \big]$ over some random variable $X^{\prime}$, the inequality states,

    \begin{align*}
    \textbf{E}   \big[  f  \big[ X^{\prime} \big] \big] \leq f  \big[ \textbf{E}   \big[ X^{\prime}  \big]  \big] . 
    \end{align*}

        \item[$\bullet$] \textit{Generalizing two-player asymmetry to the multiplayer game-theoretic setting}. For game-theoretic settings with more than two players, given a function,

        \begin{align*}
       \mu \big( Q_1, Q_2 , \cdots , Q_N \big)  \sim \mu \big( \mathcal{Q}_1, \mathcal{Q}_2 , \cdots , \mathcal{Q}_N \big) = \mu \big( \underset{1 \leq i \leq N}{\cup} Q_i \big) \equiv \mu \big( \underline{\mathcal{Q}} \big) ,
        \end{align*}

        \noindent one can upper bound the expected value of the above function,

        \begin{align*}
        \textbf{E}_{ Q_1 \times \cdots \times Q_N}   \big[   \mu \big( Q_1, Q_2 , \cdots , Q_N \big) \big]   
        \sim \textbf{E}_{\mathcal{Q}_1 \times \cdots \times \mathcal{Q}_N}   \big[   \mu \big( Q_1, Q_2 , \cdots , Q_N \big) \big]  , 
        \end{align*}

        \noindent through an inequality of the form,

  {\small  \begin{align*}
        \textbf{E}_{\mathcal{Q}_1 \times \cdots \times \mathcal{Q}_N}   \big[   \mu \big( Q_1, Q_2 , \cdots , Q_N \big) \big]  \leq  {\tiny C  \text{ } \mathrm{sup} \bigg\{   \epsilon \epsilon^{\prime}    ,  \epsilon \epsilon^{\prime} \epsilon^{\prime\prime}  , \cdots , \epsilon \epsilon^{\prime} \epsilon^{\prime\prime}  \times \cdots \times \epsilon^{\prime\prime\cdots \prime} ,    \epsilon \delta , \epsilon \delta + \epsilon^{\prime} \delta^{\prime}   ,  \cdots } \\ \\ {\small ,       \underset{1 \leq i \leq N}{\sum} \epsilon_i \delta_i    \bigg\}               }        , \\ \\  \tag{\textbf{2}} \\  
        \end{align*}  }

        \noindent for some $C >0$.

        \bigskip 

            \item[$\bullet$] \textit{An additional implication of asymmetry in the multiplayer game-theoretic setting}. Related to the previous item provided above, one has that the up to constants upper bound,

               {\small  \begin{align*}
        \textbf{E}_{\mathcal{Q}_1 \times \cdots \times \mathcal{Q}_N}   \big[   \mu \big( Q_1, Q_2 , \cdots , Q_N \big) \big]  \lesssim     \Psi^{-1} \bigg[     2^{-N}   \mathrm{sup}   \big\{        \textbf{E}_{\mathcal{Q}_1} \big[ \psi \big[  \textbf{E}_{\mathcal{Q}_2} \big[ \mu \big( Q_1 , Q_2 , \cdots , Q_N \big)  \big]     \big] \big]  \\ , \textbf{E}_{\mathcal{Q}_2} \big[ \psi^{\prime} \big[  \textbf{E}_{\mathcal{Q}_1} \big[ \mu \big( Q_1  , Q_2 , \cdots , Q_N \big)  \big]     \big] \big]  , \cdots ,        \textbf{E}_{\mathcal{Q}_N} \big[ \psi \big[  \textbf{E}_{\mathcal{Q}_1} \big[ \mu \big( Q_1  , Q_2 , \cdots , Q_N \big)  \big]     \big] \big]  , \\ \cdots , \textbf{E}_{\mathcal{Q}_N} \big[ \psi^{\prime\cdots \prime}_{N-1} \big[  \textbf{E}_{\mathcal{Q}_1 \times \cdots \times \mathcal{Q}_{N-1}}  \big[ \mu \big( Q_1  , Q_2  , \cdots , Q_N \big)  \big]     \big] \big]                      \big\}               \bigg]^{2^N }                        , \\ 
        \end{align*}  }

\noindent holds where $\Psi$ is the multiplayer concave function,

\begin{align*}
   \Psi_{\textit{Mult}} \equiv  \Psi =   N - \underset{1 \leq i \leq N}{\prod} \mathrm{exp} \big[ - q_i x_i \big]   ,  
\end{align*}

\noindent which one can straightforwardly verify is monotonic and concave over the unit interval $\big[ 0 , 1 \big]$. In the presence of a suitably chosen combinatorial factor, the above inequality can be sharpened to one of the following form,

         {\small  \begin{align*}
\textbf{E}_{\mathcal{Q}_1 \times \cdots \times \mathcal{Q}_N}   \big[   \mu \big( Q_1, Q_2 , \cdots , Q_N \big) \big]  \leq  C \big( N  , n \big) \cdot  \Psi^{-1}  \bigg[  2^{-N}  \mathrm{sup}   \big\{        \textbf{E}_{\mathcal{Q}_1} \big[ \psi \big[  \textbf{E}_{\mathcal{Q}_2} \big[ \mu \big( Q_1 , Q_2 ,  \cdots , Q_N \big)  \big]     \big] \big]   \\ , \textbf{E}_{\mathcal{Q}_2} \big[ \psi^{\prime} \big[  \textbf{E}_{\mathcal{Q}_1} \big[ \mu \big( Q_1  , Q_2 , \cdots , Q_N \big)  \big]     \big] \big]  , \cdots ,        \textbf{E}_{\mathcal{Q}_N} \big[ \psi \big[  \textbf{E}_{\mathcal{Q}_1} \big[ \mu \big( Q_1  , Q_2 , \cdots , Q_N \big)  \big]     \big] \big]  , \cdots \\  , \textbf{E}_{\mathcal{Q}_N} \big[ \psi^{\prime\cdots \prime}_{N-1} \big[ \textbf{E}_{\mathcal{Q}_1 \times \cdots \times \mathcal{Q}_{N-1}}  \big[ \mu \big( Q_1  , Q_2  , \cdots , Q_N \big)  \big]     \big] \big]                   \big\}               \bigg]^{2^N}         , \\ \\  \tag{\textbf{3}}  \\ 
        \end{align*}  }

\noindent where,

{\small \begin{align*}
    C \big( N , n  \big) \equiv N        \underset{i \in [ n ] }{\sum} \binom{N}{i}  , \\ 
\end{align*} }

\noindent denotes the amplification constant for sharpening the up to constants upper bound to an inequality. 

\bigskip

\item[$\bullet$] \textit{Relating the amplification constant $C \big( N \big)$ to the product of $\epsilon$ and summation of $\delta$ factors}. In comparison to the upper bounds from the two-player setting obtained in [10] (\textbf{Theorem} \textit{2}, \textbf{Corollary} \textit{1}),

{\small \begin{align*}
    \underset{i \in [ n ] }{\prod} \epsilon_i +  \underset{i \in [ n ] }{\sum} \delta_i , \\ \\ \underset{i \in [ n ] }{\prod} \delta_i  ,   \\         
 \end{align*} }

 \noindent hold we demonstrate that upper bounds of the following form hold,

{\small \begin{align*}
    \underset{i \neq i^{\prime} \neq \cdots \neq i^{\prime\cdots\prime}}{\underset{i^{\prime\cdots\prime} \in [n]}{\underset{\vdots}{\underset{i^{\prime} \in [n]}{\underset{i \in [ n ] }{\prod}}}}} \big[  \epsilon_i + \epsilon_{i^{\prime}} + \cdots + e_{i^{\prime\cdots\prime}} \big]  +  \underset{i \neq i^{\prime} \neq \cdots \neq i^{\prime\cdots\prime}}{\underset{i^{\prime\cdots\prime} \in [n]}{\underset{\vdots}{\underset{i^{\prime} \in [n]}{\underset{i \in [ n ] }{\sum}}}}} \big[ \delta_i + \delta_{i^{\prime}} + \cdots + \delta_{i^{\prime\cdots\prime}} \big]  , \\ \\  \tag{\textbf{4}} \end{align*}

    \begin{align*}      \underset{i \neq i^{\prime} \neq \cdots \neq i^{\prime\cdots\prime}}{\underset{i^{\prime\cdots\prime} \in [n]}{\underset{\vdots}{\underset{i^{\prime} \in [n]}{\underset{i \in [ n ] }{\mathrm{sup}}}}}}    {\tiny \bigg[   \underset{k , k^{\prime} \in \{ i , i^{\prime} , \cdots , i^{\prime\cdots\prime} \} }{\prod} \delta_k \epsilon_{k^{\prime}}        \bigg]^N }   ,  \\    \\ \tag{\textbf{5}}   \\    
 \end{align*} }

for $\epsilon_k, \delta_k \in \big[0,1 \big]$, and,

{\small \begin{align*}
    \textbf{E}_{ Q_1 \times \cdots \times Q_N}   \big[   \mu \big( Q_1, Q_2 , \cdots , Q_N \big) \big]   
        \sim    \textbf{E}_{\mathcal{Q}_1 \times \cdots \times \mathcal{Q}_N} \big[ \mu \big( Q_1 , \cdots , Q_N \big)  \big]     , \\ \\ \tag{\textbf{6}} \\    
\end{align*} }

\noindent the expected value of $\mu \big( Q_1 , \cdots , Q_N \big)$ supported over $\mathcal{Q}_1 \times \cdots \times \mathcal{Q}_N$. 
            
\end{itemize}

\subsection{This paper's contributions}

This paper further develops the use of concave functions introduced in [10] for amplification in game-theoretic settings with two players. While decays for the optimal value of multiplayer games have previously been obtained by the author in [21] by generalizing dependency-breaking and anchoring variables introduced in [2,3,4], it is of interest to formulate the role that concave functions assume in the absence of dependency-breaking and anchoring variables. Roughly speaking, while dependency-breaking and anchoring variables together remove correlations from strategies that Alice and Bob can use for Quantum advantage, a generalization of concave functions used for two-player games  can be used to upper bound expectation values supported over the set of questions,

{\small \begin{align*}
  \mathcal{Q}^{\otimes k} \equiv   \big\{ \big( \mathcal{Q}_1, \cdots, \mathcal{Q}_k \big) , \cdots , \big( \mathcal{Q}^{\prime\cdots\prime}_1, \cdots, \mathcal{Q}^{\prime\cdots\prime}_k \big) \big\}       \text{, } \\ 
\end{align*} }

\noindent where $k< n$ denotes the number of parallel repetition operations of a multiplayer Quantum game and $\big\{ \mathcal{Q}_1, \cdots, \mathcal{Q}^{\prime\cdots\prime}_1\big\}, \cdots, \big\{ \mathcal{Q}_1, \cdots, \mathcal{Q}^{\prime\cdots\prime}_1\big\}$ respectively denote the questions that are distributed to each of the $N$ participants after $k$ rounds of parallel repetition. To this end we: introduce previously results (specifically \textbf{Theorem} \textit{1}, \textbf{Theorem} \textit{2}, \textbf{Theorem} \textit{3}, \textbf{Corollary} \textit{4}, \textbf{Proposition} \textit{1} and \textbf{Corollary} \textit{2} from [10]) which capture how suitably defined expectation values can be upper bounded with concave functions; provide statements of results previously obtained by the author in [21] which quantitatively describe how the optimal value decays with respect to the number of rounds of parallel repetition; describe how decays on the optimal value imply several complexity-theoretic related results that are also obtained by the author in [21]; relate how Eve's forgery probability obtained in [24], which shares connections with the success probability, [17], and the false acceptance probability, [23], are related to HINT and VERIFICATION queries introduced in [10]. The arguments presented in this work demonstrate how other types of quantitative observations surrounding parallel repetition can be obtained through larger families of concave functions besides the concave, monotonic functions $\psi$ and $\psi^{\prime}$ introduced in [10].

As demonstrated through the monotonic, concave functions adapted for the multiplayer setting, the combinatorial factor $2^N$ appearing in (\textbf{3}) differs from the constant power $2$ appearing in \textbf{Corollary} \textit{2} of [10]. The factor to which the concave function is raised in the amplification result is not only dependent upon the total number of participants in the multiplayer game, but also upon $C \big( N , n \big)$. Essentially, we are interested in characterizing the behavior of,

{\small \begin{align*}
 \bigg\{  \frac{C \big( N , n \big)}{4}  \bigg\}  \Psi^{-1}  \bigg[   2^{-N}  \mathrm{sup}   \big\{        \textbf{E}_{\mathcal{Q}_1} \big[ \psi \big[  \textbf{E}_{\mathcal{Q}_2} \big[ \mu \big( Q_1 , Q_2 , \cdots , Q_N \big)  \big]     \big] \big]   , \textbf{E}_{\mathcal{Q}_2} \big[ \psi^{\prime} \big[  \textbf{E}_{\mathcal{Q}_1} \big[ \mu \big( Q_1  , Q_2 , \cdots \\ , Q_N \big)  \big]     \big] \big]  , \cdots ,      \textbf{E}_{\mathcal{Q}_N} \big[ \psi \big[  \textbf{E}_{\mathcal{Q}_1} \big[ \mu \big( Q_1  , Q_2 , \cdots , Q_N \big)  \big]     \big] \big]  , \cdots  , \textbf{E}_{\mathcal{Q}_N} \big[ \psi^{\prime\cdots \prime}_{N-1} \big[  \textbf{E}_{\mathcal{Q}_1 \times \cdots \times \mathcal{Q}_{N-1}}  \big[ \mu \big( Q_1  , Q_2  , \\  \cdots  , Q_N \big)  \big]     \big] \big]                   \big\}               \bigg]^{2^N}  \bigg\{  \xi^{-1} \bigg[ \frac{1}{2}  \mathrm{sup}  \big\{ \textbf{E}_X \big[ \psi \big[ \textbf{E}_Y  \big[ \mu \big( X , Y \big)  \big] \big] \big]   ,    \textbf{E}_Y \big[ \psi \big[ \textbf{E}_X  \big[ \mu \big( X , Y \big)  \big] \big] \big]         \big\}    \bigg]^{-2}    \bigg\}^{-1}                           , \\ 
\end{align*} }

\noindent as a function of $N$, namely comparing the value of a preimage under the pullback $\Psi^{-1} \big[ \cdot \big] $ raised to $2^N$ instead of $\xi^{-1}$ raised to the power of $2$.

\subsection{Paper organization}

\noindent Given the overview in the previous sections on amplification of the optimal value through concave functions, [10], previous work of the author on parallel repetition of multiplayer Quantum games, [21], which is related to nonlocality and SDPs, [18,19,20,23], we state the main results in the next section. By and large these results qualitatively describe how more general families of concave functions are expected to constrain the decay of the optimal value under parallel repetition. Despite the fact that estimates of decay for the optimal value have previously been generalized by the author in the multiplayer setting, [21], the absence of dependency-breaking variables in the forthcoming arguments is of independent interest to formalize. Being able to formulate arguments independently of the dependency-breaking variable illustrates the role that families of concave functions in assume in a wide variety of multiplayer settings.

\subsection{Game-theoretic objects}

\noindent We define the following objects.

\subsubsection{Amplification of the winning probability through concave functions}

\noindent We draw the attention of the reader to a previous use of concave functions in two-player Quantum games [10].

\bigskip

\noindent \textbf{Theorem} (\textbf{Theorem} \textit{1}, [10], \textit{amplification of the winning probability with two concave functions}). Fix two probability distributions $\mathcal{X}, \mathcal{Y}$ with $X \sim \mathcal{X}$ and $Y \sim \mathcal{Y}$ and $\epsilon \neq \epsilon^{\prime}, \delta \neq \delta^{\prime} \in \big[ 0, 1 \big]$. For $\mu : \mathcal{X} \times \mathcal{Y} \longrightarrow \big[ 0 , 1 \big]$ and monotonically increasing concave functions $\psi$ and $\psi^{\prime}$ such that,

\begin{align*}  
\textbf{E}_X \big[   \psi \big[ \textbf{E}_Y \big[  \mu \big( X , Y \big)  \big]   \big]        \big]  \leq   \epsilon  \psi  \big[ \delta \big]           , \\ \\\textbf{E}_Y \big[    \psi^{\prime} \big[  \textbf{E}_X \big[ \mu \big( X , Y \big) \big]  \big]   \big]  \leq  \epsilon^{\prime} \psi^{\prime}  \big[ \delta \big]      , 
\end{align*}

\noindent which respectively correspond to amplification inequalities for Alice and Bob. Then the joint expectation over $XY$ satisfies,

\begin{align*}
   \textbf{E}_{XY} \big[  \mu \big( X , Y \big) \big]       \leq \epsilon \epsilon^{\prime} + \delta + \delta^{\prime} . 
\end{align*}

\noindent \textbf{Theorem} (\textbf{Theorem} \textit{2}, [10], \textit{the joint expectation with respect to probability distributions over $\mathcal{X}$ is upper bounded from a product of $\epsilon$ factors and a summation of $\delta$ factors}). Fix $i \in \big[ n \big]$. For any function $\mu : \mathcal{X}^n \longrightarrow \big[ 0 , 1 \big]$, sequence of random variables $\big\{ X_i \big\}_{1\leq i \leq C}$ where $X_i \sim \mathcal{X}$ for every $i$, and family $\big\{ \psi_i \big\}_{1\leq i \leq C}$ of monotonically increasing functions,

\begin{align*}
  \textbf{E}_{X_i} \big[   \psi_i \big[   \textbf{E}_{X_1 \times \cdots \times X_n} \big[ \mu \big( X_1 , \cdots , X_n \big) \big]   \big]      \big] \leq \epsilon_i \psi \big[ \delta_i \big]   , 
\end{align*}

\noindent for some $\delta_i \neq \epsilon_i \in \big[ 0 , 1 \big]$. Then one has that,

\begin{align*}
    \textbf{E}_{X_1 \times \cdots \times X_n} \big[  \mu \big( X_1 , \cdots , X_n \big)     \big] \leq \underset{i \in [ n ]}{\prod} \epsilon_i + \underset{i \in [ n ]}{\sum}  \delta_i   . 
\end{align*}

\noindent \textbf{Theorem} (\textbf{Theorem} \textit{3}, [10], \textit{amplification of the winning probabilities of Alice and Bob from two concave functions}). Let $\mu : \mathcal{X} \times \mathcal{Y} \longrightarrow \big[ 0 , 1 \big]$ be any function and $X$ and $Y$ as probability distributions over $\mathcal{X}$ and $\mathcal{Y}$, respectively. Moreover let $\psi$ and $\psi^{\prime}$ be two monotonically increasing, concave functions on $\big[0,1\big]$. Suppose that,

{\small \begin{align*}
  \textbf{E}_X \big[    \psi \big[ \textbf{E}_Y \big[   \mu \big( X , Y \big)  \big] \big] \big] \leq  \epsilon \psi \big[ \delta \big]   , \\ \\   \textbf{E}_Y \big[     \psi^{\prime} \big[   \textbf{E}_X \big[  \mu \big( X , Y \big)   \big] \big]       \big] \leq    \epsilon^{\prime} \psi^{\prime} \big[ \delta^{\prime} \big]       , 
\end{align*} }

\noindent for some $\delta \neq \delta^{\prime}, \epsilon \neq \epsilon^{\prime} \in \big[ 0 , 1 \big]$. Then one has that,

\begin{align*}
   \textbf{E}_{XY } \big[   \mu \big( X , Y \big) \big] \leq         \mathrm{max} \big\{ \epsilon \epsilon^{\prime} , \epsilon \delta + \epsilon^{\prime} \delta^{\prime}  \big\}     .
\end{align*}

\noindent \textbf{Corollary} (\textbf{Corollary} \textit{4}, [10], \textit{optimality within a factor of $2$}). Fix functions $\mu$, $\psi$ and $\psi^{\prime}$ as denoted with the previous result above. Suppose that,

{\small \begin{align*}
 \textbf{E}_X \big[ \psi \big[  \textbf{E}_Y      \big]  \big] \leq  \epsilon \psi \big[  \epsilon^{\prime} \big] , \\ \\     \textbf{E}_Y \big[ \psi^{\prime} \big[ \textbf{E}_X  \big] \big]  \leq     \epsilon^{\prime} \psi^{\prime} \big[ \epsilon \big]      , 
\end{align*} } 

\noindent then one has that,

\begin{align*}
    \textbf{E}_{XY} \big[ \mu \big( X , Y \big) \big]   \leq 2 \epsilon \epsilon^{\prime} . 
\end{align*}

\noindent \textbf{Proposition} (\textbf{Proposition} \textit{1}, [10], \textit{deviation away from optimality}). Fix functions $\mu$, $\psi$, $\psi^{\prime}$ and $\epsilon \neq \epsilon^{\prime} \in \big[ 0 , 1 \big]$ as denoted with the previous result above. Suppose that,

{\small \begin{align*}
  \textbf{E}_X \big[ \psi \big[   \textbf{E}_Y \big[ \mu \big( X , Y \big) \big]      \big] \big]  \leq  \epsilon \psi \big[ \epsilon^{\prime} \big]        , \\ \\  \textbf{E}_Y \big[      \psi^{\prime} \big[ \textbf{E}_X \big[ \mu \big( X , Y \big)  \big]  \big]    \big]   \leq \epsilon^{\prime} \psi^{\prime} \big[ \epsilon  \big]    , 
\end{align*} }

\noindent then one has that,

\begin{align*}
  \textbf{E}_{XY} \big[ \mu \big( X , Y \big)  \big]  > \epsilon \epsilon^{\prime} .
\end{align*}

\bigskip

\noindent \textbf{Corollary} (\textbf{Corollary} \textit{2}, [10], \textit{upper bounding the expected value of $\mu \big( X , Y \big)$ with the square of the pullback of a concave function}). Fix $\mu$, and $\psi$ as denoted with previous results above. For the function $\xi \big( x \big) \equiv x \psi \big( x \big)$, one has that,

\begin{align*}
  \textbf{E}_{XY } \big[ \mu \big( X , Y \big) \big] \leq 4 \cdot \xi^{-1} \bigg[ \frac{1}{2}  \mathrm{sup}  \big\{ \textbf{E}_X \big[ \psi \big[ \textbf{E}_Y  \big[ \mu \big( X , Y \big)  \big] \big] \big] ,    \textbf{E}_Y \big[ \psi \big[ \textbf{E}_X  \big[ \mu \big( X , Y \big)  \big] \big] \big]         \big\}    \bigg]^2   .
\end{align*}

\subsubsection{The optimal value under parallel repetition}

\noindent \textbf{Theorem} (\textbf{Theorem} \textit{5.6}, [3], \textit{polynomial function for the rate of decay of the optimal value for the $\alpha$-anchored $k$-player game}). For an $\alpha$-anchored game $G$ such that the optimal value of the game is $\leq 1 - \epsilon$ for $\epsilon$ taken to be sufficiently small, for a game with at least $n > 1$ players,

\begin{align*}
    \big[ 1 - \frac{\gamma^9}{2} \big]^{c \alpha^{8k} \frac{n}{s}}     \text{,}
\end{align*}

\noindent corresponds to the probability of winning more than $\big( 1 - \epsilon + \gamma \big) n$ games, for $s$ being the length of all responses from each participant, and for a universal constant, $c$.

\bigskip

\noindent \textbf{Theorem} (\textbf{Theorem} \textit{4}, [20], \textit{sharpening the exponential rate of decay for parallel repetition of the optimal value}). Fix $\epsilon_{\mathrm{Multiplayer}}$ strictly positive, $s_{\mathrm{Multiplayer}} \equiv \mathrm{max} \big\{ \underset{1 \leq i \leq N}{\prod} \mathrm{log} \big| \mathcal{Q}_i \big| ,1\big\}$, $0 < \alpha_{\mathrm{Multiplayer}} \leq 1$, a multiplayer $\alpha$-anchored game $G$, $\epsilon_{\mathrm{Multiplayer}}$ strictly positive, and universal constant $c_{\mathrm{Multiplayer}}$, where,

\begin{align*}
  0 < c_{\mathrm{Multiplayer}} < \frac{1}{N^{2N} \mathrm{log} \big( e \big)}  .
\end{align*}

\noindent One has that,

{\small \begin{align*}
   \omega_{\mathrm{Multiplayer}} \big( \big( G_{\bot} \big)    \big)^{\otimes n} \leq  \frac{10}{\epsilon_{\mathrm{Multiplayer}}} \mathrm{exp} \big[  -  \frac{c_{\mathrm{Multiplayer}} \alpha^{20N + 1}_{\mathrm{Multiplayer}}        \epsilon^{6N}_{\mathrm{Multiplayer}}  n }{s_{\mathrm{Multiplayer}}  }\big]          \text{, }
\end{align*} }

\noindent under the assumption that,

{\small \begin{align*}
    \omega_{\mathrm{Multiplayer}} \big( \big( G_{\bot} \big)    \big)^{\otimes n}   \precsim  \mathrm{exp} \big[  -  \frac{c_{\mathrm{Multiplayer}} \alpha^{20N + 1}_{\mathrm{Multiplayer}}        \epsilon^{6N}_{\mathrm{Multiplayer}}  n }{s_{\mathrm{Multiplayer}}  }\big]         \text{. }
\end{align*} }

\bigskip

\begin{itemize}
\item[$\bullet$] \textit{Optimal values for expanded games under anchoring}, [2,3,4]. Fix $0 < \alpha < 1$. Then the optimal value, $\omega^*$, of an expanded game $G$ under parallel repetition,  in comparison to the original optimal value, $\omega$, of $G$ before performing parallel repetition, and subsequently, anchoring, satisfies,

\begin{align*}
  \omega^{*} \big( G_{\bot} \big) = 1 - \big( 1 - \alpha \big)^2 \big[ 1 - \omega^* \big( G \big) \big]     \text{, }
\end{align*}

\noindent for the $\alpha$-anchored game, $G_{\bot}$.

\bigskip

\item[$\bullet$] \textit{Parallel repetition of} $\alpha$-\textit{anchoring}. Fix the same choice of $\alpha$ in the previous result, as well as $\epsilon$ sufficiently small so that $\omega^{*} \big( G_{\bot} \big) < 1 - \epsilon$. After providing the estimate described in the previous item, the authors of [2,3,4,30,56] also describe how performing the parallel repetition operation of $\alpha$-anchoring impacts the estimate. In particular, parallel repetition of $\alpha$-anchored expanded games yields the following exponentially decaying estimate,

{\small \begin{align*}
 \omega^{*} \big( G^n_{\bot} \big) \leq \frac{4}{\epsilon}         \mathrm{exp} \big[  - \frac{c \alpha^{48} \epsilon^{17} n}{s}   \big]    \text{, }
\end{align*} } 

\noindent for $s \equiv \mathrm{sup} \big\{ \mathrm{log} \big( \big| \mathcal{A} \mathcal{B} \big| , 1 \big\}$, ie the supremum of the natural logarithm of the product of the alphabets from Alice and Bob and $1$, as well as strictly positive $c$.

\item[$\bullet$] \textit{Generalizations of the previous item for expressions of optimal values under anchoring}. Following the authors of [2,3,4], the authors of [30] argued that a generalization of the anchored parallel repetition on the optimal value satisfies,

{\small \begin{align*}
  \omega^* \big( {\widetilde{G}}^k \big) = \big[ 1 - \big[ 1 - \omega^* \big( \widetilde{G} \big) \big]^5 \big]^{\Omega ( k ( \mathrm{log} ( | \mathcal{A} | | \mathcal{B} | ) )^{-1} )}  \text{, }
\end{align*} }

\noindent for the sets, and strictly positive parameter,

{\small \begin{align*}
 \mathcal{A} \equiv \textit{Set of answers from first player}   \text{, } \end{align*}
 
 \begin{align*} \mathcal{B} \equiv \textit{Set of answers from second player}  \text{, } \end{align*}
 
 \begin{align*}  k \equiv \textit{Number of parallel repetition operations} > 0  \text{, } \\ 
\end{align*}
 }

\noindent as well as the space of functions,

\begin{align*}
  \Omega \equiv \underset{a \in \mathcal{A}, b \in \mathcal{B} }{\bigcup} \big\{ \text{functions} \big( a , b \big) \big\}  \equiv \text{functions} \big( \mathcal{A} , \mathcal{B} \big)         \text{. }
\end{align*}

\end{itemize}

\subsubsection{Dependency breaking relations}

\noindent Denote the collection of unitaries,

{\small \[  \left\{\!\begin{array}{ll@{}>{{}}l} 
 \mathcal{U}_{q_1} \equiv \text{\textit{First player unitary}}   \text{, } \\  \\ \mathcal{U}_{q_2} \equiv \text{\textit{Second player unitary}}  \text{, } \\ \vdots \\ \mathcal{U}_{q_N} \equiv \text{\textit{N th player unitary}}  \text{, } \\ \\ 
\end{array}\right. \equiv   \left\{\!\begin{array}{ll@{}>{{}}l} 
 \mathcal{U}_1 \equiv \text{\textit{First player unitary}}   \text{, } \\  \\ \mathcal{U}_2 \equiv \text{\textit{Second player unitary}}  \text{, } \\ \vdots \\ \mathcal{U}_N \equiv \text{\textit{N th player unitary}}  \text{, }  \\ \\ 
\end{array}\right. 
\] }

\noindent introduced for each player, as well as the collection of dependency-breaking unitaries,

{\small \[   \left\{\!\begin{array}{ll@{}>{{}}l} 
 \mathcal{U}_{1,r_{-i}} \equiv \text{\textit{First player dependency-breaking unitary}}   \text{, } \\  \\ \mathcal{U}_{2,r_{-i}} \equiv \text{\textit{Second player  dependency-breaking unitary}}  \text{, } \\ \vdots \\ \mathcal{U}_{N,r_{-i}} \equiv \text{\textit{N th player dependency-breaking unitary}}  \text{. } 
\end{array}\right. 
\]  }

\begin{itemize}
  \item[$\bullet$] \textit{Constructing statistically close conditional probabilities with responses from the anchored set}. For,

  \begin{align*}
     \Omega_{\mathrm{Multiplayer}} \equiv \textit{Action space of a Multiplayer Quantum game} ,
  \end{align*}

  \noindent one has that,

\begin{align*}
     \textbf{P} \big[ \Omega_{\mathrm{Multiplayer}} \big| \textbf{X}_{1,2,\cdots,N-1} \equiv x \backslash \big\{ N \big\} , \textbf{X}_N \equiv \bot ,  W_C  \big]  \approx       \textbf{P} \big[ \Omega_{\mathrm{Multiplayer}} \\ \big| \textbf{X}_{1,2,\cdots,N-2} \equiv x \backslash \big\{ N -1 , N \big\} , \textbf{X}_{N-1} \equiv \bot  , \textbf{X}_N \equiv \bot ,  W_C  \big] \\ \vdots \\  \approx  \textbf{P} \big[ \Omega_{\mathrm{Multiplayer}}  \big| \textbf{X}_{1 } \equiv x \backslash \big\{ 2, \cdots , N -1 , N \big\} , \textbf{X}_2 \dots \equiv \textbf{X}_N \equiv \bot ,  W_C  \big]         \text{. }
\end{align*}

\noindent Iterating further along the lines of the observations above, one also has that,

\begin{align*}
        \textbf{P} \big[ \Omega_{\mathrm{Multiplayer}}  \big| \textbf{X}_{1 } \equiv x \backslash \big\{ 2, \cdots , N -1 , N \big\} , \textbf{X}_2 \dots \equiv \textbf{X}_N \equiv \bot ,  W_C  \big] \\  \\  \approx    \textbf{P} \big[ \Omega_{\mathrm{Multiplayer}}  \big| \textbf{X}_{1 } \equiv x \backslash \big\{ 2, \cdots , N -1 , N \big\} , \textbf{X}_3 \dots \equiv \textbf{X}_N \equiv \bot ,  W_C  \big]   \\ \vdots \\ \approx   \textbf{P} \big[ \Omega_{\mathrm{Multiplayer}}  \big| \textbf{X}_{1 } \equiv x \backslash \big\{ 2, \cdots , N -1 , N \big\} , \textbf{X}_N \equiv \bot ,  W_C  \big]         \text{. }
\end{align*}

\item[$\bullet$] \textit{Putting it all together: Relating the conditional probability with anchored questions to each other}. Altogether, one has,

\begin{align*}
 \big\{  \textbf{P} \big[ \Omega_{\mathrm{Expanded}} \big| \textbf{X}_i \equiv x, W_C \big] \approx \textbf{P} \big[ \Omega_{\mathrm{Expanded}} \big| \textbf{X}_i \equiv x, \textbf{Y}_i \equiv y,  W_C  \big] \big\}   \Longrightarrow  \big\{   \textbf{P} \big[ \Omega_{\mathrm{Multiplayer}} \\  \big| \textbf{X}_{1 } \equiv x \backslash \big\{ 3,  \cdots  , N -1 , N \big\} , \textbf{X}_{N-1} \equiv  \textbf{X}_N \equiv \bot ,  W_C  \big]   \\  \\     \approx  \textbf{P} \big[ \Omega_{\mathrm{Multiplayer}}  \big| \textbf{X}_{1 } \equiv x \backslash \big\{ 2, \cdots , N -1 , N \big\} , \textbf{X}_N \equiv \bot ,  W_C  \big]  \big\}      \text{. }
\end{align*}

\end{itemize}

\noindent For the following result below, denote,

{\small \begin{align*}
\underset{R_{-i}| W_C}{\textbf{E}} \big[ \cdot \big]   \equiv \textit{Expectation of a dependency breaking} \\ \textit{variable $R$ occurring at coordinate $i$, conditionally} \\ \textit{upon $W_C$} , \end{align*}

\begin{align*} \underset{X}{\textbf{E}} \big[ \cdot \big]  \equiv \textit{Expectation of the random variable $X$} ,  \\ \\  \underset{Y}{\textbf{E}} \big[ \cdot \big]   \equiv \textit{Expectation of the random variable $Y$}  , \\ \\ \underset{XY}{\textbf{E}} \big[ \cdot \big]   \equiv \textit{Joint expectation of the random vari-} \\ \textit{able $XY$}   . 
\end{align*} }

\bigskip

\noindent \textbf{Lemma} (\textbf{Lemma} \textit{5.12}, [2,3], \textit{computing expected values of dependency breaking variables conditionally upon winning in a fixed coordinate}). For all $r_{-i}$, $x$, and $y$, there exists unitaries $U_{r_{-i}.x}$ acting on $E_A$, along with unitaries $V_{r_{-i},y}, V_{r_{-i},x,y}$ acting on $E_B$ such that with probability at least $1 - \mathrm{O
} \big( \delta^{\frac{1}{16}} \big)$, for $\delta$ taken to be sufficiently small, over the choice of uniformly random $\big\{ i \in [n] \backslash C \big\}  \equiv  \big\{ i \in \big\{ 1, \cdots , n \big\} \backslash C \big\}$,

{\small \begin{align*}
  \underset{R_{-i}| W_C}{\textbf{E}} \underset{X}{\textbf{E}} \big| \big|  \big( U_{r_{-i}, x } \otimes \textbf{I} \big) \ket{\Psi_{r_{-i}, \bot, \bot}} - \ket{\Psi_{r_{-i}, x, \bot}}     \big| \big|   = \mathrm{O} \bigg(  \frac{\delta^{\frac{1}{16}}}{\alpha^{\frac{5}{4}}} \bigg)     \text{,} \\ \\  \underset{R_{-i}| W_C}{\textbf{E}} \underset{Y}{\textbf{E}} \big| \big|  \big(\textbf{I} \otimes    V_{r_{-i}, x }  \big) \ket{\Psi_{r_{-i}, \bot, \bot}} - \ket{\Psi_{r_{-i}, \bot, y}}     \big| \big|   = \mathrm{O} \bigg(  \frac{\delta^{\frac{1}{16}}}{\alpha^{\frac{5}{4}}} \bigg)    \text{, } \end{align*}
  
  \begin{align*} \underset{R_{-i}| W_C}{\textbf{E}} \underset{XY}{\textbf{E}} \big| \big|  \big(\textbf{I} \otimes    V_{r_{-i}, x,y }  \big) \ket{\Psi_{r_{-i}, \bot \backslash x, y}} - \ket{\Psi_{r_{-i}, \bot, \bot}}     \big| \big|        = \mathrm{O} \bigg(  \frac{\delta^{\frac{1}{16}}}{\alpha^{\frac{5}{4}}} \bigg)     \text{. } \\ 
\end{align*} }

\noindent  \textbf{Lemma} (\textbf{Lemma} \textit{5.5}, [2,3,4], \textit{big} $\mathrm{O}$ \textit{values of the expected value for games with three participants under $\alpha$-anchoring, a generalization of ordinary anchoring}). For all $\big( x,y,z \big) \in \mathcal{X} \times \mathcal{Y} \times \mathcal{Z}$, namely the set of all possible questions for all three players, and the number of parallel repetition operations $k$, there exists unitaries $U_x, V_y, W_z$ acting on $E_A, E_B, E_C$, respectively, such that,

\begin{align*}
   \underset{XYZ}{\textbf{E}} \big| \big| \big( U_x \otimes V_y \otimes W_z \big) \ket{\Psi_{\bot,\bot,\bot}} - \ket{\Psi_{x,y,z}}\big| \big|  = \mathrm{O} \bigg( \frac{\delta^{\frac{1}{4}}}{\alpha^{2k}} \bigg)        \text{. }
\end{align*}

\noindent For the following result below, denote,

{\small \begin{align*}
  W_C \equiv \big\{ \text{Players win the expanded game for all coordinates } C \big\}   \text{,}   \end{align*}
  
  \begin{align*} \underset{\mathcal{Q}_1}{\textbf{E}} \big[ \cdot \big]  \equiv \textit{Expectation of the random variable $\mathcal{Q}_1$} ,  \\ \\    \underset{\mathcal{Q}_2}{\textbf{E}} \big[ \cdot \big]  \equiv \textit{Expectation of the random variable $\mathcal{Q}_2$} , \\ \vdots  \\ \underset{\mathcal{Q}_N}{\textbf{E}} \big[ \cdot \big] \equiv \textit{Expectation of the random variable $\mathcal{Q}_N$} , \\ \vdots \end{align*}
  
  \begin{align*}   \underset{\mathcal{Q}_1 \times \cdots \times \mathcal{Q}_N}{\textbf{E}} \big[ \cdot \big] \equiv \textit{Joint expectation of the random variable $\mathcal{Q}_1 \times$} \\  \textit{$\cdots \times \mathcal{Q}_N$} , \end{align*}
  
  \begin{align*}    \ket{\psi_{r_{-i}}, \bot , \cdots, \bot} \equiv \textit{Dependency-breaking strategy for a Quantum} \\ \textit{multiplayer game with $r_{-i} \sim R_{-i}$ for the first coordinate and } \\ \textit{anchoring $\bot$ for the remaining coordinates} , \\ \vdots \\    \ket{\psi_{\bot,\cdots,\bot,r_{-i}}} \equiv \textit{Dependency-breaking strategy for a Quantum multiplayer} \\ \textit{game with $r_{-i} \sim R_{-i}$ for the last coordinate and } \\ \textit{anchoring $\bot$ for the remaining coordinates}   , \end{align*}
  
  \begin{align*}     \ket{\psi_{r_{-i}, q_1 \backslash \bot, \bot, \cdots , \bot}} \equiv   \textit{Dependency-breaking strategy for a Quantum multiplayer game with} \\ \textit{ $r_{-i} \sim R_{-i}$ for the first coordinate, $q_1$ or $\bot$ with probabilities $p_1 > p_2$, where $0 < p_2 < $} \\ \textit{$p_1 < 1$ for the second coordinate, and anchoring $\bot$ for the  remaining coordinates} , \\          \vdots \\ \ket{\psi_{r_{-i}, q_1 \backslash \bot , q_2 , \cdots, q_N}} \equiv        \textit{Dependency-breaking strategy for a Quantum multiplayer game with} \\ \textit{ $r_{-i} \sim R_{-i}$ for the first coordinate, $q_1$ or $\bot$ with probabilities $p_1 > p_2$, where $0 < p_2 < $} \\ \textit{$p_1 < 1$ for the second coordinate, and $q_2, \cdots, q_N$ for the remaining coordinates} . \\  
\end{align*} }

\noindent \textbf{Lemma} (\textbf{Lemma} \textit{1}, [20], \textit{computing expectation values with respect to suitable unitaries, as provided in the system of relations} $\mathscr{E}$). For all $\mathscr{r}_{-i}$, $q_1, \cdots, q_N$, there exists unitaries  $\mathcal{U}_1, \cdots, \mathcal{U}_N \equiv \mathcal{U}_{1,\mathscr{r}_{-i}}, \cdots, \mathcal{U}_{N,\mathscr{r}_{-i}}$, such that,

{\small \[   \left\{\!\begin{array}{ll@{}>{{}}l} 
   \underset{R_{-i} | W_C}{\textbf{E}} \underset{\mathcal{Q}_1}{\textbf{E}}       \big| \big| \big( \mathcal{U}_{1,r_{-i}} \otimes \textbf{I}^{\otimes N-1} \big) \ket{\psi_{r_{-i},\bot , \cdots , \bot }}   - \ket{\psi_{r_{-i},q_1 , \bot \cdots , \bot}}    \big| \big| \text{, } \\   \\  \underset{R_{-i} | W_C}{\textbf{E}} \underset{\mathcal{Q}_2}{\textbf{E}}      \big| \big|\big( \textbf{I} \otimes \mathcal{U}_{2,r_{-i}} \otimes \textbf{I}^{\otimes N-2} \big)\ket{\psi_{r_{-i},\bot , \cdots , \bot }}    -  \ket{\psi_{r_{-i}, \bot , q_2, \bot , \cdots , \bot}}    \big| \big|\text{, } \\ \vdots 
\\    \underset{R_{-i} | W_C}{\textbf{E}} \underset{\mathcal{Q}_N}{\textbf{E}}     \big| \big| \big( \textbf{I}^{\otimes N-1 } \otimes \mathcal{U}_{N,r_{-i}}  \big) \ket{\psi_{r_{-i},\bot , \cdots , \bot }}  - \ket{\psi_{r_{-i},\bot , \cdots , \bot,  q_N }}    \big| \big| \text{, }  \\  \underset{R_{-i} | W_C}{\textbf{E}} \underset{\mathcal{Q}_1 \mathcal{Q}_2 }{\textbf{E}}    \big| \big| \big( \textbf{I}^{\otimes N-1 } \otimes \mathcal{U}_{N,r_{-i}}  \big) \ket{\psi_{r_{-i},\bot \backslash q_1 , q_2, \cdots, q_N}}  - \ket{\psi_{r_{-i},\bot \backslash q_1 ,\bot, q_3, \cdots, q_N}}   \big| \big|  \text{, }
\\  \vdots \\ \underset{R_{-i} | W_C}{\textbf{E}} \underset{\mathcal{Q}_1 \times \cdots \times \mathcal{Q}_N}{\textbf{E}}    \big| \big|\big( \textbf{I}^{\otimes N-1 } \otimes \mathcal{U}_{N,r_{-i}}  \big) \ket{\psi_{r_{-i}, q_1 \backslash \bot, q_2 , \cdots ,  q_N} } \\  - \ket{\psi_{r_{-i}, q_1 \backslash \bot, q_2 \backslash \bot, \cdots , q_{N-1} \backslash \bot, q_N} }  \big| \big| , \\ \\ 
\end{array}\right. 
\] } 

   

\noindent for dependency-breaking unitaries,

{\small \[   \left\{\!\begin{array}{ll@{}>{{}}l} 
 \mathcal{U}_{1,r_{-i}} \equiv \text{\textit{First player  dependency-breaking unitary}}   \text{, } \\  \\ \mathcal{U}_{2,r_{-i}} \equiv \text{\textit{Second player dependency-breaking unitary}}  \text{, } \\ \vdots \\ \mathcal{U}_{N,r_{-i}} \equiv \text{\textit{N th player dependency-breaking unitary}}  \text{, } 
\end{array}\right. 
\]  }

\noindent where,

{\small \begin{align*}
   \ket{\psi_{r_{-i},\bot , \cdots , \bot }}  , \ket{\psi_{r_{-i},q_1 , \bot \cdots , \bot}}  ,   \ket{\psi_{r_{-i}, \bot , q_2, \bot , \cdots , \bot}}  ,  \ket{\psi_{r_{-i},q_1 , \bot \cdots , \bot}}   , \ket{\psi_{r_{-i},\bot , \cdots , \bot,  q_N }}  ,   \ket{\psi_{r_{-i},\bot \backslash q_1 , q_2, \cdots, q_N}} \\  ,  \ket{\psi_{r_{-i},\bot \backslash q_1 ,\bot, q_3, \cdots, q_N}} ,    \cdots      ,   \ket{\psi_{r_{-i}, q_1 \backslash \bot, q_2 , \cdots ,  q_N} }  ,   \ket{\psi_{r_{-i}, q_1 \backslash \bot, q_2 \backslash \bot, \cdots , q_{N-1} \backslash \bot, q_N}}             \in \ket{\Psi^{\prime}}  \text{, } \\ 
\end{align*} } 

\noindent given the collection of states,

\begin{align*}
 \ket{\Psi^{\prime}} \equiv \underset{1 \leq i \leq N}{\underset{\textit{dependency-breaking on answer from the i th player}}{\bigcup}} \ket{\Psi^{\prime}_i}  \text{. }
\end{align*}

\noindent For the following result below, denote,

\begin{align*}
  \delta \equiv \textit{First Quantum multiplayer threshold} > 0   ,  \\ \\  \alpha \equiv \textit{Second Quantum multiplayer threshold}  > 0   , 
\end{align*}

\noindent subject to the condition that $\delta > \alpha$.

\bigskip

\noindent \textbf{Lemma} (\textbf{Lemma} \textit{4.6}, [3], \textit{determining the value of the multiplayer $\delta$ threshold for differences of probabilities}). For,

\begin{align*}
  W_C \equiv \big\{ \text{Players win the expanded game for all coordinates } C \big\}   \text{,}
\end{align*}

\noindent one has that:

\begin{itemize}
\item[$\bullet$] \textit{The average of the probabilities of sampling dependency-breaking variables is strictly upper bounded by a fraction of $\delta_{\mathrm{Multiplayer}}$}:

{\tiny  \begin{align*}
\frac{1}{N} \underset{1 \leq i \leq N}{\sum}  \big| \big| \textbf{P} _{\Omega_i \mathcal{Q}_1 \times \cdots \times \mathcal{Q}_N | W_C} - \textbf{P}_{\Omega_i \mathcal{Q}_1 \times \cdots \times \mathcal{Q}_N}  \big| \big|       \leq \sqrt{\delta^{\frac{N}{300}}_{\mathrm{Multiplayer}}} .
\end{align*}}

\item[$\bullet$] \textit{The average of the probabilities of sampling $R$ is strictly upper bounded by a fraction of $\delta_{\mathrm{Multiplayer}}$}:

{\tiny  \begin{align*}
   \frac{1}{N} \underset{1 \leq i \leq N}{\sum}  \big| \big| \textbf{P}_{R \mathcal{Q}_1 \times \cdots \times \mathcal{Q}_N | W_C} - \textbf{P}_{R \mathcal{Q}_1 \times \cdots \times \mathcal{Q}_N}  \big| \big|       \leq \sqrt{\delta^{\frac{N}{300}}_{\mathrm{Multiplayer}}} .
\end{align*} }

\item[$\bullet$] \textit{The average of the probabilities of sampling $R$, conditioned upon all player responding with anchored responses $\bot$, is strictly upper bounded by a fraction of $\frac{\delta_{\mathrm{Multiplayer}}}{\alpha_{\mathrm{Multiplayer}}}$}:

{\tiny \begin{align*}
   \frac{1}{N} \underset{1 \leq i \leq N}{\sum}  \big| \big| \textbf{P}_{\Omega_i | W_C} \textbf{P}_{R_{-i} | \mathcal{Q}_1 \equiv \bot, \cdots, \mathcal{Q}_N \equiv \bot , W_C} -  \textbf{P}_{\Omega_i | W_C} \textbf{P}_{R_{-i} | \Omega_{-i} W_C }  \big| \big|        \leq  \mathrm{O} \bigg(  \frac{\sqrt{\delta^{\frac{N}{300}}_{\mathrm{Multiplayer}}}}{\alpha^{N+1}_{\mathrm{Multiplayer}}} \bigg)   . \\ 
\end{align*} }

\item[$\bullet$] \textit{The average of the probabilities of sampling $\mathcal{Q}_1, \cdots, \mathcal{Q}_N$, conditioned upon all player responding with anchored responses $\bot$, is strictly upper bounded by a fraction of $\frac{\delta_{\mathrm{Multiplayer}}}{\alpha_{\mathrm{Multiplayer}}}$}:

{\tiny  \begin{align*}
   \frac{1}{N} \underset{1 \leq i \leq N}{\sum}  \big| \big| \textbf{P}_{\Omega_i | W_C} \textbf{P}_{R_{-i} | \mathcal{Q}_1 \equiv \bot, \cdots, \mathcal{Q}_N \equiv \bot , W_C} -  \textbf{P}_{\mathcal{Q}_1 \times \cdots \times \mathcal{Q}_N | W_C} \textbf{P}_{R_{-i} | \mathcal{Q}_1 \times \cdots \times \mathcal{Q}_N W_C }  \big| \big|       \leq                    \mathrm{O} \bigg(  \frac{\sqrt{\delta^{\frac{N}{300}}_{\mathrm{Multiplayer}}}}{\alpha^{N+1}_{\mathrm{Multiplayer}}} \bigg)   . \\ 
\end{align*} }

\end{itemize}

\noindent For the following results below, denote,

{\small \[   \left\{\!\begin{array}{ll@{}>{{}}l}  
  \textbf{1} \equiv \textit{First player's quantum state} , \\ \textbf{2} \equiv \textit{Second player's quantum state} ,  \\ \vdots  \\  \textbf{N} \equiv \textit{N th player's quantum state} ,    \\ 
\end{array}\right. 
\] }

\noindent corresponding to the collection of players,

{\small   \[  \mathscr{E} \equiv  \left\{\!\begin{array}{ll@{}>{{}}l} 
   \underset{R_{-i} | W_C}{\textbf{E}} \underset{\mathcal{Q}_1}{\textbf{E}}       \big| \big|  \big[ \big( \mathcal{U}_1 \bigotimes \textbf{I}^{\otimes N-1} \big) - \textbf{I} \big] \ket{\psi}    \big| \big|\text{, } \\   \\  \underset{R_{-i} | W_C}{\textbf{E}} \underset{\mathcal{Q}_2}{\textbf{E}}      \big| \big|  \big[ \big( \textbf{I} \bigotimes \mathcal{U}_2 \bigotimes \textbf{I}^{\otimes N-2} \big) - \textbf{I} \big] \ket{\psi}    \big| \big|\text{, } \\ \vdots \\    \underset{R_{-i} | W_C}{\textbf{E}} \underset{\mathcal{Q}_N}{\textbf{E}}     \big| \big|  \big[ \big( \textbf{I}^{\otimes N-1 } \bigotimes \mathcal{U}_N  \big) - \textbf{I} \big] \ket{\psi}   \big| \big|  \text{, } \\ \\ \underset{R_{-i} | W_C}{\textbf{E}} \underset{\mathcal{Q}_1 \mathcal{Q}_2 }{\textbf{E}}     \big| \big|   \big[ \big( \textbf{I}^{\otimes N-1 } \bigotimes \mathcal{U}_N  \big) - \textbf{I} \big] \ket{\psi}    \big| \big| \text{, } \\ \vdots \\   \underset{R_{-i} | W_C}{\textbf{E}} \underset{\mathcal{Q}_1 \times \cdots \times \mathcal{Q}_N}{\textbf{E}}    \big| \big|  \big[ \big( \textbf{I}^{\otimes N-1 } \bigotimes \mathcal{U}_N  \big) - \textbf{I} \big] \ket{\psi}   \big| \big|  , \\ \\ 
\end{array}\right. 
\] }

\noindent corresponding to the collection of conditionally defined expectation values over the dependency breaking variables $R_{-i}$, conditionally upon $W_C$, and also upon $\mathcal{Q}_i$,

\begin{align*}
  \ket{\Psi} \equiv \underset{1 \leq k \leq N}{\underset{\textit{dependency-breaking among the first}\text{ }  k\text{ }  \textit{players}}{\bigcup}} \ket{\Psi_k}     \text{, }
\end{align*}

\noindent corresponding to the strategies of any of the $N$ players that are dependent upon dependency breaking variables introduced along any of the $N$ coordinates,

\[   \left\{\!\begin{array}{ll@{}>{{}}l}  \textit{Dependency-breaking amongst the first player} \equiv \ket{\Psi_1} , \\ \vdots \\  \textit{Dependency-breaking amongst all players} \equiv \ket{\Psi_N} , 
\end{array}\right. 
\]

\noindent corresponding to the collection of dependency-breaking strategies,

{\small \begin{align*}
  \widetilde{\ket{\psi} }  \equiv \frac{\ket{\psi}}{\big| \big| \ket{\psi}\big| \big|} \propto \ket{\psi}  \text{. }
\end{align*} } 

\noindent corresponding to the dependency-breaking strategy at some coordinate $C$ normalized with itself, and,

\begin{align*}
 \Gamma_i \equiv   \bigg\{            \frac{ \big| \big|                                   \textbf{1}_{\ \{         {\textit{the first i th answers are anchored}\} }}       \ket{\psi_{r_{-i}, \bot , \cdots , \bot, q_{i+1} , \cdots, q_N}}         \big| \big|}{\big| \big|  \textbf{1}_{\{          \textit{all answers are anchored}\} }         \ket{\psi_{r_{-i}, \bot , \cdots , \bot }}       \big| \big| }    \bigg\}_
   {1 \leq i \leq N}\text{, } \\ 
\end{align*}

\noindent corresponding to a series of constants, over $1 \leq i \leq N$, for which the first $i$ coordinates, versus all coordinates, of a multiplayer strategy are anchored.

\bigskip

\noindent \textbf{Theorem} (\textbf{Theorem} \textit{1}, [20], \textit{quantifying the largest order 90
of the system of expected values with normalization factors}). For the sequence of normalization factors $\Gamma_i$,

{\small \[  \widetilde{\mathscr{E}_{\Gamma_i}} \equiv  \left\{\!\begin{array}{ll@{}>{{}}l} 
   \underset{R_{-i} | W_C}{\textbf{E}} \underset{\mathcal{Q}_1}{\textbf{E}}       \big| \big|  \big[ \big( \mathcal{U}_1 \bigotimes \textbf{I}^{\otimes N-1} \big) - \Gamma_1  \big] \widetilde{\ket{\psi} }   \big| \big|\text{, } \\   \\  \underset{R_{-i} | W_C}{\textbf{E}} \underset{\mathcal{Q}_2}{\textbf{E}}      \big| \big|  \big[ \big( \textbf{I} \bigotimes \mathcal{U}_2 \bigotimes \textbf{I}^{\otimes N-2} \big) - \Gamma_2 \big] \widetilde{\ket{\psi} }     \big| \big|\text{, } \\ \vdots \\    \underset{R_{-i} | W_C}{\textbf{E}} \underset{\mathcal{Q}_N}{\textbf{E}}     \big| \big|  \big[ \big( \textbf{I}^{\otimes N-1 } \bigotimes \mathcal{U}_N  \big) - \Gamma_N  \big] \widetilde{\ket{\psi} }    \big| \big|  \text{, } \\ \\ \underset{R_{-i} | W_C}{\textbf{E}} \underset{\mathcal{Q}_1 \mathcal{Q}_2 }{\textbf{E}}     \big| \big|  \big[ \big( \textbf{I}^{\otimes N-1 } \bigotimes \mathcal{U}_N  \big) - \Gamma_1 \Gamma_2  \big] \widetilde{\ket{\psi} }   \big| \big| \text{, } \\  \vdots \\ \underset{R_{-i} | W_C}{\textbf{E}} \underset{\mathcal{Q}_1 \times \cdots \times \mathcal{Q}_N}{\textbf{E}}    \big| \big|  \big[ \big( \textbf{I}^{\otimes N-1 } \bigotimes \mathcal{U}_N  \big) - \big[ \underset{1 \leq i \leq N}{\prod} \Gamma_i \big] \big] \widetilde{\ket{\psi} }  \big| \big|  . \\ 
\end{array}\right. 
\]  }



\noindent $\widetilde{\mathscr{E}_{\Gamma_i}} \equiv    \mathrm{O} \bigg(  \frac{\delta^{\frac{N}{300}}_{\mathrm{Multiplayer}}}{\alpha^{N+1}_{\mathrm{Multiplayer}}} \bigg)    $.

\bigskip

\noindent The result below generalizes a computation of a probability measure conditioned upon the occurence of some $r_{-i} \sim R_{-i}$.

\bigskip

\noindent \textbf{Theorem} (\textbf{Theorem} \textit{2}, [20], \textit{quantifying the largest order of the POVM system of expected values}). For the following Positive Operator Valued Measurements (POVMs),

{\small \[   \left\{\!\begin{array}{l}
\mathscr{P}\mathscr{O}\mathscr{V}\mathscr{M}_1 \equiv  \mathrm{Tr} \big[    \big[  \bar{\big( \mathcal{Q}_1 \big)_{q_1} \big( a_1 \big) } \otimes    \bar{\big( \mathcal{Q}_2 \big)_{q_2} \big( a_2 \big)  \big] \widetilde{\Psi_{r_{-i}, q_1,q_2}}}  \big]  \equiv \textbf{P}_{\mathcal{Q}_1\mathcal{Q}_2 \mathcal{Q}_3| r_{-i}, q_1 q_2 q_3 } \big( q_1, q_2 , q_3 \big) \text{, } \\ \\ \mathscr{P}\mathscr{O}\mathscr{V}\mathscr{M}_2 \equiv   \mathrm{Tr} \big[ \big[  \bar{\big( \mathcal{Q}_1 \big)_{q_1} \big( a_1 \big) } \otimes    \bar{\big( \mathcal{Q}_2 \big)_{q_2} \big( a_2 \big)}  \otimes    \bar{\big( \mathcal{Q}_3 \big)_{q_3} \big( a_3 \big)}  \big]   \widetilde{\Psi_{r_{-i}, q_1,q_2}}    \big] \\   \equiv \textbf{P}_{\mathcal{Q}_1\mathcal{Q}_2 \mathcal{Q}_3 \mathcal{Q}_4| r_{-i}, q_1 q_2 q_3 q_4 }  \big( q_1, q_2  , q_3, q_4  \big)  \text{,} \\  \vdots \\   \mathscr{P}\mathscr{O}\mathscr{V}\mathscr{M}_N \equiv   \mathrm{Tr} \big[  \big[  \bar{\big( \mathcal{Q}_1 \big)_{q_1} \big( a_1 \big) } \otimes    \bar{\big( \mathcal{Q}_2 \big)_{q_2} \big( a_2 \big)}  \otimes    \bar{\big( \mathcal{Q}_3 \big)_{q_3} \big( a_3 \big)} \otimes \cdots  \otimes   \bar{\big( \mathcal{Q}_N \big)_{q_N} \big( a_N \big)} \big] \\ \times  \widetilde{\Psi_{r_{-i}, q_1,q_2, \cdots, q_N}} \big] \\   \equiv \textbf{P}_{\mathcal{Q}_1\mathcal{Q}_2 \mathcal{Q}_3 \times \cdots \times \mathcal{Q}_N| r_{-i}, q_1 q_2 q_3 \times \cdots \times q_N} \big( q_1, q_2 , q_3, q_4, \cdots, q_N  \big)  \text{,} \\ \\ 
\end{array}\right. 
\]  }

\noindent $\mathscr{P}\mathscr{O}\mathscr{V}\mathscr{M}_1, \cdots, \mathscr{P}\mathscr{O}\mathscr{V}\mathscr{M}_N$, $ \mathscr{P}\mathscr{O}\mathscr{V} \mathscr{M} \mathscr{E} \equiv \mathrm{O} \bigg( \frac{\sqrt{\sqrt{\delta^{\frac{N}{300}}_{\mathrm{Multiplayer}}}}}{\alpha^{N+5}_{\mathrm{Multiplayer}}} \bigg) $.

\bigskip

\noindent \textbf{Theorem} (\textbf{Theorem} \textit{3}, [20], \textit{quantifying the larges order of the system of expected values with Pinsker's inequality}). Given the quantum states $\textbf{I}, \cdots, \textbf{N}$,

\[ \mathcal{P} \mathcal{I} \equiv   \left\{\!\begin{array}{ll@{}>{{}}l} 
 \underset{I}{\textbf{E}} \underset{R_{-i} | W_C}{\textbf{E}} \underset{\mathcal{Q}_1 \mathcal{Q}_2 \times \cdots \times \mathcal{Q}_N}{\textbf{E}}  \big| \big| \textbf{I}^{E_1}_{r_{-i}, q_1,q_2, \cdots, q_N}  -  \textbf{I}^{E_1}_{r_{-i}, q_1,\bot,q_3, \cdots, q_N} \big| \big|^2_1   , \\ \\ \underset{I}{\textbf{E}} \underset{R_{-i} | W_C}{\textbf{E}} \underset{\mathcal{Q}_1 \mathcal{Q}_2 \times \cdots \times \mathcal{Q}_N}{\textbf{E}}  \big| \big| \textbf{2}^{E_2}_{r_{-i}, q_1,q_2, \cdots, q_N}  -  \textbf{2}^{E_2}_{r_{-i}, \bot, q_2, q_3, \cdots, q_N} \big| \big|^2_1  ,  \\ \vdots \\ \underset{I}{\textbf{E}} \underset{R_{-i} | W_C}{\textbf{E}} \underset{\mathcal{Q}_1 \mathcal{Q}_2 \times \cdots \times \mathcal{Q}_N}{\textbf{E}}  \big| \big| \textbf{N}^{E_N}_{r_{-i}, q_1,q_2, \cdots, q_N}  -  \textbf{N}^{E_N}_{r_{-i}, q_1,\cdots, q_{N-2}, \bot, q_N} \big| \big|^2_1  , \\ \\ 
\end{array}\right. 
\]

\noindent $\mathcal{P}\mathcal{I} \equiv    \mathrm{O} \bigg(  \frac{\sqrt{\delta^{\frac{N}{300}}_{\mathrm{Multiplayer}}}}{\alpha^{2(N+1)}_{\mathrm{Multiplayer}}} \bigg) $.

\bigskip

\noindent \textbf{Theorem} (\textbf{Theorem} \textit{4}, [20], \textit{rate of decay of the optimal value under anchored parallel repetition}). Fix $s_{\mathrm{Multiplayer}} \equiv \mathrm{max} \big\{ \underset{1 \leq i \leq N}{\prod} \mathrm{log} \big| \mathcal{Q}_i \big| ,1\big\}$, $0 < \alpha_{\mathrm{Multiplayer}} \leq 1$, a multiplayer $\alpha$-anchored game $G$, $\epsilon_{\mathrm{Multiplayer}}$ strictly positive, and universal constant $c_{\mathrm{Multiplayer}}$, where,

{\small \begin{align*}
  0 < c_{\mathrm{Multiplayer}} < \frac{1}{N^{2N} \mathrm{log} \big( e \big)}  . \\ 
\end{align*}} 

\noindent The decay of the multiplayer optimal value, $\omega_{\mathrm{Multiplayer}} \big( G \big)$, under $\alpha$-anchored parallel repetition, $\omega_{\mathrm{Multiplayer}} \big( G_{\bot} \big)^{\otimes k} $, satisfies,

{\small \begin{align*}
 \omega_{\mathrm{Multiplayer}} \big( G_{\bot} \big)^{\otimes k} \equiv   \omega_{\mathrm{Multiplayer}} \big( G^k_{\bot} \big) \leq \frac{10}{\epsilon_{\mathrm{Multiplayer}}} \mathrm{exp} \big[  -  \frac{c_{\mathrm{Multiplayer}} \alpha^{20N + 1}_{\mathrm{Multiplayer}}        \epsilon^{6N}_{\mathrm{Multiplayer}}  k }{s_{\mathrm{Multiplayer}}  }\big]  .
\end{align*} }

\subsubsection{Eve's forgery probability}

\noindent Given the fact that amplification results under parallel repetition described in [] are related to verification and hint queries, we describe several results from the author relating to the forgery probability.

\bigskip

\noindent \textbf{Theorem} (\textbf{Theorem} \textit{1}, [24], \textit{Holevo information of weak correlated randomness between Alice and Bob and a unified security threshold}). Let Alice and Bob share weak correlated randomness through the candidate Quantum secret keys $X_A$ and $X_B$, which Eve gathers information about through $E$. Given an authentication tag $T$ that Alice or Bob produce when generating $X_A$, or $X_B$, respectively, there exists a protocol that is composed of the following steps:

\begin{itemize}
    \item[$\bullet$] \textit{Error correction}. Given the candidate secret keys $X_A$ and $X_B$, the probability that Eve can learn a key for forging over the noisy Quantum channel has the upper bound $f_{\mathrm{DP}} \big[ \chi_{E,C} \big[ X_A \textit{ or } X_B \big] \big] + p_{\mathrm{FA},\mathrm{E}}$.

    \bigskip

        \item[$\bullet$] \textit{Privacy amplification}. To decrease Eve's forgery probability as much as possible, Alice and Bob can implement a privacy amplification protocol, for the purposes of obtaining composable security.

\bigskip

            \item[$\bullet$] \textit{Authentication based on hashing from two-universal functions}. Fix a secrecy threshold $\epsilon_S \neq 0$ taken sufficiently small. Alice and Bob can authenticate that the secret key that they wish to agree upon has length $l >0$ satisfying,

                \begin{align*}
                    l   \lesssim n - \chi_E \big[ X_A \textit{ or } X_B] - \mathrm{log} \big[ \epsilon^{-1}_S \big]      .
                \end{align*}
\end{itemize}

\noindent \textbf{Theorem} (\textbf{Theorem} \textit{2}, [24], \textit{the Holevo-gap threshold}). If Alice generates $X_A$ with Bob who wants to generate $X_B$ that they agree upon if $X_A \approx X_B$ at the end of the protocol, from the Holevo gap,

\begin{align*}
  \Delta \equiv H \big[ X_A \textit{ or } X_B \big] - \chi_E \big[ X_A \textit{ or } X_B \big]  , 
\end{align*}

\noindent one has that:

\begin{itemize}
    \item[$\bullet$] Suppose that $\Delta >0$. There exists an information-theoretic protocol with the corresponding security threshold $2^{-\Delta}$ with positive probability.

    \item[$\bullet$] Suppose that $\Delta \leq 0$. There exists an information-theoretic protocol with the corresponding security threshold $2^{-\Delta}$ with vanishing probability, and thus, $p_{\mathrm{forge}} \geq \Omega \big[ 1 \big]$.
\end{itemize}

\noindent \textbf{Corollary} (\textbf{Corollary} \textit{2}, [24], \textit{the extracted key length and the Holevo information}). Fix $l > k >0$. If $\chi_E \big[ X_A \textit{ or } X_B \big] \leq H \big[ X_A \textit{ or } X_B \big] - k - l $, then there exists a protocol for which:

\begin{itemize}
    \item[$\bullet$] The probability that Eve makes an error when authenticating $X_A$ or $X_B$ is less than or equal to $2^{-k}$.
    \item[$\bullet$] The probability that Eve extracts a Quantum secret key shared between Alice or Bob, which is dependent upon that probability that $X_A \approx X_B$, with secrecy error, is less than or equal to $2^{-l}$.
\end{itemize}

\noindent \textbf{Lemma} (\textbf{Lemma}, [24], \textit{upper bounding Eve's guessing probability with the Relative min-entropy and her Holevo information of $X_A$ or $X_B$}). Fix $X_A \neq X_B \in \big\{ 0 , 1 \big\}^n$ corresponding to random variables in Quantum secret keys held by Alice and Bob, respectively and $\delta >0$ sufficiently small. Suppose that, $\chi_E \big[ X_A \textit{ or } X_B \big] \leq \delta$, and also that, $p_{\mathrm{guess}} \big[ X_A \textit{ or } X_B \big| E \big] \leq 2^{-H_{\infty} [ X_A \textit{ or } X_B \big| E ] }$. With the Relative-min entropy and Eve's Holevo information of $X_A$ or $X_B$, her guessing and forgery probabilities satisfy,

\begin{align*}
   p_{\mathrm{forge}}   \leq  p_{\mathrm{guess}} \big[ X_A \textit{ or } X_B \big| E \big] \leq 2^{-         H [ X_A \textit{ or } X_B ] - \delta      } . 
\end{align*}

\noindent As a result $H_{\infty} \big[ X_A \textit{ or } X_B \big| E \big] \geq H \big[ X_A \textit{ or } X_B \big] - \delta$.

\section{Main Results}

\subsection{Statement}

\noindent We state each main result.

\bigskip

\noindent \textbf{Theorem} \textit{1} (\textit{families of $N$ concave functions for amplification in the multiplayer game-theoretic setting}). Fix strictly positive, distinct parameters $\epsilon, \epsilon^{\prime}, \cdots,\epsilon^{\prime\cdots\prime}, \delta, \delta^{\prime}, \cdots, \delta^{\prime\cdots\prime}$. Suppose that $q_i <      n \times \big( n - 1 \big) \times \cdots \times  \mathrm{ln} \big[ \epsilon^{-1}_i \big]        \bigg[   \underset{j \neq i}{\underset{j \in [ n ]}{\prod}} \delta_j             \bigg] \bigg[   \underset{j \neq j^{\prime} \neq i}{\underset{j^{\prime} \in [ n ]}{\prod}} \delta_{j^{\prime}}                 \bigg]   \times \cdots \times \bigg[         \underset{j \neq j^{\prime} \neq \cdots \neq j^{\prime\cdots\prime} \neq i}{\underset{j^{\prime\cdots\prime} \in [ n ]}{\prod}} \delta_{j^{\prime\cdots\prime}}                     \bigg]         $ and that (\textbf{1}) holds. The optimal value satisfies,

{\small \begin{align*}
   \omega_{\mathrm{Multiplayer}} \big(  G    \big)^{\otimes n} \geq                    \frac{\big( 1 - \epsilon_i \big) \delta_i}{\mathrm{exp} \big[ - N q_i \big] + N H_i                       }                                                      
   \end{align*} }

\bigskip

\noindent \textbf{Theorem} \textit{2} (\textit{a generalization of $\textbf{Theorem}$ 2 from [10] for relating the expected value of the function $\mu$ to $\epsilon$ and $\delta$ parameters}). Under the same choice of parameters provided in \textbf{Theorem} \textit{1}, (\textbf{2}) holds.

\bigskip

\noindent \textbf{Corollary} \textit{1} (\textit{a generalization of $\textbf{Corollary}$ 2 from [10] for relating the expected value of the function $\mu$ to the square of the pullback of a concave function}). Under the same choice of parameters provided in \textbf{Theorem} \textit{1}, (\textbf{3}) holds.

\bigskip

\noindent \textbf{Theorem} \textit{3} (\textit{upper bounding the expected value of the multiplayer function $\mu$}). Under the same choice of parameters provided in \textbf{Theorem} \textit{1}, $(\textbf{6}) < (\textbf{4})$.

\bigskip

\noindent \textbf{Corollary} \textit{2} (\textit{upper bounding the expected value of the multiplayer function $\mu$}). Under the same choice of parameters provided in \textbf{Theorem} \textit{1}, $(\textbf{6}) < (\textbf{5})$.

\bigskip

\noindent \textbf{Corollary} \textit{3} (\textit{the system of expected values from multiplayer concave functions holds}). Under the choice of parameters introduced in \textbf{Theorem} \textit{1}, (\textbf{1}) holds.

\subsection{Arguments}

\noindent We present the arguments for each main result below.

\bigskip

\noindent \textit{Proof of Theorem 1}. We make use of observations previously provided in the arguments for $\textbf{Theorem}$ \textit{5} of [10], from which we provide discuss how generalizations of the multiplayer concave function imply that (\textbf{2}) and (\textbf{3}) hold, from the total number of HINT queries $h_i$ which can be issued by winning players for a chosen message. To this end, introduce,

{\small
\begin{align*}
{\small      \Psi \big[  \omega_{\mathrm{Multiplayer}} \big( G_i  \big) \big]  = N - \underset{1 \leq i \leq N}{\prod} \mathrm{exp} \big[ - q_i \omega_{\mathrm{Multiplayer}} \big( G_i \big)  \big]                       , 
} \\ 
\end{align*}
}

\noindent corresponding to the evaluation of the multiplayer concave function $\Psi$ at the multiplayer winning probability $\omega_{\mathrm{Multiplayer}} \big( G_i \big)$, ie at the $i$ th instance of $G$ for $i < n$ rounds of parallel repetition. To upper bound the expected value,

{\small 

\begin{align*}
       {\tiny    \textbf{E} \bigg\{    \Psi \bigg[                      \omega_{\mathrm{Multiplayer}}     \bigg[   \frac{\big\{   G_1 \cdots G_{(i-1)} G_i G_{(i+1)} \cdots G_n  \big\} , W_C    }{  \mathrm{exp} \big[ - N q_i \big] + N H_i                                                                }   \bigg]                                   \bigg]         \bigg\} \sim          \textbf{E} \bigg\{    \Psi \bigg[                      \omega_{\mathrm{Multiplayer}}     \bigg[   \frac{\big\{   \mathcal{G}_1 \cdots \mathcal{G}_{(i-1)} \mathcal{G}_i \mathcal{G}_{(i+1)} \cdots \mathcal{G}_n  \big\} , W_C    }{  \mathrm{exp} \big[ - N q_i \big] + N H_i                                                                }   \bigg]                                   \bigg]         \bigg\}       ,    }                                                  \\ 
\end{align*}

}

\noindent from the fact that,

{\small
\begin{align*}
   {\small      \omega_{\mathrm{Multiplayer}} \big( \mathcal{G} \big) \sim \omega_{\mathrm{Multiplayer}} \big( G \big)  \leq    \big(     \mathrm{exp} \big[ - q_i x_i \big]       +      N                                                              \big)                                   \hat{\omega}_{\mathrm{Multiplayer}} \big( G , W_i  \big)              \leq     \big(   \mathrm{exp} \big[ - N q_i \big] + N H_i                 \big)                                   \hat{\omega}_{\mathrm{Multiplayer}} \big( G    } \\ \\ {\small , W_i , \textit{restrictions}^{\prime\cdots\prime}  \big)  \sim          \big(         \mathrm{exp} \big[ - N q_i \big] + N H_i                \big)                                   \hat{\omega}_{\mathrm{Multiplayer}} \big( \mathcal{G} , W_i , \textit{restrictions}^{\prime\cdots\prime} \big)                                       , } \\ 
\end{align*}

}

\noindent given a deterministic instance $G_i$ of $\mathcal{G}_i$ satisfying,

{\small
\begin{align*}
    {\small   - \underset{1 \leq i \leq N}{\prod} \mathrm{exp} \big[ - q_i \omega_{\mathrm{Multiplayer}} \big( G_i \big)  \big]   \sim - \underset{1 \leq i \leq N}{\prod} \mathrm{exp} \big[ - q_i \omega_{\mathrm{Multiplayer}} \big( \mathcal{G}_i \big)  \big]         ,       }  \\ 
 \end{align*}

}

\noindent along with the mapping,

{\small

\begin{align*}
    {\small    \varphi :   \bigcup \omega_{\mathrm{Multiplayer}}      \longrightarrow    \bigcup \hat{\omega}_{\mathrm{Multiplayer}}  \mapsto    \omega_{\mathrm{Multiplayer}}  \big\{ \textit{restrictions} \big\}         , } \\ 
 \end{align*}

}

\noindent from multiplayer optimal values before and after enforcing restrictions one writes,

\begin{align*}
   {\small          \frac{\Psi \big[                      \omega_{\mathrm{Multiplayer}}     \big[  \big\{   G_1 \cdots G_{(i-1)} G_i G_{(i+1)} \cdots G_n  \big\} , W_i       \big]                                   \big]}{                    \mathrm{exp} \big[ - N q_i \big] + N H_i                                          }       \leq   \Psi \big[                      \omega_{\mathrm{Multiplayer}}     \big[  \big\{   G_1 \cdots G_{(i-1)} G_i G_{(i+1)} \cdots G_n  \big\} , W_i       \big]                                   \big] } \\ \\ {\small \times   \big\{ \mathrm{exp} \big[ - q_i x_i \big]       +      N    \big\}^{-1}                                                     }                               \end{align*}

   \begin{align*} {\small                 \leq        \frac{\Psi \big[                      {\omega}_{\mathrm{Multiplayer}}     \big[  \big\{   G_1 \cdots G_{(i-1)} G_i G_{(i+1)} \cdots G_n  \big\} , W_i       \big]                                   \big]}{      \omega_{\mathrm{Multiplayer}}     \big[  \big\{   G_1 \cdots G_{(i-1)} G_i G_{(i+1)} \cdots G_n  \big\} , W_i       \big]                                                }                                                    }            \end{align*}

   \begin{align*} {\small =               \frac{\Psi \big[                      {\omega}_{\mathrm{Multiplayer}}     \big[  \big\{   G_1 \cdots G_{(i-1)} G_i G_{(i+1)} \cdots G_n  \big\} , W_i       \big]                                   \big]}{      \omega_{\mathrm{Multiplayer}}     \big[  \big\{   G_1 \cdots G_{(i-1)} G_i G_{(i+1)} \cdots G_n  \big\} , W_i       \big]                                                }            \frac{{\omega}_{\mathrm{Multiplayer}}     \big[  \big\{   G_1 \cdots G_{(i-1)} G_i G_{(i+1)} \cdots G_n  \big\} , W_i       \big]}{{\omega}_{\mathrm{Multiplayer}}     \big[  \big\{   G_1 \cdots G_{(i-1)} G_i G_{(i+1)} \cdots G_n  \big\} , W_i       \big]      }                                                                                                              }          \\  \end{align*}

   \begin{align*} {\small    \leq               \frac{\Psi \big[                      {\omega}_{\mathrm{Multiplayer}}     \big[  \big\{   G_1 \cdots G_{(i-1)} G_i G_{(i+1)} \cdots G_n  \big\} , W_i       \big]                                   \big]}{      \omega_{\mathrm{Multiplayer}}     \big[  \big\{   G_1 \cdots G_{(i-1)} G_i G_{(i+1)} \cdots G_n  \big\} ,  W_i       \big]                                                }            {\omega}_{\mathrm{Multiplayer}}     \big[  \big\{   G_1 \cdots G_{(i-1)} G_i G_{(i+1)} \cdots G_n  \big\}  } \\ \\ {\small ,  W_i       \big]                                                                                                              }        \\ \\  {\small =            \frac{\Psi \big[                      {\omega}_{\mathrm{Multiplayer}}     \big[  \big\{   G_1 \cdots G_{(i-1)} G_i G_{(i+1)} \cdots G_n  \big\} , W_i       \big]                                   \big]}{      \omega_{\mathrm{Multiplayer}}     \big[  \big\{   G_1 \cdots G_{(i-1)} G_i G_{(i+1)} \cdots G_n  \big\} , W_i       \big]                                                }       \textbf{E}_F  \big[ {\omega}_{\mathrm{Multiplayer}}     \big[  \big\{   G_1 \cdots G_{(i-1)} G_i G_{(i+1)} \cdots G_n  \big\}  } \\ \\ {\small , W_i       \big] \big] } \\ \\ {\small      =    \textbf{E}_F  \bigg\{           \frac{\Psi \big[                      {\omega}_{\mathrm{Multiplayer}}     \big[  \big\{   G_1 \cdots G_{(i-1)} G_i G_{(i+1)} \cdots G_n  \big\} , W_i       \big]                                   \big]}{      \omega_{\mathrm{Multiplayer}}     \big[  \big\{   G_1 \cdots G_{(i-1)} G_i G_{(i+1)} \cdots G_n  \big\} , W_i       \big]                                                }        {\omega}_{\mathrm{Multiplayer}}     \big[  \big\{   G_1 \cdots G_{(i-1)} G_i G_{(i+1)} \cdots  } \\ \\ {\small G_n  \big\} ,  W_i       \big]  \bigg\}                  }   \\ \\ \\  {\small           \leq            \textbf{E}_F  \bigg\{           \frac{\Psi \big[                      {\omega}_{\mathrm{Multiplayer}}     \big[  \big\{   G_1 \cdots G_{(i-1)} G_i G_{(i+1)} \cdots G_n  \big\} , W_C       \big]                                   \big]}{      \omega_{\mathrm{Multiplayer}}     \big[  \big\{   G_1 \cdots G_{(i-1)} G_i G_{(i+1)} \cdots G_n  \big\} , W_C       \big]                                                }        \hat{\omega}_{\mathrm{Multiplayer}}     \big[  \big\{   G_1 \cdots G_{(i-1)} G_i G_{(i+1)} \cdots } \\ \\ {\small   G_n  \big\}  , W_i  , \textit{restrictions}^{\prime\cdots\prime}        \big]  \bigg\}                                                                                               }   \end{align*}

   \begin{align*} {\small              =   \textbf{E}_F  \big\{         \psi \big[       \hat{\omega}_{\mathrm{Multiplayer}}     \big[  \big\{   G_1 \cdots G_{(i-1)} G_i G_{(i+1)} \cdots G_n  \big\}  , W_i  , \textit{restrictions}^{\prime\cdots\prime}        \big] \big]  \big\}              } \\ \\  \\ {\small             \leq    \hat{\omega}_{\mathrm{Multiplayer}}     \big[  \big\{   G_1 \cdots G_{(i-1)} G_i G_{(i+1)} \cdots G_n  \big\}  , W_i   ,   \textit{restrictions}^{\prime\cdots\prime}     \big] \big]  \big\}                ,                                                                                              }  \\ 
\end{align*}

\noindent where $F$ denotes a filter for which $F \in \mathrm{support} \big\{   \textit{restrictions}^{\prime\cdots\prime} \big\}$.

\bigskip

\noindent Given an instance $g \sim \mathcal{G}$, the above computations providing an upper bound in terms of the image of an optimal value $\omega$ under $\varphi$ straightforwardly imply,

{\small

\begin{align*}
    {\small          \frac{\Psi \big[                      \omega_{\mathrm{Multiplayer}}     \big[  \big\{   g_1 \cdots g_{(i-1)} g_i g_{(i+1)} \cdots g_n  \big\} , W_i        \big]                                   \big]}{                   \mathrm{exp} \big[ - N q_i \big] + N H_i                                                 }       \leq   \hat{\omega}_{\mathrm{Multiplayer}}     \big[  \big\{   g_1 \cdots g_{(i-1)} g_i g_{(i+1)} \cdots g_n  \big\}  , W_i                   } \\ \\ {\small  , \textit{restrictions}^{\prime\cdots\prime} \big] \big]  \big\}   \sim  \hat{\omega}_{\mathrm{Multiplayer}}     \big[  \big\{   \mathcal{G}_1 \cdots \mathcal{G}_{(i-1)} \mathcal{G}_i \mathcal{G}_{(i+1)} \cdots \mathcal{G}_n  \big\}  , W_i  ,    } \\ \\ {\small \textit{restrictions}^{\prime\cdots\prime}    \big] \big]  \big\}              .  }     \\ 
\end{align*}

}

\noindent Under the identification between the multiplayer optimal value $\omega$ with the multiplayer optimal value $\hat{\omega}$ under restrictions,

{\small

\begin{align*}
    {\small  \omega_{\mathrm{Multiplayer}} \big[ \big\{ G_1 \cdots G_{(i-1)} G_i G_{(i+1)} \cdots G_n         \big\}         ,   W_i  \big]    = \omega_{\mathrm{Multiplayer} } \big( G \big)^{\otimes i }       ,      } \\  
\end{align*}

}

\noindent one readily obtains the that desired lower bound for parallel repetition of the optimal value holds from the fact that,

{\small

\begin{align*}
   {\small                \big\{     \mathrm{exp} \big[ - N q_i \big] + N H_i               \big\}     \omega_{\mathrm{Multiplayer}}  \big( G_i ,                                                   \big) \equiv   \big\{     \mathrm{exp} \big[ - N q_i \big] + N H_i               \big\}       \textbf{E}_{G_i } \big[ \omega_{\mathrm{Multiplayer}}  \big( G_i ,                                                             \big)  \big]                                                                                                                                     , } \\  
\end{align*}

}

\noindent one also has that,

{\small
\begin{align*}
   {\small               \omega \big[ \big\{     G_1 \cdots G_{(i-1)} G_i G_{(i+1)} \cdots G_n                     \big\} , W_i  \big]   \geq       \frac{\big( 1 - \epsilon_i \big) \delta_i}{\mathrm{exp} \big[ - N q_i \big] + N H_i                       }                                            .  } \\ 
\end{align*}
}

\noindent Straightforwardly adapting the above computations for obtaining the lower bound for parallel repetition of the multiplayer optimal value for $n$, instead of for $i$ rounds of parallel repetition readily follows, from which we conclude the argument. \boxed{}

\bigskip

\noindent \textit{Proof of Theorem 2}. To demonstrate that the desired system of inequalities holds, introduce,

{\small \[  \big\{ \mathcal{Q}_i \big\}_{1 \leq i \leq N} \equiv  \left\{\!\begin{array}{ll@{}>{{}}l} 
   \big[ \mathcal{Q}_1  \big]_{\geq \delta} \equiv \big\{ Q_1 \in \mathcal{Q}_1 : \textbf{E}_{\mathcal{Q}_2 \times \cdots \times \mathcal{Q}_N}  \big[  \mu \big( Q_1 , \mathcal{Q}_2 , \cdots , \mathcal{Q}_N      \big)   \big] \geq \delta \big\}      , \\ \vdots \\      \big[ \mathcal{Q}_N  \big]_{\geq \delta} \equiv \big\{ Q_N \in \mathcal{Q}_N : \textbf{E}_{\mathcal{Q}_1 \times \cdots \times \mathcal{Q}_{N-1}}   \big[  \mu \big( \mathcal{Q}_1 , \mathcal{Q}_2 , \cdots , Q_N      \big)   \big] \geq \delta \big\}      ,  
  \end{array}\right. 
\]   }

 {\small \[  \left\{\!\begin{array}{ll@{}>{{}}l}   
   \big[ \mathcal{Q}_1 \mathcal{Q}_2 \big]_{\geq \delta} \equiv  \big\{ Q_1 \in \mathcal{Q}_1 : \textbf{E}_{\mathcal{Q}_2 \times \cdots \times \mathcal{Q}_N}  \big[  \mu \big( Q_1 , \mathcal{Q}_2 , \mathcal{Q}_3  \cdots , \mathcal{Q}_N      \big)  \mu \big( \mathcal{Q}_1 \\ , Q_2  , \mathcal{Q}_3  \cdots , \mathcal{Q}_N      \big)     \big] \geq \delta \big\}   , \\ \vdots \\       \big[ \mathcal{Q}_1 \mathcal{Q}_2  \times \cdots \times \mathcal{Q}_N \big]_{\geq \delta} \equiv  \big\{ Q_1 \in \mathcal{Q}_1 : \textbf{E}_{\mathcal{Q}_2 \times \cdots \times \mathcal{Q}_N}  \big[  \mu \big( Q_1 , \mathcal{Q}_2  , \mathcal{Q}_3  \cdots \\ , \mathcal{Q}_N      \big)  \mu \big( \mathcal{Q}_1  , Q_2  , \mathcal{Q}_3  \cdots , \mathcal{Q}_N      \big)     \big] \times \cdots \times \mu \big(  \mathcal{Q}_1  , \cdots , \mathcal{Q}_{N-1} ,  Q_N \big)  \geq \delta \big\}   ,   \\ 
\end{array}\right. 
\]   }

\noindent given some strictly positive $\delta$. Hence it suffices to compute conditional expectations of the form,

{\small \[ \textbf{E}_{\mathcal{Q}} \big[ \Psi \big[ \cdot \big] \big] \equiv \left\{\!\begin{array}{ll@{}>{{}}l} 
   \textbf{E}_{Q_1} \big[ \psi \big[ \textbf{E}_{\mathcal{Q}_2 \times \cdots \times \mathcal{Q}_N} \big[          \mu \big[ Q_1 , \mathcal{Q}_2 , \cdots , \mathcal{Q}_N          \big]     \big] \big| Q_1 \not\in \big[ \mathcal{Q}_1 \big]_{\geq \delta } \big]   , \\ \vdots \\   \textbf{E}_{Q_1 Q_N } \big[ \psi^{\prime\cdots\prime} \big[ \textbf{E}_{\mathcal{Q}_2 \times \cdots \times \mathcal{Q}_{N-1}} \big[          \mu \big[  Q_1 , \mathcal{Q}_2 , \cdots , \mathcal{Q}_N     \mu \big[  \mathcal{Q}_1 , \mathcal{Q}_2 , \\ \cdots , Q_N          \big]            \big]     \big]  \big| Q_1 \not\in \big[ \mathcal{Q}_1 \big]_{\geq \delta } \big[ \mathcal{Q}_N \big]_{\geq \delta }   \big]       , \\  \\ \textbf{E}_{Q_2} \big[ \psi \big[ \textbf{E}_{\mathcal{Q}_1 \mathcal{Q}_3  \times \cdots \times \mathcal{Q}_N} \big[          \mu \big[ Q_1 , \mathcal{Q}_2 , \cdots , \mathcal{Q}_N          \big]     \big] \big| Q_2  \not\in \big[ \mathcal{Q}_2 \big]_{\geq \delta } \big]    , \\ \vdots \\ \textbf{E}_{Q_2 Q_N } \big[ \psi \big[ \textbf{E}_{\mathcal{Q}_1 \mathcal{Q}_3  \times \cdots \times \mathcal{Q}_N} \big[          \mu \big[ Q_1 , \mathcal{Q}_2 , \cdots , \mathcal{Q}_N          \big]  \mu \big[ \mathcal{Q}_1 , Q_2 , \mathcal{Q}_3  \\ , \cdots , \mathcal{Q}_N          \big]      \big] \big| Q_2  \not\in \big[ \mathcal{Q}_2  \big]_{\geq \delta }\big[ \mathcal{Q}_N  \big]_{\geq \delta }  \big] , \\ \vdots \\  \textbf{E}_{Q_1 \times \cdots \times Q_{N-1} } \big[ \psi \big[ \textbf{E}_{\mathcal{Q}_1 \mathcal{Q}_2  \times \cdots \times \mathcal{Q}_{N-1} Q_N } \big[          \mu \big[ Q_1 , \mathcal{Q}_2 , \cdots , \mathcal{Q}_N          \big]  \times  \\  \cdots  \times       \mu \big[ \mathcal{Q}_1 , \cdots \mathcal{Q}_{N-2}   ,  Q_{N-1} , {Q}_N           \big]    \big] \big| Q_2  \not\in \big[ \mathcal{Q}_2 \big]_{\geq \delta } \big[ \mathcal{Q}_3 \big]_{\geq \delta }  \times \cdots \\  \times \big[ \mathcal{Q}_{N-1} \big]_{\geq \delta }  \big]    .      
\end{array}\right. 
\]   }

\noindent Denote the symmetric group of $N$ letters with $S_N$. By direct computation write,

{\small \begin{align*}
  \underset{i \neq j}{\underset{\sigma_j \in S_N}{\underset{\sigma_i \in S_N}{\sum}}}      \textbf{E}_{\mathcal{Q}_1 \times \cdots \times \mathcal{Q}_{i-1} \mathcal{Q}_{i+1} \times \cdots \times \mathcal{Q}_N} \big[ \textbf{E}_{\mathcal{Q}_i} \big[ \psi^{\prime\cdots\prime}_i \big[       \mu  \big( Q_1 , Q_2 , \cdots , Q_N \big)     \big]    \big]   \big|    Q_j \not\in \mathcal{Q}_j        \big] \end{align*}

  \begin{align*} =    \underset{i \neq j}{\underset{\sigma_j \in S_N}{\underset{\sigma_i \in S_N}{\sum}}}      \textbf{E}_{\mathcal{Q}_1 \times \cdots \times \mathcal{Q}_{i-1} \mathcal{Q}_{i+1} \times \cdots \times \mathcal{Q}_N} \bigg[ \underset{1 \leq i \leq N}{\prod} \delta_i \bigg\{ \underset{1 \leq i \leq N}{\prod} \delta_i  \bigg\}^{-1}  \textbf{E}_{\mathcal{Q}_i} \big[ \psi^{\prime\cdots\prime}_i \big[       \mu  \big( Q_1   , Q_2 , \cdots   , Q_N \big)    \big]     \big] \\  \big|    Q_j \not\in \mathcal{Q}_j        \bigg] \end{align*}

  \begin{align*} =  \underset{i \neq j}{\sum}       \textbf{E}_{\mathcal{Q}_1 \times \cdots \times \mathcal{Q}_{i-1} \mathcal{Q}_{i+1} \times \cdots \times \mathcal{Q}_N} \bigg[ \underset{\sigma_i \in S_N}{\sum} \bigg\{ \underset{1 \leq i \leq N}{\prod} \delta_i \bigg\{ \underset{1 \leq i \leq N}{\prod} \delta_i  \bigg\}^{-1} \textbf{E}_{\mathcal{Q}_i} \big[ \psi^{\prime\cdots\prime}_i \big[       \mu  \big( Q_1 , Q_2 , \cdots \\ , Q_N \big)     \big]    \big] \bigg\}  \bigg|  \underset{\sigma_i \in S_N}{\sum}  \big\{ Q_j :   Q_j  \not\in \mathcal{Q}_j \big\}        \bigg]               \\  \\   \geq     \underset{i \neq j}{\sum}       \textbf{E}_{\mathcal{Q}_1 \times \cdots \times \mathcal{Q}_{i-1} \mathcal{Q}_{i+1} \times \cdots \times \mathcal{Q}_N} \bigg[ \underset{\sigma_i \in S_N}{\sum} \bigg\{ \underset{1 \leq i \leq N}{\prod} \psi^{\prime\cdots\prime}_i \big[ \delta_i \big]  \bigg\{ \underset{1 \leq i \leq N}{\prod} \delta_i \bigg\}^{-1}  \textbf{E}_{\mathcal{Q}_i} \big[  \big[       \mu  \big( Q_1 , Q_2 , \cdots \\ , Q_N \big)    \big]     \big] \bigg\}  \bigg|  \underset{\sigma_i \in S_N}{\sum}  \big\{ Q_j :   Q_j  \not\in \mathcal{Q}_j \big\}        \bigg] \end{align*}

  \begin{align*} = \underset{\sigma_i \in S_N}{\sum} \bigg\{ \underset{1 \leq i \leq N}{\prod} \psi^{\prime\cdots\prime}_i \big[ \delta_i \big]  \bigg\{ {\underset{1 \leq i \leq N}{\prod} \delta_i }  \bigg\}^{-1} \underset{i \neq j}{\sum}       \textbf{E}_{\mathcal{Q}_1 \times \cdots \times \mathcal{Q}_{i-1} \mathcal{Q}_{i+1} \times \cdots \times \mathcal{Q}_N} \bigg[ \underset{\sigma_i \in S_N}{\sum}  \big\{  \textbf{E}_{\mathcal{Q}_i} \big[  \big[       \mu  \big( Q_1 \\ , Q_2 , \cdots  , Q_N \big)    \big]     \big] \big\}   \bigg|  \underset{\sigma_i \in S_N}{\sum}  \big\{ Q_j  :   Q_j  \not\in \mathcal{Q}_j \big\}        \bigg]           \bigg\}  . \\ \\ \tag{\textbf{0}} \\ 
\end{align*} }

\noindent Proceeding, from the observation that the probability measure over each $Q_i \in \mathcal{Q}_i$ can be expressed as,

{\small \begin{align*}
      \underset{ i \neq j}{\underset{ \mathcal{Q}_i , \mathcal{Q}_j \in \{ \mathcal{Q}_1 , \cdots , \mathcal{Q}_N  \}  }{\prod}}  \bigg\{      \bigg\{   \underset{\sigma_i \in S_N}{\sum } \textbf{P}  \big[   Q_i \not\in \mathcal{Q}_i     \big]     \big|_{Q_i}      \bigg\} \bigg\{        \underset{j \neq i}{\underset{\sigma_j \in S_N}{\sum}}    \textbf{E}_{\mathcal{Q}_1 \times \cdots \times \mathcal{Q}_{j-1} \mathcal{Q}_{j+1} \times \cdots \times \mathcal{Q}_N} \big[          \textbf{E}_{\mathcal{Q}_j} \big[ \mu \big( Q_1 , \\ \cdots , Q_N \big)  \big|   Q_i \not\in \mathcal{Q}_i    \big]               \big] \bigg\} \bigg\}      \leq  C \big( N , n \big)               \underset{i \neq i^{\prime} \neq \cdots \neq i^{\prime\cdots\prime}}{\underset{1 \leq i^{\prime \cdots \prime} \leq N}{\underset{\vdots}{\underset{1 \leq i^{\prime}\leq N}{\underset{1 \leq i \leq N}{\prod}} }}}  \big[  \delta_i \delta_{i^{\prime}} \times \cdots \times \delta_{i^{\prime\cdots\prime}} \big]   \bigg[               \underset{i \neq i^{\prime} \neq \cdots \neq i^{\prime\cdots\prime}}{\underset{1 \leq i^{\prime \cdots \prime} \leq N}{\underset{\vdots}{\underset{1 \leq i^{\prime}\leq N}{\underset{1 \leq i \leq N}{\prod}} }}}     \big[   \epsilon_i \epsilon_{i^{\prime}} \\ \times \cdots \times \epsilon_{i^{\prime\cdots\prime}} \big]  - \underset{\sigma_i \in S_N}{\prod} \textbf{P} \big[ Q_i \not\in \mathcal{Q}_i \big] \big|_{Q_i}     \bigg]                                                     . 
\end{align*} }

\noindent Furthermore,

{\small \begin{align*}
    (\textbf{0}) \leq               \underset{i \neq j}{\underset{\sigma_j \in S_N}{\underset{\sigma_i \in S_N}{\prod}}}  \big[    \textbf{P} \big[ Q_i \not\in \mathcal{Q}_i \big] \big|_{Q_i}  \textbf{P} \big[ Q_j \not\in \mathcal{Q}_j \big] \big|_{Q_j}       \big]      +     \underset{i \neq i^{\prime} \neq \cdots \neq i^{\prime\cdots\prime}}{\underset{1 \leq i^{\prime \cdots \prime} \leq N}{\underset{\vdots}{\underset{1 \leq i^{\prime}\leq N}{\underset{1 \leq i \leq N}{\sum}} }}}   \big[ \delta_i \times \cdots \times \delta_{i^{\prime\cdots\prime}}  \big]  \epsilon      +   \underset{j \neq \cdots \neq j^{\prime\cdots\prime} \neq i \neq \cdots \neq i^{\prime\cdots\prime}}{\underset{j \neq j^{\prime} \neq \cdots \neq j^{\prime\cdots\prime}}{\underset{1 \leq j^{\prime \cdots \prime} \leq N}{\underset{\vdots}{\underset{1 \leq j^{\prime}\leq N}{\underset{1 \leq j \leq N}{\sum}} }}}}   \big[ \delta_j \\ \times \cdots \times  \delta_{j^{\prime\cdots\prime}}  \big]   \epsilon             -     \underset{i \neq i^{\prime} \neq \cdots \neq i^{\prime\cdots\prime}}{\underset{1 \leq i^{\prime \cdots \prime} \leq N}{\underset{\vdots}{\underset{1 \leq i^{\prime}\leq N}{\underset{1 \leq i \leq N}{\sum}} }}}   \big[ \delta_i \times \cdots \times \delta_{i^{\prime\cdots\prime}}  \big]  \bigg[   \underset{\sigma_i \in S_N}{\sum}    \textbf{P} \big[ Q_i \not\in \mathcal{Q}_i  \big] \big|_{Q_i}     \bigg]      -  \underset{j \neq \cdots \neq j^{\prime\cdots\prime} \neq i \neq \cdots \neq i^{\prime\cdots\prime}}{\underset{j \neq j^{\prime} \neq \cdots \neq j^{\prime\cdots\prime}}{\underset{1 \leq j^{\prime \cdots \prime} \leq N}{\underset{\vdots}{\underset{1 \leq j^{\prime}\leq N}{\underset{1 \leq j \leq N}{\sum}} }}}}   \big[ \delta_j \\ \\ \times \cdots \times \delta_{j^{\prime\cdots\prime}}  \big]  \bigg[     \underset{\sigma_j \in S_N}{\sum}  \textbf{P} \big[ Q_j \not\in \mathcal{Q}_j  \big] \big|_{Q_j}\bigg]    +  \underset{i \neq i^{\prime} \neq \cdots \neq i^{\prime\cdots\prime}}{\underset{1 \leq i^{\prime \cdots \prime} \leq N}{\underset{\vdots}{\underset{1 \leq i^{\prime}\leq N}{\underset{1 \leq i \leq N}{\sum}} }}}   \big[ \delta_i \times \cdots \times \delta_{i^{\prime\cdots\prime}}  \big]    \underset{j \neq \cdots \neq j^{\prime\cdots\prime} \neq i \neq \cdots \neq i^{\prime\cdots\prime}}{\underset{j \neq j^{\prime} \neq \cdots \neq j^{\prime\cdots\prime}}{\underset{1 \leq j^{\prime \cdots \prime} \leq N}{\underset{\vdots}{\underset{1 \leq j^{\prime}\leq N}{\underset{1 \leq j \leq N}{\sum}} }}}}   \big[ \delta_j \\ \\ \times \cdots \delta_{j^{\prime\cdots\prime}}  \big]     \bigg[ \underset{\sigma_i \in S_N}{\sum}    \textbf{P} \big[ Q_i \not\in \mathcal{Q}_i  \big] \big|_{Q_i} \bigg]   \bigg[ \underset{j \neq i}{\underset{\sigma_j \in S_N}{\sum}}    \textbf{P} \big[ Q_j \not\in \mathcal{Q}_j  \big] \big|_{Q_j} \bigg]            +       \underset{1 \leq j \leq N}{\underset{1 \leq i \leq N}{\underset{i \neq j}{\sum}} }   \big( - 1 \big)^{i+j} \binom{N}{i}  \binom{N}{j}    \\ \\ \times               \bigg[      \textbf{P} \big[ Q_i \not\in \mathcal{Q}_i  \big] \big|_{Q_i}  \textbf{P} \big[ Q_j \not\in \mathcal{Q}_j  \big] \big|_{Q_j}          \bigg]                                                                                        , \\ 
\end{align*} }

\noindent where in the last line, the contribution from the last term alternates been $\pm 1$ according to the conditions,

{\small \[   \left\{\!\begin{array}{ll@{}>{{}}l} 
  \big\{  \big( - 1 \big)^{i+j}  = 1 \big\} \Longleftrightarrow  \big\{ i+j \text{ } \textit{even}  \big\}   , \\ \\   \big\{  \big( - 1 \big)^{i+j}  =  - 1 \big\} \Longleftrightarrow  \big\{ i+j \text{ } \textit{odd}  \big\}  . \\ 
\end{array}\right. 
\] \\  }

\noindent Altogether, the desired upper bound takes the form, from the observation,

{\small
\begin{align*}
 (\textbf{0}^{*}) \equiv    \underset{i \neq j}{\underset{\sigma_j \in S_N}{\underset{\sigma_i \in S_N}{\prod}}}  \big[    \textbf{P} \big[ Q_i \not\in \mathcal{Q}_i \big] \big|_{Q_i}  \textbf{P} \big[ Q_j \not\in \mathcal{Q}_j \big] \big|_{Q_j}       \big]     \leq  \mathrm{sup} \big\{ \epsilon_i \epsilon_j , \epsilon_i \epsilon_j \epsilon^{\prime}_i \epsilon^{\prime}_j , \cdots , \epsilon_i \epsilon_j \epsilon^{\prime}_i \epsilon^{\prime}_j  \times \cdots \times \epsilon^{\prime\cdots\prime}_i \epsilon^{\prime\cdots\prime}_j \big\}      ,  \end{align*}
     
   \noindent corresponding to the first term,  
     \begin{align*} \\ (\textbf{0}^{**}) \equiv  \underset{i \neq i^{\prime} \neq \cdots \neq i^{\prime\cdots\prime}}{\underset{1 \leq i^{\prime \cdots \prime} \leq N}{\underset{\vdots}{\underset{1 \leq i^{\prime}\leq N}{\underset{1 \leq i \leq N}{\sum}} }}}   \big[ \delta_i \times \cdots \times \delta_{i^{\prime\cdots\prime}}  \big]  \epsilon    \leq  \mathrm{sup} \big\{ \delta_i , \delta_{i^{\prime}} , \cdots , \delta_{i^{\prime\cdots\prime}} \big\} \epsilon   , \\  \end{align*}
     
    \noindent corresponding to the second term,

    {\small \begin{align*}  \\ (\textbf{0}^{***}) \equiv      \underset{j \neq \cdots \neq j^{\prime\cdots\prime} \neq i \neq \cdots \neq i^{\prime\cdots\prime}}{\underset{j \neq j^{\prime} \neq \cdots \neq j^{\prime\cdots\prime}}{\underset{1 \leq j^{\prime \cdots \prime} \leq N}{\underset{\vdots}{\underset{1 \leq j^{\prime}\leq N}{\underset{1 \leq j \leq N}{\sum}} }}}}   \big[ \delta_j \times \cdots   \times \delta_{j^{\prime\cdots\prime}}  \big]   \epsilon           \leq \mathrm{sup} \big\{ \delta_j , \delta_{j^{\prime}}  , \cdots  , \delta_{j^{\prime\cdots\prime}} \big\} \epsilon     , \\ \end{align*} } 

  \noindent corresponding to the third term,

     {\small \begin{align*} (\textbf{0}^{****}) \equiv         -     \underset{i \neq i^{\prime} \neq \cdots \neq i^{\prime\cdots\prime}}{\underset{1 \leq i^{\prime \cdots \prime} \leq N}{\underset{\vdots}{\underset{1 \leq i^{\prime}\leq N}{\underset{1 \leq i \leq N}{\sum}} }}}   \big[ \delta_i \times \cdots \times \delta_{i^{\prime\cdots\prime}}  \big]  \bigg[   \underset{\sigma_i \in S_N}{\sum}    \textbf{P} \big[ Q_i \not\in \mathcal{Q}_i  \big] \big|_{Q_i}     \bigg]   \leq      {\tiny \mathrm{sup} \big\{   \big[ \delta_i \times \cdots  \times \delta_{i^{\prime\cdots\prime}}  \big]  \epsilon_i ,  \big[ \delta_i \times   \cdots }  \\ {\tiny \times \delta_{i^{\prime\cdots\prime}}  \big]  \epsilon_i \epsilon_{i^{\prime}} , \cdots  ,   \big[ \delta_i \times \cdots \times \delta_{i^{\prime\cdots\prime}}  \big]  \epsilon_i \epsilon_{i^{\prime}}  \times \cdots \times \epsilon_{i^{\prime\cdots\prime}}          \big\}                  ,   } \\ \end{align*} }

   \noindent corresponding to the fourth term,

     {\small \begin{align*}   (\textbf{0}^{*****}) \equiv  -  \underset{j \neq \cdots \neq j^{\prime\cdots\prime} \neq i \neq \cdots \neq i^{\prime\cdots\prime}}{\underset{j \neq j^{\prime} \neq \cdots \neq j^{\prime\cdots\prime}}{\underset{1 \leq j^{\prime \cdots \prime} \leq N}{\underset{\vdots}{\underset{1 \leq j^{\prime}\leq N}{\underset{1 \leq j \leq N}{\sum}} }}}}   \big[ \delta_j \times \cdots \times \delta_{j^{\prime\cdots\prime}}  \big]  \bigg[ \underset{\sigma_j \in S_N}{\sum}  \textbf{P} \big[ Q_j \not\in \mathcal{Q}_j  \big] \big|_{Q_j} \bigg]  \leq         \mathrm{sup} \big\{        \big[ \delta_j \times  \cdots \times \delta_{j^{\prime\cdots\prime}}  \big]    \epsilon_j ,    \big[ \delta_j \times \cdots \\ \times \delta_{j^{\prime\cdots\prime}}  \big]    \epsilon_j \epsilon_{j^{\prime}} , \cdots   ,  \big[ \delta_j \times \cdots \times \delta_{j^{\prime\cdots\prime}}  \big]    \epsilon_j \epsilon_{j^{\prime}} \times \cdots \times \epsilon_{j^{\prime\cdots\prime}}          \big\}          ,  \\  \end{align*}  }

       \noindent corresponding to the fifth term,  
     
    {\small  \begin{align*} (\textbf{0}^{******}) \equiv   \underset{i \neq i^{\prime} \neq \cdots \neq i^{\prime\cdots\prime}}{\underset{1 \leq i^{\prime \cdots \prime} \leq N}{\underset{\vdots}{\underset{1 \leq i^{\prime}\leq N}{\underset{1 \leq i \leq N}{\sum}} }}}   \big[ \delta_i \times \cdots \times \delta_{i^{\prime\cdots\prime}}  \big]   \underset{j \neq \cdots \neq j^{\prime\cdots\prime} \neq i \neq \cdots \neq i^{\prime\cdots\prime}}{\underset{j \neq j^{\prime} \neq \cdots \neq j^{\prime\cdots\prime}}{\underset{1 \leq j^{\prime \cdots \prime} \leq N}{\underset{\vdots}{\underset{1 \leq j^{\prime}\leq N}{\underset{1 \leq j \leq N}{\sum}} }}}}   \big[ \delta_j \times \cdots \times \delta_{j^{\prime\cdots\prime}}  \big]  \bigg[ \underset{\sigma_i \in S_N}{\sum}    \textbf{P} \big[ Q_i \not\in \mathcal{Q}_i  \big] \big|_{Q_i} \bigg]   \\ \times     \bigg[ \underset{j \neq i}{\underset{\sigma_j \in S_N}{\sum}}    \textbf{P} \big[ Q_j \not\in \mathcal{Q}_j  \big] \big|_{Q_j} \bigg]    \leq  \mathrm{sup} \big\{      \big[ \delta_i \times \cdots \times \delta_{i^{\prime\cdots\prime}}  \big]    \big[ \delta_j \times \cdots \times \delta_{j^{\prime\cdots\prime}}  \big]   \epsilon_i \epsilon_j , \cdots  ,          \big[ \delta_i \times \cdots \\ \times \delta_{i^{\prime\cdots\prime}}  \big]    \big[ \delta_j \times \cdots \times \delta_{j^{\prime\cdots\prime}}  \big]   \epsilon_i \epsilon_j   \times \epsilon_{i^{\prime}} \epsilon_{j^{\prime}} \times \cdots \times   \epsilon_{i^{\prime\cdots\prime}} \epsilon_{j^{\prime\cdots\prime}}     \big\}      ,  \\  \end{align*} } 

       \noindent corresponding to the sixth term,  and,
     
  {\small    \begin{align*} (\textbf{0}^{*******}) \equiv           \underset{1 \leq j \leq N}{\underset{1 \leq i \leq N}{\underset{i \neq j}{\sum}} }   \big( - 1 \big)^{i+j} \binom{N}{i}  \binom{N}{j}                    \bigg[    \textbf{P} \big[ Q_i \not\in \mathcal{Q}_i  \big] \big|_{Q_i}     \textbf{P} \big[ Q_j \not\in \mathcal{Q}_j  \big] \big|_{Q_j}  \bigg]   \leq       \underset{1 \leq j \leq N}{\underset{1 \leq i \leq N}{\underset{i \neq j}{\mathrm{sup}}}} \bigg\{          \binom{N}{i}  , \binom{N}{j} ,   \epsilon_i , \\ \cdots , \epsilon_{i^{\prime\cdots\prime}} , \epsilon_j  , \cdots , \epsilon_{j^{\prime\cdots\prime}} , \epsilon_i \epsilon_j ,  \cdots , \epsilon_i \epsilon_j \epsilon_{i^{\prime}} \epsilon_{j^{\prime}} \times \cdots \times \epsilon_{i^{\prime\cdots\prime}}  \epsilon_{j^{\prime\cdots\prime}}     \bigg\}   \\ \\ \propto       \underset{1 \leq j \leq N}{\underset{1 \leq i \leq N}{\underset{i \neq j}{\mathrm{sup}}}} \bigg\{          \binom{N}{i} ,  \binom{N}{j} \bigg\}    {\mathrm{sup}} \big\{ \epsilon_i , \cdots , \epsilon_{i^{\prime\cdots\prime}} , \epsilon_j  , \cdots , \epsilon_{j^{\prime\cdots\prime}} , \epsilon_i \epsilon_j ,  \cdots , \epsilon_i \epsilon_j \epsilon_{i^{\prime}} \epsilon_{j^{\prime}} \times \cdots \times \epsilon_{i^{\prime\cdots\prime}}  \\ \\ \times \epsilon_{j^{\prime\cdots\prime}}     \big\}           , \\ 
\end{align*} }  }

\noindent   \noindent corresponding to the seventh term, for,

{\small \begin{align*}
  (\textbf{0}^*) \leq \mathrm{sup} \big\{ \mathcal{C}_1 \big\}   , \\ \vdots  \end{align*} \begin{align*} (\textbf{0}^{*******}) \leq \mathrm{sup} \big\{ \mathcal{C}_7 \big\} , \\ 
\end{align*} } 

\noindent as well as for,

{\small \begin{align*}
 \underset{1 \leq i \leq 7}{\bigcup} \mathrm{sup} \big\{ \mathcal{C}_i \big\}   =  \mathrm{sup} \big\{ \mathcal{C}_1 , \mathcal{C}_2 , \cdots , \mathcal{C}_7 \big\} , \\ 
\end{align*}}

\noindent together imply that,

{\small \begin{align*}
   C \text{ } {\tiny  \mathrm{sup} \bigg\{   \epsilon \epsilon^{\prime}    ,  \epsilon \epsilon^{\prime} \epsilon^{\prime\prime}  , \cdots , \epsilon \epsilon^{\prime} \epsilon^{\prime\prime}  \times \cdots \times \epsilon^{\prime\prime\cdots \prime} ,    \epsilon \delta , \epsilon \delta + \epsilon^{\prime} \delta^{\prime}   , \cdots ,      \underset{1 \leq i \leq N}{\sum} \epsilon_i \delta_i    \bigg\} \geq \mathrm{sup} \big\{ \mathcal{C}_1 , \mathcal{C}_2 , \cdots , \mathcal{C}_7 \big\}    }     , \\ 
\end{align*} }

\noindent and also that,

{\small
\begin{align*}
  \underset{1 \leq i \leq N}{\prod} \textbf{P} \big[ Q_i \in \mathcal{Q}_i \big] \big|_{Q_i} \leq \epsilon \epsilon^{\prime} \times \cdots \times \epsilon^{\prime\cdots\prime}  , \\ 
\end{align*}
}

\noindent from the fact that, 

{\small \begin{align*}
   C  \text{ } {\tiny \mathrm{sup} \bigg\{   \epsilon \epsilon^{\prime}    ,  \epsilon \epsilon^{\prime} \epsilon^{\prime\prime}  , \cdots , \epsilon \epsilon^{\prime} \epsilon^{\prime\prime}  \times \cdots \times \epsilon^{\prime\prime\cdots \prime} ,    \epsilon \delta , \epsilon \delta + \epsilon^{\prime} \delta^{\prime}   , \cdots ,      \underset{1 \leq i \leq N}{\sum} \epsilon_i \delta_i    \bigg\} }  \overset{(*)}{\geq}    \mathcal{T}_1 \equiv  \mathrm{sup} \big\{ \mathcal{C}_1 , \mathcal{C}_2 , \cdots , \mathcal{C}_7 \big\}         \\ \\   \geq  (\textbf{0}^{*}) + \cdots + (\textbf{0}^{*******})      =         (\textbf{0})           , 
\end{align*} }

\noindent where, for,

{\small \begin{align*}
    \mathcal{T}_1 = \mathcal{T}^{\prime}_1 \mathrm{sup} \big\{ \mathcal{C}_1 , \mathcal{C}_2 , \mathcal{C}_3 , \mathcal{C}_4 ,   \mathcal{C}_5 , \mathcal{C}_6     \big\}     , \\ 
\end{align*} }

\noindent in $(*)$,

\begin{align*}
     \mathcal{T}_1  \leq     \underset{1 \leq j \leq N}{\underset{1 \leq i \leq N}{\underset{i \neq j}{\mathrm{sup}}}} \bigg\{          \binom{N}{i} ,  \binom{N}{j} \bigg\}   \mathcal{T}_1 \leq C_1 \mathcal{T}^{\prime}_1   \leq  C \text{ }  {\tiny \mathrm{sup} \bigg\{   \epsilon \epsilon^{\prime}    ,  \epsilon \epsilon^{\prime} \epsilon^{\prime\prime}  , \cdots , \epsilon \epsilon^{\prime} \epsilon^{\prime\prime}  \times \cdots \times \epsilon^{\prime\prime\cdots \prime} ,    \epsilon \delta , \epsilon \delta + \epsilon^{\prime} \delta^{\prime}   , \cdots }  \\ ,     {\tiny   \underset{1 \leq i \leq N}{\sum} \epsilon_i \delta_i    \bigg\} }  \\ \\ \\  \propto \bigg[ \binom{N}{i} +  \binom{N}{j}    \bigg] {\tiny  \mathrm{sup} \bigg\{   \epsilon \epsilon^{\prime}    ,  \epsilon \epsilon^{\prime} \epsilon^{\prime\prime}  , \cdots , \epsilon \epsilon^{\prime} \epsilon^{\prime\prime}  \times \cdots \times \epsilon^{\prime\prime\cdots \prime} ,    \epsilon \delta , \epsilon \delta + \epsilon^{\prime} \delta^{\prime}   , \cdots  ,      \underset{1 \leq i \leq N}{\sum} \epsilon_i \delta_i    \bigg\} }  , \\  
\end{align*}

\noindent for some $C>C_1>0$ upon setting,

{\small \begin{align*}
   \delta \equiv \delta_{i} \delta_{i^{\prime}} \times \cdots \times \delta_{i^{\prime\cdots\prime}}  , \\ \\ \delta^{\prime} \equiv \delta_{j} \delta_{j^{\prime}} \times \cdots \times \delta_{j^{\prime\cdots\prime}}  , \end{align*}
   
   \begin{align*} \epsilon \equiv \epsilon_i \epsilon_{i^{\prime}} , \\ \vdots  \end{align*}
   
   \begin{align*}   \epsilon^{\prime\prime\cdots\prime} \equiv \epsilon_i \epsilon_{i^{\prime}} \times \cdots  \times \epsilon_{i^{\prime\prime\cdots\prime}}, \end{align*}
   
   \begin{align*}
   \epsilon \equiv \epsilon_j \epsilon_{j^{\prime}}  , \end{align*}
   
   \begin{align*} \epsilon^{\prime\cdots\prime} \equiv \epsilon_j \epsilon_{j^{\prime}}  \times \cdots \times \epsilon_{j^{\prime\prime\cdots\prime}} , \\ 
\end{align*} }

\noindent from which we conclude the argument. \boxed{}

\bigskip

\noindent \textit{Proof of Corollary 1}. To demonstrate that the desired inequality with respect to the pullback $\Psi^{-1} \big[ \cdot \big]$ holds, observe,

{\small \begin{align*}
    C \big( N , n \big)                \Psi^{-1} \bigg[  \frac{1}{2} \mathrm{sup} \big\{    \textbf{E}_{\mathcal{Q}_1 \mathcal{Q}_3 \times \cdots \times \mathcal{Q}_N} \big[ \psi \big[  \textbf{E}_{\mathcal{Q}_2 } \big[ \mu \big( Q_1 , Q_2 ,   \cdots , Q_N \big)  \big]     \big] \big]        , \textbf{E}_{\mathcal{Q}_2\mathcal{Q}_3 \times \cdots \times \mathcal{Q}_N } \big[ \psi \big[  \textbf{E}_{\mathcal{Q}_1} \big[ \mu \big( Q_1 , \\  Q_2 ,    \cdots  , Q_N \big)  \big]     \big] \big]  \big\}    \bigg]^{2} \\  \leq      C \big( N , n \big)                \Psi^{-1} \bigg[  \frac{1}{3} \mathrm{sup} \big\{   \textbf{E}_{\mathcal{Q}_1 \mathcal{Q}_3 \times \cdots \times \mathcal{Q}_N} \big[ \psi \big[  \textbf{E}_{\mathcal{Q}_2 } \big[ \mu \big( Q_1 , Q_2 ,   \cdots , Q_N \big)  \big]     \big] \big]         , \textbf{E}_{\mathcal{Q}_2\mathcal{Q}_3 \times \cdots \times \mathcal{Q}_N } \big[ \psi \big[  \textbf{E}_{\mathcal{Q}_1} \big[ \mu \big( Q_1 \\ , Q_2 ,    \cdots  , Q_N \big)  \big]     \big] \big]    ,  \textbf{E}_{\mathcal{Q}_1 \mathcal{Q}_3\mathcal{Q}_4 \times \cdots \times \mathcal{Q}_N } \big[ \psi \big[  \textbf{E}_{\mathcal{Q}_2} \big[ \mu \big( Q_1 , Q_2 ,    \cdots  , Q_N \big)  \big]     \big] \big]    \big\}    \bigg]^{2^2}     \\ \\        \leq    C \big( N , n \big)  \Psi^{-1}  \bigg[  \bigg[ \underset{i \in [ n ] }{\sum} \binom{N}{i}  \bigg]^{-1}   \mathrm{sup}   \big\{        \textbf{E}_{\mathcal{Q}_1} \big[ \psi \big[  \textbf{E}_{\mathcal{Q}_2} \big[ \mu \big( Q_1 , Q_2 ,   \cdots , Q_N \big)  \big]     \big] \big]   , \textbf{E}_{\mathcal{Q}_2} \big[ \psi^{\prime} \big[  \textbf{E}_{\mathcal{Q}_1} \big[ \mu \big( Q_1  , Q_2 , \cdots \\  , Q_N \big)  \big]     \big] \big]  ,   \cdots ,        \textbf{E}_{\mathcal{Q}_N} \big[ \psi \big[  \textbf{E}_{\mathcal{Q}_1} \big[ \mu \big( Q_1  , Q_2 ,  \cdots , Q_N \big)  \big]     \big] \big]  , \cdots  , \textbf{E}_{\mathcal{Q}_N} \big[ \psi^{\prime\cdots \prime}_{N-1} \big[  \textbf{E}_{\mathcal{Q}_1 \times \cdots \times \mathcal{Q}_{N-1}} \big[ \mu \big( Q_1  , Q_2  , \\  \cdots , Q_N \big)  \big]     \big] \big]                   \big\}               \bigg]^{2^N}                   , \\ \\ \equiv        C \big( N , n \big)  \Psi^{-1}  \bigg[ 2^{-N}   \mathrm{sup}   \big\{        \textbf{E}_{\mathcal{Q}_1} \big[ \psi \big[  \textbf{E}_{\mathcal{Q}_2} \big[ \mu \big( Q_1 , Q_2 ,   \cdots , Q_N \big)  \big]     \big] \big]   , \textbf{E}_{\mathcal{Q}_2} \big[ \psi^{\prime} \big[  \textbf{E}_{\mathcal{Q}_1} \big[ \mu \big( Q_1  , Q_2 , \cdots  , Q_N \big)  \big]     \big] \big]  ,  \\  \cdots ,        \textbf{E}_{\mathcal{Q}_N} \big[ \psi \big[  \textbf{E}_{\mathcal{Q}_1} \big[ \mu \big( Q_1  , Q_2 ,  \cdots , Q_N \big)  \big]     \big] \big]  , \cdots  , \textbf{E}_{\mathcal{Q}_N} \big[ \psi^{\prime\cdots \prime}_{N-1} \big[  \textbf{E}_{\mathcal{Q}_1 \times \cdots \times \mathcal{Q}_{N-1}} \big[ \mu \big( Q_1  , Q_2  , \\  \cdots , Q_N \big)  \big]     \big] \big]                   \big\}               \bigg]^{2^N}            . \\ \end{align*}   }

\noindent Hence it suffices to argue,

{\small \begin{align*}
{\tiny C \big( N , n \big)                \Psi^{-1} \bigg[  \frac{1}{2} \mathrm{sup} \big\{    \textbf{E}_{\mathcal{Q}_1 \mathcal{Q}_3 \times \cdots \times \mathcal{Q}_N} \big[ \psi \big[  \textbf{E}_{\mathcal{Q}_2 } \big[ \mu \big( Q_1 , Q_2 ,   \cdots , Q_N \big)  \big]     \big] \big]        , \textbf{E}_{\mathcal{Q}_2\mathcal{Q}_3 \times \cdots \times \mathcal{Q}_N } \big[ \psi \big[  \textbf{E}_{\mathcal{Q}_1} \big[ \mu \big( Q_1 , Q_2 ,    \cdots  , Q_N \big)  \big]     \big] \big]  \big\}    \bigg]^{2^2} } \\ {\tiny  \geq   \textbf{E}_{\mathcal{Q}_1 \times \cdots \times \mathcal{Q}_N} \big[ \mu \big( Q_1 , \cdots , Q_N \big)  \big]    }  , \\ 
\end{align*}}

\noindent or alternatively, that,

{\small \begin{align*}
   {\tiny \Psi^{-1} \bigg[  \frac{1}{2} \mathrm{sup} \big\{    \textbf{E}_{\mathcal{Q}_1 \mathcal{Q}_3 \times \cdots \times \mathcal{Q}_N} \big[ \psi \big[  \textbf{E}_{\mathcal{Q}_2 } \big[ \mu \big( Q_1 , Q_2 ,   \cdots , Q_N \big)  \big]     \big] \big]        , \textbf{E}_{\mathcal{Q}_2\mathcal{Q}_3 \times \cdots \times \mathcal{Q}_N } \big[ \psi \big[  \textbf{E}_{\mathcal{Q}_1} \big[ \mu \big( Q_1 , Q_2 ,    \cdots  , Q_N \big)  \big]     \big] \big]  \big\}    \bigg]^{2^2} }  \\ {\tiny \times    \textbf{E}_{\mathcal{Q}_1 \times \cdots \times \mathcal{Q}_N} \big[ \mu \big( Q_1 ,  \cdots , Q_N \big)  \big]^{-1} \gtrsim 1 }   . 
\end{align*} }

\noindent For shorthand denote,

{\small \begin{align*}
 \mathrm{sup} \big\{    \textbf{E}_{\mathcal{Q}_1 \mathcal{Q}_3 \times \cdots \times \mathcal{Q}_N} \big[ \psi \big[  \textbf{E}_{\mathcal{Q}_2 } \big[ \mu \big( Q_1 , Q_2 ,   \cdots , Q_N \big)  \big]     \big] \big]        , \textbf{E}_{\mathcal{Q}_2\mathcal{Q}_3 \times \cdots \times \mathcal{Q}_N } \big[ \psi \big[  \textbf{E}_{\mathcal{Q}_1} \big[ \mu \big( Q_1 , Q_2 ,    \cdots  , Q_N \big)  \big]     \big] \big]  \big\}    \equiv  \mathrm{sup} \big\{ \textbf{E}_1 \\ , \textbf{E}_2 \big\}   ,  \\ 
\end{align*} }

\noindent from which, by direct computation, write,

{\small \begin{align*}
    \frac{ \mathrm{sup} \big\{ \textbf{E}_1 , \textbf{E}_2 \big\} }{\textbf{E}_{\mathcal{Q}_1 \times \cdots \times \mathcal{Q}_N} \big[ \mu \big( Q_1 , \cdots , Q_N \big)  \big]}    = \mathrm{sup} \bigg\{       \frac{\textbf{E}_1}{\textbf{E}_{\mathcal{Q}_1 \times \cdots \times \mathcal{Q}_N} \big[ \mu \big( Q_1 , \cdots , Q_N \big)  \big]}  ,  \frac{\textbf{E}_2 }{\textbf{E}_{\mathcal{Q}_1 \times \cdots \times \mathcal{Q}_N} \big[ \mu \big( Q_1 , \cdots , Q_N \big)  \big]}      \bigg\} \\ \\ \leq \mathrm{sup} \big\{ C_1 \big( \textbf{E}_1 \big)  , C_2 \big( \textbf{E}_2 \big)  \big\} \equiv \mathrm{sup} \big\{ C_1   , C_2  \big\}  \leq C_1 + C_2 \lesssim 1         .
\end{align*}}

\noindent Straightforwardly, the computation of the supremum amongst countably many expected values implies,

{\small \begin{align*}
    \frac{ \mathrm{sup} \big\{ \textbf{E}_1 , \cdots , \textbf{E}_N  \big\} }{\textbf{E}_{\mathcal{Q}_1 \times \cdots \times \mathcal{Q}_N} \big[ \mu \big( Q_1 , \cdots , Q_N \big)  \big]}    = \mathrm{sup} \bigg\{        \frac{\textbf{E}_1}{\textbf{E}_{\mathcal{Q}_1 \times \cdots \times \mathcal{Q}_N} \big[ \mu \big( Q_1 , \cdots , Q_N \big)  \big]} , \cdots , \frac{\textbf{E}_N }{\textbf{E}_{\mathcal{Q}_1 \times \cdots \times \mathcal{Q}_N} \big[ \mu \big( Q_1 , \cdots , Q_N \big)  \big]}     \bigg\} \\ \\ \leq \mathrm{sup} \big\{ C_1 \big( \textbf{E}_1 \big)  , C_2 \big( \textbf{E}_2 \big) , \cdots , C_2 \big( \textbf{E}_N \big)  \big\} \equiv \mathrm{sup} \big\{ C_1   , C_2 , \cdots , C_N  \big\}  \leq C_1 + C_2 + \cdots + C_N  \lesssim 1         . \\ 
\end{align*}}

\noindent Moreover,

{\small \begin{align*}
{\tiny \frac{1}{C \big( N , n \big)} }  \end{align*}

\begin{align*} {\tiny \times      \frac{ \Psi^{-1} \bigg[   2^{-1}   \mathrm{sup} \big\{  \textbf{E}_{\mathcal{Q}_1 \mathcal{Q}_3 \times \cdots \times \mathcal{Q}_N} \big[ \psi \big[  \textbf{E}_{\mathcal{Q}_2 } \big[ \mu \big( Q_1 , Q_2 ,   \cdots , Q_N \big)  \big]     \big] \big]        , \textbf{E}_{\mathcal{Q}_2\mathcal{Q}_3 \times \cdots \times \mathcal{Q}_N } \big[ \psi \big[  \textbf{E}_{\mathcal{Q}_1} \big[ \mu \big( Q_1 ,   Q_2 ,    \cdots  , Q_N \big)  \big]     \big] \big]       \big\}    \bigg]^{2} }{ \Psi^{-1} \bigg[    2^{-N}       \mathrm{sup} \big\{  \textbf{E}_{\mathcal{Q}_1 \mathcal{Q}_3 \times \cdots \times \mathcal{Q}_N} \big[ \psi \big[  \textbf{E}_{\mathcal{Q}_2 } \big[ \mu \big( Q_1 , Q_2 ,   \cdots , Q_N \big)  \big]     \big] \big]      , \cdots ,          \textbf{E}_{\mathcal{Q}_N} \big[ \psi^{\prime\cdots \prime}_{N-1} \big[  \textbf{E}_{\mathcal{Q}_1 \times \cdots \times \mathcal{Q}_{N-1}} \big[ \mu \big( Q_1  , Q_2  ,   \cdots , Q_N \big)  \big]     \big] \big]                 \big\}                 \bigg]^{2^N}  }    } \end{align*}

\begin{align*} \approx  {\tiny  \frac{1}{C \big( N , n \big)}     \Psi^{-1} \bigg[  2^{-(N+1)}     \mathrm{sup} \big\{  \textbf{E}_{\mathcal{Q}_1 \mathcal{Q}_3 \times \cdots \times \mathcal{Q}_N} \big[ \psi \big[  \textbf{E}_{\mathcal{Q}_2 } \big[ \mu \big( Q_1 , Q_2 ,  \cdots , Q_N \big)  \big]     \big] \big]      , \cdots ,      \textbf{E}_{\mathcal{Q}_N} \big[ \psi^{\prime\cdots \prime}_{N-1} \big[  \textbf{E}_{\mathcal{Q}_1 \times \cdots \times \mathcal{Q}_{N-1}} \big[ \mu \big( Q_1  , Q_2  ,  } \\ \\ {\tiny         \cdots , Q_N \big)  \big]     \big] \big]                 \big\}                                 \bigg]^{2^{N-1}} }  \overset{(\text{LR})}{\underset{\textit{n taken sufficiently large}}{\underset{}{\underset{N \longrightarrow + \infty}{\longrightarrow}}}} 0                    \lesssim 1  , \\ 
\end{align*}}

\noindent from the fact that, in (LR),

{\small
\begin{align*}
  {\tiny  \underset{n \text{ } \textit{sufficiently small}}{\underset{N \longrightarrow + \infty}{\mathrm{lim}}}    \bigg\{ \frac{1}{C \big( N , n \big)}     \Psi^{-1} \bigg[  2^{-(N+1)}     \mathrm{sup} \big\{  \textbf{E}_{\mathcal{Q}_1 \mathcal{Q}_3 \times \cdots \times \mathcal{Q}_N} \big[ \psi \big[  \textbf{E}_{\mathcal{Q}_2 } \big[ \mu \big( Q_1 , Q_2 ,  \cdots , Q_N \big)  \big]     \big] \big]      , \cdots ,  } \\ \\ {\tiny       \textbf{E}_{\mathcal{Q}_N} \big[ \psi^{\prime\cdots \prime}_{N-1} \big[  \textbf{E}_{\mathcal{Q}_1 \times \cdots \times \mathcal{Q}_{N-1}} \big[ \mu \big( Q_1  , Q_2  ,         \cdots , Q_N \big)  \big]     \big] \big]                 \big\}                                 \bigg]^{2^{N-1}}  \bigg\}  }    \\    \\ \\        \approx   -  \underset{n \text{ } \textit{sufficiently small}}{\underset{N \longrightarrow + \infty}{\mathrm{lim}}}      {\tiny \bigg\{  \frac{C^{\prime} \big( N , n \big)}{C \big( N , n \big)^2 }     \Psi^{-1} \bigg[  2^{-(N+1)}     \mathrm{sup} \big\{  \textbf{E}_{\mathcal{Q}_1 \mathcal{Q}_3 \times \cdots \times \mathcal{Q}_N} \big[ \psi \big[  \textbf{E}_{\mathcal{Q}_2 } \big[ \mu \big( Q_1 , Q_2 ,  \cdots , Q_N \big)  \big]     \big] \big]      , \cdots  } \\ \\ {\tiny     ,      \textbf{E}_{\mathcal{Q}_N} \big[ \psi^{\prime\cdots \prime}_{N-1} \big[  \textbf{E}_{\mathcal{Q}_1 \times \cdots \times \mathcal{Q}_{N-1}} \big[ \mu \big( Q_1  , Q_2  ,      \cdots , Q_N \big)  \big]     \big] \big]                 \big\}                                 \bigg]^{2^{N-1}}  \bigg\}  }    \end{align*}

  \begin{align*} +   {\tiny   \underset{n \text{ } \textit{sufficiently small}}{\underset{N \longrightarrow + \infty}{\mathrm{lim}}}  \bigg\{ \frac{1}{C \big( N , n \big)} \big( N-1 \big)  2^{N-1} \mathrm{log}_2 \big( 2 \big)  \bigg\{    \Psi^{-1} \bigg[  2^{-(N+1)}     \mathrm{sup} \big\{  \textbf{E}_{\mathcal{Q}_1 \mathcal{Q}_3 \times \cdots \times \mathcal{Q}_N} \big[ \psi \big[  \textbf{E}_{\mathcal{Q}_2 } \big[ \mu \big( Q_1 , Q_2 ,  \cdots   } \\ \\ {\tiny    , Q_N \big)  \big]     \big] \big]      , \cdots ,      \textbf{E}_{\mathcal{Q}_N} \big[ \psi^{\prime\cdots \prime}_{N-1} \big[  \textbf{E}_{\mathcal{Q}_1 \times \cdots \times \mathcal{Q}_{N-1}} \big[ \mu \big( Q_1  , Q_2  ,       \cdots , Q_N \big)  \big]     \big] \big]                 \big\}   \big\}                              \bigg]^{2^{N-2}}  \big\} \bigg\{    2^{-(N+1)} \mathrm{log}_2 \big( 2 \big)   2^{-N}   }  \\ \\ {\tiny            \times       \mathrm{sup} \big\{  \textbf{E}_{\mathcal{Q}_1 \mathcal{Q}_3 \times \cdots \times \mathcal{Q}_N} \big[ \psi \big[  \textbf{E}_{\mathcal{Q}_2 } \big[ \mu \big( Q_1 , Q_2 ,  \cdots ,    Q_N \big)  \big]     \big] \big]      , \cdots ,      \textbf{E}_{\mathcal{Q}_N} \big[ \psi^{\prime\cdots \prime}_{N-1} \big[  \textbf{E}_{\mathcal{Q}_1 \times \cdots \times \mathcal{Q}_{N-1}} \big[ \mu \big( Q_1  , Q_2  ,          } \\ \\ {\tiny   \cdots , Q_N \big)  \big]     \big] \big]                 \big\}                 } \\ \\  {\tiny    +      2^{-(N+1)} \mathrm{sup}  \big\{            \big\{  \big\{ \textbf{E}_{\mathcal{Q}_1 \mathcal{Q}_3 \times \cdots \times \mathcal{Q}_N} \big[ \psi \big[  \textbf{E}_{\mathcal{Q}_2 } \big[ \mu \big( Q_1 , Q_2 ,  \cdots ,    Q_N \big)  \big]     \big] \big]   \big\}^{\prime}   , \cdots ,      \big\{ \textbf{E}_{\mathcal{Q}_N} \big[ \psi^{\prime\cdots \prime}_{N-1} \big[  \textbf{E}_{\mathcal{Q}_1 \times \cdots \times \mathcal{Q}_{N-1}} \big[ \mu \big( Q_1  , Q_2  ,          } \\ \\ {\tiny   \cdots , Q_N \big)  \big]     \big] \big]          \big\}^{\prime}       \big\}            \bigg\}      } \\ \end{align*}

  \begin{align*} \approx    -  {\tiny \bigg\{   \underset{n \text{ } \textit{sufficiently small}}{\underset{N \longrightarrow + \infty}{\mathrm{lim}}}      \frac{C^{\prime} \big( N , n \big)}{C \big( N , n \big)^2 } \bigg\} }  {\tiny  \bigg\{  \underset{n \text{ } \textit{sufficiently small}}{\underset{N \longrightarrow + \infty}{\mathrm{lim}}}   \bigg\{                       \underset{1 \leq i \leq N}{\prod} \bigg\{  - q^{-1}_i \mathrm{log}_2 \bigg[  - N     +   2^{-(N+1)}    } \\ \\ {\tiny  \times   \mathrm{sup} \big\{  \textbf{E}_{\mathcal{Q}_1 \mathcal{Q}_3 \times \cdots \times \mathcal{Q}_N} \big[ \psi \big[  \textbf{E}_{\mathcal{Q}_2 } \big[ \mu \big( Q_1 , Q_2 ,  \cdots , Q_N \big)  \big]     \big] \big]     ,   \cdots     ,      \textbf{E}_{\mathcal{Q}_N} \big[ \psi^{\prime\cdots \prime}_{N-1} \big[  \textbf{E}_{\mathcal{Q}_1 \times \cdots \times \mathcal{Q}_{N-1}} \big[ \mu \big( Q_1  , Q_2 } \\ \\ {\tiny  ,      \cdots , Q_N \big)  \big]     \big] \big]                 \big\}    \bigg]         \bigg\}  \bigg\}  \bigg\}  }   \\  \\ {\tiny   + \underset{n \text{ } \textit{sufficiently small}}{\underset{N \longrightarrow + \infty}{\mathrm{lim}}}  \bigg\{ \frac{1}{C \big( N , n \big)} \big( N-1 \big)  2^{N-1} \mathrm{log}_2 \big( 2 \big)  \bigg\{   \underset{1 \leq i \leq N}{\prod} \bigg\{  - q^{-1}_i  \mathrm{log}_2  \bigg[ -  N  +    2^{-(N+1)}     \mathrm{sup} \big\{  \textbf{E}_{\mathcal{Q}_1 \mathcal{Q}_3 \times \cdots \times \mathcal{Q}_N} \big[  } \\ \\ {\tiny  \psi \big[  \textbf{E}_{\mathcal{Q}_2 } \big[ \mu \big( Q_1 ,     Q_2 ,  \cdots   , Q_N \big)  \big]     \big] \big]      , \cdots ,      \textbf{E}_{\mathcal{Q}_N} \big[ \psi^{\prime\cdots \prime}_{N-1} \big[  \textbf{E}_{\mathcal{Q}_1 \times \cdots \times \mathcal{Q}_{N-1}} \big[ \mu \big( Q_1  , Q_2  ,       \cdots , Q_N \big)  \big]     \big] \big]                 \big\}   \big\}                              \bigg]^{2^{N-2}} \bigg\}  \bigg\}  }  \\ \\ {\tiny \times \bigg\{    2^{-(N+1)} \mathrm{log}_2 \big( 2 \big)    2^{-N}       \mathrm{sup} \big\{  \textbf{E}_{\mathcal{Q}_1 \mathcal{Q}_3 \times \cdots \times \mathcal{Q}_N} \big[ \psi \big[  \textbf{E}_{\mathcal{Q}_2 } \big[ \mu \big( Q_1 , Q_2 ,           \cdots ,    Q_N \big)  \big]     \big] \big]      ,  }  \\ \\  {\tiny  \cdots ,      \textbf{E}_{\mathcal{Q}_N} \big[ \psi^{\prime\cdots \prime}_{N-1} \big[  \textbf{E}_{\mathcal{Q}_1 \times \cdots \times \mathcal{Q}_{N-1}} \big[ \mu \big( Q_1  , Q_2  ,     \cdots , Q_N \big)  \big]     \big] \big]                 \big\}                 } \\ \\  {\tiny    +      2^{-(N+1)} \mathrm{sup}  \big\{            \big\{  \big\{ \textbf{E}_{\mathcal{Q}_1 \mathcal{Q}_3 \times \cdots \times \mathcal{Q}_N} \big[ \psi \big[  \textbf{E}_{\mathcal{Q}_2 } \big[ \mu \big( Q_1 , Q_2 ,  \cdots ,    Q_N \big)  \big]     \big] \big]   \big\}^{\prime}   , \cdots  } \\ \\ {\tiny ,      \big\{ \textbf{E}_{\mathcal{Q}_N} \big[ \psi^{\prime\cdots \prime}_{N-1} \big[  \textbf{E}_{\mathcal{Q}_1 \times \cdots \times \mathcal{Q}_{N-1}} \big[ \mu \big( Q_1  , Q_2  ,             \cdots , Q_N \big)  \big]     \big] \big]          \big\}^{\prime}       \big\}            \bigg\}      }    ,     \\ \\      
\end{align*} }

\noindent by L'Hopital's rule, where,

{\small
\begin{align*}
  {\tiny    \underset{n \text{ } \textit{sufficiently small}}{\underset{N \longrightarrow + \infty}{\mathrm{lim}}}      \frac{C^{\prime} \big( N , n \big)}{C \big( N , n \big)^2 }    =      \underset{n \text{ } \textit{sufficiently small}}{\underset{N \longrightarrow + \infty}{\mathrm{lim}}}     \frac{ \bigg\{ N        \underset{i \in [ n ] }{\sum} \binom{N}{i}  \bigg\}^{\prime}            }{ \bigg\{ N        \underset{i \in [ n ] }{\sum} \binom{N}{i}  \bigg\}^2 }  =        \underset{n \text{ } \textit{sufficiently small}}{\underset{N \longrightarrow + \infty}{\mathrm{lim}}}      \frac{ \frac{\partial}{\partial N} \bigg\{ N        \underset{i \in [ n ] }{\sum} \binom{N}{i}  \bigg\}           }{ \bigg\{ N        \underset{i \in [ n ] }{\sum} \binom{N}{i}  \bigg\}^2 }              }  \\ \end{align*}

  \begin{align*}   {\tiny  =        \underset{n \text{ } \textit{sufficiently small}}{\underset{N \longrightarrow + \infty}{\mathrm{lim}}}      \frac{           \underset{i \in [ n ] }{\sum} \binom{N}{i} + N \bigg\{  \frac{\partial}{\partial N} \underset{i \in [ n ] }{\sum} \binom{N}{i}     \bigg\}       }{ \bigg\{ N        \underset{i \in [ n ] }{\sum} \binom{N}{i}  \bigg\}^2 }        }    \\  \end{align*}

  \begin{align*}   {\tiny  =   \underset{n \text{ } \textit{sufficiently small}}{\underset{N \longrightarrow + \infty}{\mathrm{lim}}}      \frac{           \underset{i \in [ n ] }{\sum} \binom{N}{i} + N \bigg\{   \underset{i \in [ n ] }{\sum} \binom{N}{i}  \bigg\{        \frac{\partial}{\partial N}      \bigg\{     \mathrm{log}_2                \big[  N  ! \big]                    \bigg\}     -        \frac{\partial}{\partial N}     \bigg\{       \mathrm{log}_2              \big[        \big( N - i \big) !      \big]     \bigg\}       \bigg\}    \bigg\}       }{ \bigg\{ N        \underset{i \in [ n ] }{\sum} \binom{N}{i}  \bigg\}^2 }            } \\ \\ \\  {\tiny =      \underset{n \text{ } \textit{sufficiently small}}{\underset{N \longrightarrow + \infty}{\mathrm{lim}}}      \frac{           \underset{i \in [ n ] }{\sum} \binom{N}{i} + N \bigg\{   \underset{i \in [ n ] }{\sum} \binom{N}{i}  \bigg\{          \bigg\{     \frac{1}{N!} \frac{\partial}{\partial N} \big[ N! \big]                   \bigg\}     -          \bigg\{          \frac{1}{ \big( N - i \big) ! }  \frac{\partial}{\partial N} \big[ \big( N - i \big) !  \big]      \big]     \bigg\}       \bigg\}    \bigg\}       }{ \bigg\{ N        \underset{i \in [ n ] }{\sum} \binom{N}{i}  \bigg\}^2 }                                                                               } \\ \end{align*}

  \begin{align*} {\tiny =      \underset{n \text{ } \textit{sufficiently small}}{\underset{N \longrightarrow + \infty}{\mathrm{lim}}}      \frac{           \underset{i \in [ n ] }{\sum} \binom{N}{i}}{ \bigg\{ N        \underset{i \in [ n ] }{\sum} \binom{N}{i}  \bigg\}^2 } +    \underset{n \text{ } \textit{sufficiently small}}{\underset{N \longrightarrow + \infty}{\mathrm{lim}}}    \frac{N \bigg\{   \underset{i \in [ n ] }{\sum} \binom{N}{i}  \bigg\{          \bigg\{     \frac{1}{N!} \frac{\partial}{\partial N} \big[ N! \big]                   \bigg\}     -          \bigg\{          \frac{1}{ \big( N - i \big) ! }  \frac{\partial}{\partial N} \big[ \big( N - i \big) !  \big]      \big]     \bigg\}       \bigg\}    \bigg\}       }{ \bigg\{ N        \underset{i \in [ n ] }{\sum} \binom{N}{i}  \bigg\}^2 }                                                                       }  \\ \\ \\  {\tiny \overset{(*)}{\sim}        \underset{n \text{ } \textit{sufficiently small}}{\underset{N \longrightarrow + \infty}{\mathrm{lim}}}      \bigg\{ N        \underset{i \in [ n ] }{\sum} \binom{N}{i}  \bigg\}^{-1}       +     \underset{n \text{ } \textit{sufficiently small}}{\underset{N \longrightarrow + \infty}{\mathrm{lim}}}    \frac{N \bigg\{   \underset{i \in [ n ] }{\sum} \binom{N}{i}  \bigg\{          \bigg\{     \frac{1}{N!} \frac{\partial}{\partial N} \big[ N! \big]                   \bigg\}     -          \bigg\{          \frac{1}{ \big( N - i \big) ! }  \frac{\partial}{\partial N} \big[ \big( N - i \big) !  \big]      \big]     \bigg\}       \bigg\}    \bigg\}       }{ \bigg\{ N        \underset{i \in [ n ] }{\sum} \binom{N}{i}  \bigg\}^2 }                                                                                  }      \\ \end{align*}

  \begin{align*} {\tiny \approx         \underset{n \text{ } \textit{sufficiently small}}{\underset{N \longrightarrow + \infty}{\mathrm{lim}}}    \frac{N \bigg\{   \underset{i \in [ n ] }{\sum} \binom{N}{i}  \bigg\{          \bigg\{     \frac{1}{N!} \frac{\partial}{\partial N} \big[ N! \big]                   \bigg\}     -          \bigg\{          \frac{1}{ \big( N - i \big) ! }  \frac{\partial}{\partial N} \big[ \big( N - i \big) !  \big]      \big]     \bigg\}       \bigg\}    \bigg\}       }{ \bigg\{ N        \underset{i \in [ n ] }{\sum} \binom{N}{i}  \bigg\}^2 }                                                                                         } \\ \end{align*}

  {\small \begin{align*}     {\tiny     \approx           \underset{n \text{ } \textit{sufficiently small}}{\underset{N \longrightarrow + \infty}{\mathrm{lim}}}                         \bigg\{ N        \underset{i \in [ n ] }{\sum} \binom{N}{i}  \bigg\}^{-1}                                    \frac{1}{N!} \frac{\partial}{\partial N} \big[ N! \big]         -      \underset{n \text{ } \textit{sufficiently small}}{\underset{N \longrightarrow + \infty}{\mathrm{lim}}}                         \bigg\{ N        \underset{i \in [ n ] }{\sum} \binom{N}{i}  \bigg\}^{-2}      \frac{1}{ \big( N - i \big) ! }  \frac{\partial}{\partial N} \big[ \big( N - i \big) !  \big]      \big]                                                                                                 } \\ \end{align*}

  \begin{align*} {\tiny     =     \underset{n \text{ } \textit{sufficiently small}}{\underset{N \longrightarrow + \infty}{\mathrm{lim}}}                         \bigg\{ N        \underset{i \in [ n ] }{\sum} \binom{N}{i}  \bigg\}^{-1}                                    \frac{1}{N!}                      \bigg\{   N!      \frac{  \frac{\partial}{\partial N} \bigg\{ \int_0^{+\infty}   t^{N+1} e^{-t} \mathrm{d} t   \bigg\}  }{    \int_0^{+\infty}   t^{N+1} e^{-t} \mathrm{d} t        }             \bigg\}      -      \underset{n \text{ } \textit{sufficiently small}}{\underset{N \longrightarrow + \infty}{\mathrm{lim}}}                         \bigg\{ N        \underset{i \in [ n ] }{\sum} \binom{N}{i}  \bigg\}^{-2}      \frac{1}{ \big( N - i \big) ! }      } \\ \\ {\tiny \times    \bigg\{ \big( N - i \big) !                                        \frac{  \frac{\partial}{\partial N} \bigg\{ \int_0^{+\infty}   t^{N-i+1} e^{-t} \mathrm{d} t   \bigg\}  }{    \int_0^{+\infty}   t^{N-i+1} e^{-t} \mathrm{d} t        }                         \bigg\}                                                                                                    }   \\ \end{align*}

  \begin{align*} {\tiny =          \underset{n \text{ } \textit{sufficiently small}}{\underset{N \longrightarrow + \infty}{\mathrm{lim}}}                         \bigg\{ N        \underset{i \in [ n ] }{\sum} \binom{N}{i}  \bigg\}^{-1}                                                      \bigg\{      \frac{    t^{N+1} e^{-t} \mathrm{d} t    }{    \int_0^{+\infty}   t^{N+1} e^{-t} \mathrm{d} t        }             \bigg\}      -      \underset{n \text{ } \textit{sufficiently small}}{\underset{N \longrightarrow + \infty}{\mathrm{lim}}}                         \bigg\{ N        \underset{i \in [ n ] }{\sum} \binom{N}{i}  \bigg\}^{-2}       \bigg\{                                     \frac{    t^{N-i+1} e^{-t} \mathrm{d} t   }{    \int_0^{+\infty}   t^{N-i+1} e^{-t} \mathrm{d} t        }                         \bigg\}                                                                                                    }   \\ \end{align*}

 \begin{align*} {\tiny  =            \bigg\{ \underset{n \text{ } \textit{sufficiently small}}{\underset{N \longrightarrow + \infty}{\mathrm{lim}}}                         \bigg\{ N        \underset{i \in [ n ] }{\sum} \binom{N}{i}  \bigg\}^{-1}                                     \bigg\}      \bigg\{      \underset{n \text{ } \textit{sufficiently small}}{\underset{N \longrightarrow + \infty}{\mathrm{lim}}}                    \frac{    t^{N+1} e^{-t} \mathrm{d} t    }{    \int_0^{+\infty}   t^{N+1} e^{-t} \mathrm{d} t        }         \bigg\}     -   \bigg\{    \underset{n \text{ } \textit{sufficiently small}}{\underset{N \longrightarrow + \infty}{\mathrm{lim}}}                         \bigg\{ N        \underset{i \in [ n ] }{\sum} \binom{N}{i}  \bigg\}^{-2}  \bigg\}  }   \\ \\ {\tiny \times  \bigg\{   \underset{n \text{ } \textit{sufficiently small}}{\underset{N \longrightarrow + \infty}{\mathrm{lim}}}                                          \frac{    t^{N-i+1} e^{-t} \mathrm{d} t   }{    \int_0^{+\infty}   t^{N-i+1} e^{-t} \mathrm{d} t   }                              \bigg\}      }    \\                    \end{align*}

 \begin{align*} {\tiny  \overset{(\mathrm{LR})}{=}            \bigg\{ \underset{n \text{ } \textit{sufficiently small}}{\underset{N \longrightarrow + \infty}{\mathrm{lim}}}                         \bigg\{ N        \underset{i \in [ n ] }{\sum} \binom{N}{i}  \bigg\}^{-1}                                     \bigg\}      \bigg\{      \underset{n \text{ } \textit{sufficiently small}}{\underset{N \longrightarrow + \infty}{\mathrm{lim}}}             \frac{\partial}{\partial N}     \bigg\{      \frac{    t^{N+1} e^{-t} \mathrm{d} t    }{    \int_0^{+\infty}   t^{N+1} e^{-t} \mathrm{d} t        }             \bigg\}  \bigg\}      -   \bigg\{    \underset{n \text{ } \textit{sufficiently small}}{\underset{N \longrightarrow + \infty}{\mathrm{lim}}}                         \bigg\{ N      }   \\ \\ {\tiny \times     \underset{i \in [ n ] }{\sum} \binom{N}{i}  \bigg\}^{-2}  \bigg\}    \bigg\{   \underset{n \text{ } \textit{sufficiently small}}{\underset{N \longrightarrow + \infty}{\mathrm{lim}}}     \frac{\partial}{\partial N}   \bigg\{                                     \frac{    t^{N-i+1} e^{-t} \mathrm{d} t   }{    \int_0^{+\infty}   t^{N-i+1} e^{-t} \mathrm{d} t   }                              \bigg\}   \bigg\}      }                                                                                              \\ \\ \\  {\tiny  \approx 0   ,                                                                       }                                                                    \\                     \end{align*} }

\noindent corresponding to the first term, where in $(*)$ one makes use of the fact that,

{\small

\begin{align*}
    {\tiny  \frac{   \underset{n \text{ } \textit{sufficiently small}}{\underset{N \longrightarrow + \infty}{\mathrm{lim}}}      \frac{           \underset{i \in [ n ] }{\sum} \binom{N}{i}}{ \bigg\{ N        \underset{i \in [ n ] }{\sum} \binom{N}{i}  \bigg\}^2 }}{ \underset{n \text{ } \textit{sufficiently small}}{\underset{N \longrightarrow + \infty}{\mathrm{lim}}}      \frac{1}{  N        \underset{i \in [ n ] }{\sum} \binom{N}{i}  }         }          =          \underset{n \text{ } \textit{sufficiently small}}{\underset{N \longrightarrow + \infty}{\mathrm{lim}}}                 \frac{ \frac{           \underset{i \in [ n ] }{\sum} \binom{N}{i}}{ \bigg\{ N        \underset{i \in [ n ] }{\sum} \binom{N}{i}  \bigg\}^2 }}{  \frac{1}{  N        \underset{i \in [ n ] }{\sum} \binom{N}{i}  }        }                \approx     \underset{n \text{ } \textit{sufficiently small}}{\underset{N \longrightarrow + \infty}{\mathrm{lim}}}           \frac{ \frac{  N^{-1}         }{        \bigg\{  \underset{i \in [ n ] }{\sum} \binom{N}{i}  \bigg\}^2  }}{  \frac{1}{  N^2         \underset{i \in [ n ] }{\sum} \binom{N}{i}  }        }                                               } \\ \\ {\tiny         =     \underset{n \text{ } \textit{sufficiently small}}{\underset{N \longrightarrow + \infty}{\mathrm{lim}}}  \text{ } \bigg\{   N    \underset{i \in [ n ] }{\sum} \binom{N}{i}     \bigg\}       }        . \\ 
\end{align*}

}

\noindent Proceeding, one also computes the second desired limit from the product of,

  \begin{align*} \\  {\tiny    \underset{n \text{ } \textit{sufficiently small}}{\underset{N \longrightarrow + \infty}{\mathrm{lim}}}     \Psi^{-1} \bigg[  2^{-(N+1)}     \mathrm{sup} \big\{  \textbf{E}_{\mathcal{Q}_1 \mathcal{Q}_3 \times \cdots \times \mathcal{Q}_N} \big[ \psi \big[  \textbf{E}_{\mathcal{Q}_2 } \big[ \mu \big( Q_1 , Q_2 ,  \cdots , Q_N \big)  \big]     \big] \big]      , \cdots     } \\ \\ {\tiny    ,      \textbf{E}_{\mathcal{Q}_N} \big[ \psi^{\prime\cdots \prime}_{N-1} \big[  \textbf{E}_{\mathcal{Q}_1 \times \cdots \times \mathcal{Q}_{N-1}} \big[ \mu \big( Q_1  , Q_2  ,      \cdots , Q_N \big)  \big]     \big] \big]                 \big\}                                 \bigg]^{2^{N-1}}   } \\ \\ \\ = {\tiny   \underset{n \text{ } \textit{sufficiently small}}{\underset{N \longrightarrow + \infty}{\mathrm{lim}}}   \bigg\{                       \underset{1 \leq i \leq N}{\prod} \bigg\{  - q^{-1}_i \mathrm{log}_2 \bigg[  - N  +   2^{-(N+1)}     \mathrm{sup} \big\{  \textbf{E}_{\mathcal{Q}_1 \mathcal{Q}_3 \times \cdots \times \mathcal{Q}_N} \big[ \psi \big[  \textbf{E}_{\mathcal{Q}_2 } \big[ \mu \big( Q_1 } \\ \\ {\tiny   , Q_2 ,  \cdots   , Q_N \big)  \big]     \big] \big]     , \cdots     ,      \textbf{E}_{\mathcal{Q}_N} \big[ \psi^{\prime\cdots \prime}_{N-1} \big[  \textbf{E}_{\mathcal{Q}_1 \times \cdots \times \mathcal{Q}_{N-1}} \big[ \mu \big( Q_1  , Q_2  ,      \cdots , Q_N \big)  \big]     \big] \big]                 \big\}    \bigg]^{2^{N-1}}          \bigg\}  \bigg\}   } , \\   \\ \end{align*}
  }

\noindent with,

 {\small \begin{align*} {\tiny                         \underset{n \text{ } \textit{sufficiently small}}{\underset{N \longrightarrow + \infty}{\mathrm{lim}}}  \bigg\{ \frac{1}{C \big( N , n \big)} \big( N-1 \big)  2^{N-1} \mathrm{log}_2 \big( 2 \big)                                                                                                                                                      \bigg\}            =      \underset{n \text{ } \textit{sufficiently small}}{\underset{N \longrightarrow + \infty}{\mathrm{lim}}}  \bigg\{ \frac{1}{  N        \underset{i \in [ n ] }{\sum} \binom{N}{i}                                           } \big( N-1 \big)  2^{N-1} \mathrm{log}_2 \big( 2 \big)                                                                                                                                                      \bigg\}                                                                                                   } \\ \\ \\ {\tiny \equiv   \mathrm{log}_2 \big( 2 \big)           \bigg\{           \underset{n \text{ } \textit{sufficiently small}}{\underset{N \longrightarrow + \infty}{\mathrm{lim}}}   \bigg\{  N        \underset{i \in [ n ] }{\sum} \binom{N}{i}                 \bigg\}    \frac{1}{ \big( N-1 \big) } 2^{- \big( N - 1 \big) }                                             \bigg\}^{-1}                                  ,                 }  \\ \\                
\end{align*}

}

\noindent and also with,

{\small

\begin{align*}
     {\tiny \underset{n \text{ } \textit{sufficiently small}}{\underset{N \longrightarrow + \infty}{\mathrm{lim}}}     \bigg\{    2^{-(N+1)} \mathrm{log}_2 \big( 2 \big)    2^{-N}           \mathrm{sup} \big\{  \textbf{E}_{\mathcal{Q}_1 \mathcal{Q}_3 \times \cdots \times \mathcal{Q}_N} \big[ \psi \big[  \textbf{E}_{\mathcal{Q}_2 } \big[ \mu \big( Q_1 , Q_2 ,           \cdots ,    Q_N \big)  \big]     \big] \big]      ,  \cdots }  \\ \\ {\tiny         , \textbf{E}_{\mathcal{Q}_N} \big[ \psi^{\prime\cdots \prime}_{N-1} \big[  \textbf{E}_{\mathcal{Q}_1 \times \cdots \times \mathcal{Q}_{N-1}} \big[ \mu \big( Q_1  , Q_2  ,     \cdots , Q_N \big)  \big]     \big] \big]                 \big\}                +      2^{-(N+1)} \mathrm{sup}  \big\{            \big\{  \big\{ \textbf{E}_{\mathcal{Q}_1 \mathcal{Q}_3 \times \cdots \times \mathcal{Q}_N} \big[ \psi \big[  \textbf{E}_{\mathcal{Q}_2 } \big[ \mu \big( Q_1 ,   } \\ \\  {\tiny  Q_2 ,     \cdots ,    Q_N \big)  \big]     \big] \big]   \big\}^{\prime}   , \cdots  ,      \big\{ \textbf{E}_{\mathcal{Q}_N} \big[ \psi^{\prime\cdots \prime}_{N-1} \big[  \textbf{E}_{\mathcal{Q}_1 \times \cdots \times \mathcal{Q}_{N-1}} \big[ \mu \big( Q_1  , Q_2  ,             \cdots , Q_N \big)  \big]     \big] \big]          \big\}^{\prime}       \big\}            \bigg\}      }                            , \\ \\  
\end{align*}

}

\noindent through the following computation,

{\small

\begin{align*}
       {\tiny \mathrm{log}_2 \big( 2 \big)^2   \underset{n \text{ } \textit{sufficiently small}}{\underset{N \longrightarrow + \infty}{\mathrm{lim}}}   \bigg\{    \bigg\{                 \frac{1}{C \big( N , n \big)     \frac{1}{\big( N-1 \big)  }  2^{N+2}  }            \bigg\{                       \underset{1 \leq i \leq N}{\prod} \bigg\{  - q^{-1}_i \mathrm{log}_2 \bigg[  - N  +   2^{-(N+1)}     \mathrm{sup} \big\{  \textbf{E}_{\mathcal{Q}_1 \mathcal{Q}_3 \times \cdots \times \mathcal{Q}_N} \big[ \psi \big[  \textbf{E}_{\mathcal{Q}_2 } \big[ } \\ \\ {\tiny   \mu \big( Q_1   , Q_2 ,  \cdots   , Q_N \big)  \big]     \big] \big]     , \cdots     ,      \textbf{E}_{\mathcal{Q}_N} \big[ \psi^{\prime\cdots \prime}_{N-1} \big[  \textbf{E}_{\mathcal{Q}_1 \times \cdots \times \mathcal{Q}_{N-1}} \big[ \mu \big( Q_1  , Q_2  ,      \cdots , Q_N \big)  \big]     \big] \big]                 \big\}    \bigg]^{2^{N-1}}          \bigg\}  \bigg\}                                                                                                                                                                                                                                                                                                                                                                                                                      \bigg\}                                       \bigg\}   \bigg\{  \mathrm{sup} \big\{  \textbf{E}_{\mathcal{Q}_1 \mathcal{Q}_3 \times \cdots \times \mathcal{Q}_N} \big[ \cdots     }   \\ \\ {\tiny            \cdots          \psi \big[  \textbf{E}_{\mathcal{Q}_2 } \big[ \mu \big( Q_1 , Q_2 ,           \cdots ,    Q_N \big)  \big]     \big] \big]      ,   \cdots ,      \textbf{E}_{\mathcal{Q}_N} \big[ \psi^{\prime\cdots \prime}_{N-1} \big[  \textbf{E}_{\mathcal{Q}_1 \times \cdots \times \mathcal{Q}_{N-1}} \big[ \mu \big( Q_1  , Q_2   }  \\ \\ {\tiny ,     \cdots , Q_N \big)  \big]     \big] \big]                 \big\}                +      2^{-(N+1)} \mathrm{sup}  \big\{            \big\{  \big\{ \textbf{E}_{\mathcal{Q}_1 \mathcal{Q}_3 \times \cdots \times \mathcal{Q}_N} \big[ \psi \big[  \textbf{E}_{\mathcal{Q}_2 } \big[ \mu \big( Q_1 , Q_2      } \\ \\  {\tiny  ,  \cdots ,    Q_N \big)  \big]     \big] \big]   \big\}^{\prime}   , \cdots  ,      \big\{ \textbf{E}_{\mathcal{Q}_N} \big[ \psi^{\prime\cdots \prime}_{N-1} \big[  \textbf{E}_{\mathcal{Q}_1 \times \cdots \times \mathcal{Q}_{N-1}} \big[ \mu \big( Q_1  , Q_2  ,            } \\ \\ {\tiny   \cdots , Q_N \big)  \big]     \big] \big]          \big\}^{\prime}       \big\}            \bigg\}                                                }     ,    \\ \end{align*}

       \begin{align*} {\tiny  \overset{(\mathrm{LR})}{= }        \mathrm{log}_2 \big( 2 \big)^2   \underset{n \text{ } \textit{sufficiently small}}{\underset{N \longrightarrow + \infty}{\mathrm{lim}}}   \bigg\{                             \frac{1}{ \frac{\partial}{\partial N} \bigg\{ C \big( N , n \big)     \frac{1}{\big( N-1 \big)  }  2^{N+2} \bigg\}  }        \frac{\partial}{\partial N}      \bigg\{                        \underset{1 \leq i \leq N}{\prod} \bigg\{  - q^{-1}_i \mathrm{log}_2 \bigg[  - N  +   2^{-(N+1)}  } \\ \\ {\tiny \times    \mathrm{sup} \big\{  \textbf{E}_{\mathcal{Q}_1 \mathcal{Q}_3 \times \cdots \times \mathcal{Q}_N} \big[    \psi \big[  \textbf{E}_{\mathcal{Q}_2 } \big[ \mu \big( Q_1   , Q_2 ,  \cdots   , Q_N \big)  \big]     \big] \big]     , \cdots     ,      \textbf{E}_{\mathcal{Q}_N} \big[ \psi^{\prime\cdots \prime}_{N-1} \big[  \textbf{E}_{\mathcal{Q}_1 \times \cdots \times \mathcal{Q}_{N-1}} \big[ \mu \big( Q_1  , Q_2  ,      }   \\ \\  {\tiny           \cdots , Q_N \big)  \big]     \big] \big]                 \big\}    \bigg]^{2^{N-1}}          \bigg\}  \bigg\}                                                                                                                                                                                                                                                                                                                                                                                                                      \bigg\}                                       \bigg\}   \bigg\{  \mathrm{sup} \big\{  \textbf{E}_{\mathcal{Q}_1 \mathcal{Q}_3 \times \cdots \times \mathcal{Q}_N} \big[ \cdots       \cdots          \psi \big[  \textbf{E}_{\mathcal{Q}_2 } \big[ \mu \big( Q_1 , Q_2 ,           \cdots ,    Q_N \big)  \big]     \big] \big]      ,   \cdots }  \\ \\ {\tiny   ,      \textbf{E}_{\mathcal{Q}_N} \big[ \psi^{\prime\cdots \prime}_{N-1} \big[  \textbf{E}_{\mathcal{Q}_1 \times \cdots \times \mathcal{Q}_{N-1}} \big[ \mu \big( Q_1  , Q_2  ,     \cdots , Q_N \big)  \big]     \big] \big]                 \big\}                +      2^{-(N+1)} } \\ \\ {\tiny \times \mathrm{sup}  \big\{            \big\{  \big\{ \textbf{E}_{\mathcal{Q}_1 \mathcal{Q}_3 \times \cdots \times \mathcal{Q}_N} \big[ \psi \big[  \textbf{E}_{\mathcal{Q}_2 } \big[ \mu \big( Q_1 , Q_2     ,  \cdots ,    Q_N \big)  \big]     \big] \big]   \big\}^{\prime}   , \cdots   } \\ \\ {\tiny ,      \big\{ \textbf{E}_{\mathcal{Q}_N} \big[ \psi^{\prime\cdots \prime}_{N-1} \big[  \textbf{E}_{\mathcal{Q}_1 \times \cdots \times \mathcal{Q}_{N-1}} \big[ \mu \big( Q_1  , Q_2  ,              \cdots , Q_N \big)  \big]     \big] \big]          \big\}^{\prime}       \big\}            \bigg\}                                                                                                                                                                         \bigg\}                                                                                                            }  \\ \end{align*}

       \begin{align*} {\tiny   \approx      -   \mathrm{log}_2 \big( 2 \big)^2    \underset{n \text{ } \textit{sufficiently small}}{\underset{N \longrightarrow + \infty}{\mathrm{lim}}}   \bigg\{                  \bigg\{      \bigg\{          \frac{\partial}{\partial N } C \big( N , n \big) \frac{1}{\big( N - 1 \big)} 2^{N+2}                                                                                                                       - C \big( N , n \big) \frac{1}{\big( N - 1 \big)^2} 2^{N+2}                                                 + C \big( N , n \big) \frac{1}{\big( N - 1 \big)}  2^{N+2}                                                                }                   \\ \\ {\tiny \times   \mathrm{log}_2 \big( 2 \big)    \bigg\}^{-1}    \underset{1 \leq i \leq N}{\prod} q^{-1}_i \bigg\{  2^{N-1} \mathrm{log}_2 \big( 2 \big)  \mathrm{log}_2  \bigg[                                 - N  +   2^{-(N+1)}     \mathrm{sup} \big\{  \textbf{E}_{\mathcal{Q}_1 \mathcal{Q}_3 \times \cdots \times \mathcal{Q}_N} \big[   \psi \big[  \textbf{E}_{\mathcal{Q}_2 } \big[ \mu \big( Q_1   , Q_2 ,  \cdots } \\ \\ {\tiny    , Q_N \big)  \big]     \big] \big]     , \cdots     ,      \textbf{E}_{\mathcal{Q}_N} \big[ \psi^{\prime\cdots \prime}_{N-1} \big[  \textbf{E}_{\mathcal{Q}_1 \times \cdots \times \mathcal{Q}_{N-1}} \big[ \mu \big( Q_1  , Q_2  ,      \cdots , Q_N \big)  \big]     \big] \big]                 \big\}    \bigg]                                                                           }  \\ \\ {\tiny  +    2^{N-1}    \mathrm{log}_2  \bigg[                                 - N  +   2^{-(N+1)}     \mathrm{sup} \big\{  \textbf{E}_{\mathcal{Q}_1 \mathcal{Q}_3 \times \cdots \times \mathcal{Q}_N} \big[   \psi \big[  \textbf{E}_{\mathcal{Q}_2 } \big[ \mu \big( Q_1   , Q_2 ,  \cdots } \\ \\  {\tiny    , Q_N \big)  \big]     \big] \big]     , \cdots     ,      \textbf{E}_{\mathcal{Q}_N} \big[ \psi^{\prime\cdots \prime}_{N-1} \big[  \textbf{E}_{\mathcal{Q}_1 \times \cdots \times \mathcal{Q}_{N-1}} \big[ \mu \big( Q_1  , Q_2  ,      \cdots , Q_N \big)  \big]     \big] \big]                 \big\}    \bigg]                 } \\ \\ {\tiny \times          \bigg\{      2^{-\big( N + 1 \big)}   \mathrm{log}_2 \big( 2 \big)    \mathrm{sup} \big\{  \textbf{E}_{\mathcal{Q}_1 \mathcal{Q}_3 \times \cdots \times \mathcal{Q}_N} \big[   \psi \big[  \textbf{E}_{\mathcal{Q}_2 } \big[ \mu \big( Q_1   , Q_2 ,  \cdots    , Q_N \big)  \big]     \big] \big]     , \cdots  } \\ \\ {\tiny     ,      \textbf{E}_{\mathcal{Q}_N} \big[ \psi^{\prime\cdots \prime}_{N-1} \big[  \textbf{E}_{\mathcal{Q}_1 \times \cdots \times \mathcal{Q}_{N-1}} \big[ \mu \big( Q_1  , Q_2  ,      \cdots , Q_N \big)  \big]     \big] \big]                 \big\}   \bigg\}    \bigg\}      } \\   \end{align*}

       \begin{align*} {\tiny   =       -   \mathrm{log}_2 \big( 2 \big)^2  \bigg\{   \underset{n \text{ } \textit{sufficiently small}}{\underset{N \longrightarrow + \infty}{\mathrm{lim}}}   \bigg\{       \underset{1 \leq i \leq N}{\prod} q^{-1}_i \bigg\}     \bigg\} \bigg\{                \underset{n \text{ } \textit{sufficiently small}}{\underset{N \longrightarrow + \infty}{\mathrm{lim}}}   \bigg\{               \frac{\partial}{\partial N } C \big( N , n \big) \frac{1}{\big( N - 1 \big)} 2^{N+2}                                                                                                                                           \bigg\}      -        \underset{n \text{ } \textit{sufficiently small}}{\underset{N \longrightarrow + \infty}{\mathrm{lim}}}     }                                                \\  {\tiny   \times  \bigg\{                                          C \big( N , n \big)           \frac{1}{\big( N - 1 \big)^2}         2^{N+2}                                                                                                                                                    \bigg\}       +           \underset{n \text{ } \textit{sufficiently small}}{\underset{N \longrightarrow + \infty}{\mathrm{lim}}}   \bigg\{                                                                                                                C \big( N , n \big) \frac{1}{\big( N - 1 \big)} 2^{N+2} \mathrm{log}_2 \big( 2 \big)                                                                                            \bigg\}          \bigg\}^{-1}                                                                                                                             \bigg\{  2^{N-1} \mathrm{log}_2 \big( 2 \big)  \mathrm{log}_2  \bigg[                                 - N  } \\  {\tiny   +   2^{-(N+1)}     \mathrm{sup} \big\{  \textbf{E}_{\mathcal{Q}_1 \mathcal{Q}_3 \times \cdots \times \mathcal{Q}_N} \big[   \psi \big[  \textbf{E}_{\mathcal{Q}_2 } \big[ \mu \big( Q_1   , Q_2 ,  \cdots   , Q_N \big)  \big]     \big] \big]     , \cdots     ,      \textbf{E}_{\mathcal{Q}_N} \big[ \psi^{\prime\cdots \prime}_{N-1} \big[  \textbf{E}_{\mathcal{Q}_1 \times \cdots \times \mathcal{Q}_{N-1}} \big[ \mu \big( Q_1  , Q_2  ,     }  \\ \\ {\tiny     \cdots , Q_N \big)  \big]     \big] \big]                 \big\}    \bigg]                                                                          +    2^{N-1}    \mathrm{log}_2  \bigg[                                 - N  +   2^{-(N+1)}     \mathrm{sup} \big\{  \textbf{E}_{\mathcal{Q}_1 \mathcal{Q}_3 \times \cdots \times \mathcal{Q}_N} \big[   \psi \big[  \textbf{E}_{\mathcal{Q}_2 } \big[ \mu \big( Q_1   , Q_2 ,  \cdots } \\ \\  {\tiny    , Q_N \big)  \big]     \big] \big]     , \cdots     ,      \textbf{E}_{\mathcal{Q}_N} \big[ \psi^{\prime\cdots \prime}_{N-1} \big[  \textbf{E}_{\mathcal{Q}_1 \times \cdots \times \mathcal{Q}_{N-1}} \big[ \mu \big( Q_1  , Q_2  ,      \cdots , Q_N \big)  \big]     \big] \big]                 \big\}    \bigg]                 } \\ \\ {\tiny \times          \bigg\{      2^{-\big( N + 1 \big)}   \mathrm{log}_2 \big( 2 \big)    \mathrm{sup} \big\{  \textbf{E}_{\mathcal{Q}_1 \mathcal{Q}_3 \times \cdots \times \mathcal{Q}_N} \big[   \psi \big[  \textbf{E}_{\mathcal{Q}_2 } \big[ \mu \big( Q_1   , Q_2 ,  \cdots    , Q_N \big)  \big]     \big] \big]     , \cdots  } \\ \\ {\tiny     ,      \textbf{E}_{\mathcal{Q}_N} \big[ \psi^{\prime\cdots \prime}_{N-1} \big[  \textbf{E}_{\mathcal{Q}_1 \times \cdots \times \mathcal{Q}_{N-1}} \big[ \mu \big( Q_1  , Q_2  ,      \cdots , Q_N \big)  \big]     \big] \big]                 \big\}   \bigg\}    \bigg\}                                                                                                                                           }        \\ \end{align*}

       \begin{align*} {\tiny    \leq      -     \mathrm{log}_2 \big( 2 \big)^2  \bigg\{  \mathcal{Q}_N \big( i , n \big)    \bigg\}                                                                                                                             \bigg\{                                    \underset{n \text{ } \textit{sufficiently small}}{\underset{N \longrightarrow + \infty}{\mathrm{lim}}}                \bigg\{      C \big( N , n \big)^2        \bigg(     \frac{1}{\big( N - 1 \big)      }         2^{N+2}  \bigg)^2 \mathrm{log}_2 \big( 2 \big)   \bigg\}                                                                                                                - \underset{n \text{ } \textit{sufficiently small}}{\underset{N \longrightarrow + \infty}{\mathrm{lim}}}   \bigg\{     \frac{\partial}{\partial N } C \big( N , n \big)                                                                                                                                      }                                                           \\ \\ {\tiny  \times \bigg(  \frac{1}{\big( N - 1 \big)} 2^{N+2}     \bigg)^2      C \big( N , n \big)    \mathrm{log}_2 \big( 2 \big)       \bigg\}                                                                                                                                                                                                               + \underset{n \text{ } \textit{sufficiently small}}{\underset{N \longrightarrow + \infty}{\mathrm{lim}}}                                         \bigg\{      \bigg\{  \frac{\partial}{\partial N } C \big( N , n \big)    \bigg\}     C \big( N , n \big)     \bigg(  \frac{1}{\big( N - 1 \big)} 2^{N+2}     \bigg)^2                                                                                                                                                    \bigg\}         \bigg\}^{-1}                                                                                           } \\ \\  {\tiny    \times   \underset{n \text{ } \textit{sufficiently small}}{\underset{N \longrightarrow + \infty}{\mathrm{lim}}}    \bigg\{  \bigg\{  2^{N-1} \mathrm{log}_2 \big( 2 \big)  \mathrm{log}_2  \bigg[                                 - N    +   2^{-(N+1)}     \mathrm{sup} \big\{  \textbf{E}_{\mathcal{Q}_1 \mathcal{Q}_3 \times \cdots \times \mathcal{Q}_N} \big[   \psi \big[  \textbf{E}_{\mathcal{Q}_2 } \big[ \mu \big( Q_1   , Q_2 ,  \cdots   , Q_N \big)  \big]     \big] \big]     ,   } \\ \\ {\tiny     \cdots ,      \textbf{E}_{\mathcal{Q}_N} \big[ \psi^{\prime\cdots \prime}_{N-1} \big[  \textbf{E}_{\mathcal{Q}_1 \times \cdots \times \mathcal{Q}_{N-1}} \big[ \mu \big( Q_1  , Q_2  ,        \cdots , Q_N \big)  \big]     \big] \big]                 \big\}    \bigg]                                                                          +    2^{N-1}    \mathrm{log}_2  \bigg[                                 - N     +   2^{-(N+1)}   }  \\ \\ {\tiny \times   \mathrm{sup} \big\{  \textbf{E}_{\mathcal{Q}_1 \mathcal{Q}_3 \times \cdots \times \mathcal{Q}_N} \big[   \psi \big[  \textbf{E}_{\mathcal{Q}_2 } \big[ \mu \big( Q_1   , Q_2 ,  \cdots   , Q_N \big)  \big]     \big] \big]     , \cdots     ,      \textbf{E}_{\mathcal{Q}_N} \big[ \psi^{\prime\cdots \prime}_{N-1} \big[  \textbf{E}_{\mathcal{Q}_1 \times \cdots \times \mathcal{Q}_{N-1}} \big[ \mu \big( Q_1  , } \\ \\  {\tiny    Q_2  ,      \cdots , Q_N \big)  \big]     \big] \big]                 \big\}    \bigg]      \bigg\{      2^{-\big( N + 1 \big)}   \mathrm{log}_2 \big( 2 \big)    \mathrm{sup} \big\{  \textbf{E}_{\mathcal{Q}_1 \mathcal{Q}_3 \times \cdots \times \mathcal{Q}_N} \big[   \psi \big[  \textbf{E}_{\mathcal{Q}_2 } \big[ \mu \big( Q_1                } \\ \\ {\tiny     ,       Q_2 ,  \cdots    , Q_N \big)  \big]     \big] \big]     , \cdots     ,      \textbf{E}_{\mathcal{Q}_N} \big[ \psi^{\prime\cdots \prime}_{N-1} \big[  \textbf{E}_{\mathcal{Q}_1 \times \cdots \times \mathcal{Q}_{N-1}} \big[ \mu \big( Q_1  , Q_2  ,     \cdots , Q_N \big)  \big]     \big] \big]                 \big\}   \bigg\}      } \\ \\ {\tiny  \times                          \bigg\{      \bigg\{ \frac{\partial}{\partial N } C \big( N , n \big) \frac{1}{\big( N - 1 \big)} 2^{N+2}                                                                           \bigg\}   \bigg\{                  C \big( N , n \big)           \frac{1}{\big( N - 1 \big)^2}         2^{N+2}                                                                                                                             \bigg\}         \bigg\{                                                  C \big( N , n \big) \frac{1}{\big( N - 1 \big)} } \\ \\ \\  {\tiny \times  2^{N+2} \mathrm{log}_2 \big( 2 \big)                                                   \bigg\}                   \bigg\}                                                                                                                              \bigg\}        } . \\  \end{align*} }

       \noindent Proceeding, the value of the above limit as $N  \longrightarrow + \infty$ can be determined from the following observations,

    {\small        \begin{align*} {\tiny                        -         \mathrm{log}_2 \big( 2 \big)^2  \bigg\{  \mathcal{Q}_N \big( i , n \big)    \bigg\}                     \underset{n \text{ } \textit{sufficiently small}}{\underset{N \longrightarrow + \infty}{\mathrm{lim}}}      \bigg\{    C \big( N , n \big)^2        \bigg(     \frac{1}{\big( N - 1 \big)      }         2^{N+2}  \bigg)^2 \mathrm{log}_2 \big( 2 \big)                    \bigg\}^{-1}  =     -         \mathrm{log}_2 \big( 2 \big)  \bigg\{  \mathcal{Q}_N \big( i , n \big)    \bigg\} }  \end{align*}

    \begin{align*} {\tiny \times                             \underset{n \text{ } \textit{sufficiently small}}{\underset{N \longrightarrow + \infty}{\mathrm{lim}}}                                \frac{1}{   \frac{1}{  C \big( N , n \big)^{-2} } \frac{1}{ { \bigg( \big( N - 1 \big)               2^{N+2}  \bigg)^{-2}}       }}       }  \\  \\  {\tiny =      -         \mathrm{log}_2 \big( 2 \big)  \bigg\{  \mathcal{Q}_N \big( i , n \big)    \bigg\}            \underset{n \text{ } \textit{sufficiently small}}{\underset{N \longrightarrow + \infty}{\mathrm{lim}}}    \frac{\partial}{\partial N}  \bigg\{      \bigg(        C \big( N , n \big)         \frac{1}{\big( N - 1 \big)      }         2^{N+2}        \bigg)^{-2}                                     \bigg\}                                     }  \\ \end{align*}

    \begin{align*} {\tiny          =        -         \mathrm{log}_2 \big( 2 \big)  \bigg\{  \mathcal{Q}_N \big( i , n \big)    \bigg\}            \underset{n \text{ } \textit{sufficiently small}}{\underset{N \longrightarrow + \infty}{\mathrm{lim}}} \frac{\partial}{\partial N}     \bigg\{                 \frac{1}{   \frac{1}{  C \big( N , n \big)^{-2} } \frac{1}{ { \bigg( \big( N - 1 \big)               2^{N+2}  \bigg)^{-2}}       }}                                                                                                        \bigg\}                                                                                                                                  } \\ \\  {\tiny =          2       \mathrm{log}_2 \big( 2 \big)  \bigg\{  \mathcal{Q}_N \big( i , n \big)    \bigg\}            \underset{n \text{ } \textit{sufficiently small}}{\underset{N \longrightarrow + \infty}{\mathrm{lim}}}         \bigg(   C \big( N , n \big)         \frac{1}{\big( N - 1 \big)      }         2^{N+2}     \bigg)^{-3} \bigg(      \frac{\partial}{\partial N} C \big( N , n \big) } \\ \\  {\tiny  \times   \frac{1}{\big( N - 1 \big)} 2^{N+2}     -                                              C \big( N , n \big) \frac{1}{\big( N - 1 \big)^2} 2^{N+2}   +     \frac{\partial}{\partial N} C \big( N , n \big) \frac{1}{\big( N - 1 \big)} 2^{N+2}   \mathrm{log}_2 \big( 2 \big)       \bigg)                                                           }                                                             \\ \\ \\ {\tiny                =         2       \mathrm{log}_2 \big( 2 \big)  \bigg\{  \mathcal{Q}_N \big( i , n \big)    \bigg\}            \underset{n \text{ } \textit{sufficiently small}}{\underset{N \longrightarrow + \infty}{\mathrm{lim}}}     \frac{       \frac{\partial}{\partial N} C \big( N , n \big)    \frac{1}{\big( N - 1 \big)} 2^{N+2}            }{ \bigg(   C \big( N , n \big)         \frac{1}{\big( N - 1 \big)      }         2^{N+2}     \bigg)^{3} }                                   -             2       \mathrm{log}_2 \big( 2 \big)  \bigg\{  \mathcal{Q}_N \big( i , n \big)    \bigg\}                 } \\ \\ \\ { \tiny  \times    \underset{n \text{ } \textit{sufficiently small}}{\underset{N \longrightarrow + \infty}{\mathrm{lim}}}            \frac{  C \big( N , n \big) \frac{1}{\big( N - 1 \big)^2} 2^{N+2}           }{ \bigg(   C \big( N , n \big)         \frac{1}{\big( N - 1 \big)      }         2^{N+2}     \bigg)^{3} }                 +    2       \mathrm{log}_2 \big( 2 \big)  \bigg\{  \mathcal{Q}_N \big( i , n \big)    \bigg\}       } \\ \\ \\ { \tiny  \times         \underset{n \text{ } \textit{sufficiently small}}{\underset{N \longrightarrow + \infty}{\mathrm{lim}}}             \frac{\frac{\partial}{\partial N} C \big( N , n \big) \frac{1}{\big( N - 1 \big)} 2^{N+2}   \mathrm{log}_2 \big( 2 \big)                  }{ \bigg(   C \big( N , n \big)         \frac{1}{\big( N - 1 \big)      }         2^{N+2}     \bigg)^{3} }                                                                                                              }   \\ \\ \\ {\tiny   =               2       \mathrm{log}_2 \big( 2 \big)  \bigg\{  \mathcal{Q}_N \big( i , n \big)    \bigg\}            \underset{n \text{ } \textit{sufficiently small}}{\underset{N \longrightarrow + \infty}{\mathrm{lim}}}                                  \frac{\frac{\partial}{\partial N} C \big( N , n \big) }{C \big( N , n \big)^3      \bigg(    \frac{2^{N+2}  }{\big( N - 1 \big)  }   \bigg)^2        }                     - 2       \mathrm{log}_2 \big( 2 \big)  \bigg\{  \mathcal{Q}_N \big( i , n \big)    \bigg\}       } \\ \\ \\ {\tiny  \times        \underset{n \text{ } \textit{sufficiently small}}{\underset{N \longrightarrow + \infty}{\mathrm{lim}}}                                             \frac{1}{C \big( N , n \big)^2   \bigg( \frac{2^{N+2}}{\big( N - 1 \big)} \bigg)^2           }                         + 2       \mathrm{log}_2 \big( 2 \big)^2   \bigg\{  \mathcal{Q}_N \big( i , n \big)    \bigg\}     } \\ \\ \\ {\tiny   \times       \underset{n \text{ } \textit{sufficiently small}}{\underset{N \longrightarrow + \infty}{\mathrm{lim}}}        \frac{\frac{\partial}{\partial N} C \big( N , n \big) }{C \big( N , n \big)     \bigg(    \frac{2^{N+2}  }{\big( N - 1 \big)  }   \bigg)^2        }                                                                                     }         \\ \end{align*}

    \begin{align*} {\tiny    \sim                     2       \mathrm{log}_2 \big( 2 \big)  \bigg\{  \mathcal{Q}_N \big( i , n \big)    \bigg\}            \underset{n \text{ } \textit{sufficiently small}}{\underset{N \longrightarrow + \infty}{\mathrm{lim}}}                                             \frac{1}{C \big( N , n \big)^2      \bigg(    \frac{2^{N+2}  }{\big( N - 1 \big)  }   \bigg)^2        }                                                -        2       \mathrm{log}_2 \big( 2 \big)  \bigg\{  \mathcal{Q}_N \big( i , n \big)    \bigg\}       } \\ \\ \\ {\tiny \times       \underset{n \text{ } \textit{sufficiently small}}{\underset{N \longrightarrow + \infty}{\mathrm{lim}}}             \frac{1}{C \big( N , n \big)^2   \bigg( \frac{2^{N+2}}{\big( N - 1 \big)} \bigg)^2           }                 +    2       \mathrm{log}_2 \big( 2 \big)  \bigg\{  \mathcal{Q}_N \big( i , n \big)    \bigg\}    } \\ \\ \\ {\tiny \times            \underset{n \text{ } \textit{sufficiently small}}{\underset{N \longrightarrow + \infty}{\mathrm{lim}}}          \frac{1 }{C \big( N , n \big)     \bigg(    \frac{2^{N+2}  }{\big( N - 1 \big)  }   \bigg)^2        }                          } \\ \\ \\ {\tiny \approx 0 , }  \\ \end{align*} }

    \noindent corresponding to the first term,

   {\small \begin{align*} {\tiny     -         \mathrm{log}_2 \big( 2 \big)^2  \bigg\{  \mathcal{Q}_N \big( i , n \big)    \bigg\}      \underset{n \text{ } \textit{sufficiently small}}{\underset{N \longrightarrow + \infty}{\mathrm{lim}}}          \bigg\{    \frac{\partial}{\partial N} C \big( N , n \big)     \bigg(  \frac{1}{\big( N - 1 \big)} 2^{N+2}     \bigg)^2      C \big( N , n \big)    \mathrm{log}_2 \big( 2 \big)     \bigg\}^{-1} =       -         \mathrm{log}_2 \big( 2 \big)  \bigg\{  \mathcal{Q}_N \big( i , n \big)    \bigg\}                                                                                                                                                                }  \\ \\ \\ {\tiny \times        \underset{n \text{ } \textit{sufficiently small}}{\underset{N \longrightarrow + \infty}{\mathrm{lim}}}                             \frac{C \big( N , n \big)^{-1} }{         \frac{\partial}{\partial N} C \big( N , n \big)     \bigg(  \frac{1}{\big( N - 1 \big)} 2^{N+2}     \bigg)^2            }                                                                                                                                                                                                                                                                                                                          }   \\ \\ \\ {\tiny                                            \sim       -         \mathrm{log}_2 \big( 2 \big)  \bigg\{  \mathcal{Q}_N \big( i , n \big)    \bigg\}            \underset{n \text{ } \textit{sufficiently small}}{\underset{N \longrightarrow + \infty}{\mathrm{lim}}}                             \frac{1 }{          C \big( N , n \big)     \bigg(  \frac{1}{\big( N - 1 \big)} 2^{N+2}     \bigg)^2            }                                                                                                                                                                                                                                                                                } \\ \\ \\  {\tiny \approx 0                                                           }         \\  \end{align*}

  \noindent corresponding to the second term and,

 {\small \begin{align*}  
   {\tiny     -         \mathrm{log}_2 \big( 2 \big)^2  \bigg\{  \mathcal{Q}_N \big( i , n \big)    \bigg\}     \underset{n \text{ } \textit{sufficiently small}}{\underset{N \longrightarrow + \infty}{\mathrm{lim}}}                               \bigg\{       \bigg\{  \frac{\partial}{\partial N } C \big( N , n \big)    \bigg\}     C \big( N , n \big)     \bigg(  \frac{1}{\big( N - 1 \big)} 2^{N+2}     \bigg)^2                                                                  \bigg\}^{-1}  =         -         \mathrm{log}_2 \big( 2 \big)^2  \bigg\{  \mathcal{Q}_N \big( i , n \big)    \bigg\}                                                                                      }  \\ \\ \\ {\tiny \times             \underset{n \text{ } \textit{sufficiently small}}{\underset{N \longrightarrow + \infty}{\mathrm{lim}}}                    \frac{C \big( N , n \big)^{-1}}{                \bigg\{  \frac{\partial}{\partial N } C \big( N , n \big)    \bigg\}     \bigg(  \frac{1}{\big( N - 1 \big)} 2^{N+2}     \bigg)^2            }                                           }     \\ \\ \\ {\tiny  \sim        -         \mathrm{log}_2 \big( 2 \big)^2  \bigg\{  \mathcal{Q}_N \big( i , n \big)    \bigg\}                     \underset{n \text{ } \textit{sufficiently small}}{\underset{N \longrightarrow + \infty}{\mathrm{lim}}}                      \frac{1}{C \big( N , n \big)  \bigg(  \frac{1}{\big( N - 1 \big)} 2^{N+2}     \bigg)^2    }                                                                                   } \\ \\ \\ {\tiny \approx 0  , } \\       
\end{align*}

}

\noindent corresponding to the third term. Altogether,

{\small

\begin{align*}
    {\tiny     \underset{n \text{ } \textit{sufficiently small}}{\underset{N \longrightarrow + \infty}{\mathrm{lim}}} \bigg\{  1 \backslash  \bigg( \bigg\{    C \big( N , n \big)^2        \bigg(     \frac{1}{\big( N - 1 \big)      }         2^{N+2}  \bigg)^2 \mathrm{log}_2 \big( 2 \big)                    \bigg\}^{-1}   +  \bigg\{    \frac{\partial}{\partial N} C \big( N , n \big)     \bigg(  \frac{1}{\big( N - 1 \big)} 2^{N+2}     \bigg)^2      C \big( N , n \big)    \mathrm{log}_2 \big( 2 \big)     \bigg\}^{-1}  } \\  {\tiny +    \bigg\{       \bigg\{  \frac{\partial}{\partial N } C \big( N , n \big)    \bigg\}     C \big( N , n \big)     \bigg(  \frac{1}{\big( N - 1 \big)} 2^{N+2}     \bigg)^2                                                                  \bigg\}^{-1}   \bigg\} \bigg)   \bigg\}                           } \\ \end{align*}

    \begin{align*}  {\tiny      \sim          \bigg\{  1 \backslash  \bigg(              \bigg(          2       \mathrm{log}_2 \big( 2 \big)  \bigg\{  \mathcal{Q}_N \big( i , n \big)    \bigg\}            \underset{n \text{ } \textit{sufficiently small}}{\underset{N \longrightarrow + \infty}{\mathrm{lim}}}                                             \frac{1}{C \big( N , n \big)^2      \bigg(    \frac{2^{N+2}  }{\big( N - 1 \big)  }   \bigg)^2        }                                                -        2       \mathrm{log}_2 \big( 2 \big)  \bigg\{  \mathcal{Q}_N \big( i , n \big)    \bigg\}       } \\ \\ \\ {\tiny \times       \underset{n \text{ } \textit{sufficiently small}}{\underset{N \longrightarrow + \infty}{\mathrm{lim}}}             \frac{1}{C \big( N , n \big)^2   \bigg( \frac{2^{N+2}}{\big( N - 1 \big)} \bigg)^2           }                 +    2       \mathrm{log}_2 \big( 2 \big)  \bigg\{  \mathcal{Q}_N \big( i , n \big)    \bigg\}    } \\ \\ \\ {\tiny \times            \underset{n \text{ } \textit{sufficiently small}}{\underset{N \longrightarrow + \infty}{\mathrm{lim}}}          \frac{1 }{C \big( N , n \big)     \bigg(    \frac{2^{N+2}  }{\big( N - 1 \big)  }   \bigg)^2        }             \bigg)      -          \bigg(                        \mathrm{log}_2 \big( 2 \big)  \bigg\{  \mathcal{Q}_N \big( i , n \big)    \bigg\}          } \\ \\ \\ {\tiny \times     \underset{n \text{ } \textit{sufficiently small}}{\underset{N \longrightarrow + \infty}{\mathrm{lim}}}                             \frac{1 }{          C \big( N , n \big)     \bigg(  \frac{1}{\big( N - 1 \big)} 2^{N+2}     \bigg)^2            }                                                                                           \bigg)           } \\ \\ \\ {\tiny    +           \bigg(     -         \mathrm{log}_2 \big( 2 \big)^2  \bigg\{  \mathcal{Q}_N \big( i , n \big)    \bigg\}                     \underset{n \text{ } \textit{sufficiently small}}{\underset{N \longrightarrow + \infty}{\mathrm{lim}}}                      \frac{1}{C \big( N , n \big)  \bigg(  \frac{1}{\big( N - 1 \big)} 2^{N+2}     \bigg)^2    }                                                         
\bigg)      \bigg) \bigg\}  }    \approx 0 .   \\  
\end{align*}

}

\noindent To compute the derivative of various expected values with respect to the number of players $N$, observe,

{\small 
\begin{align*}
      {\tiny     \big\{  \textbf{E}_{\mathcal{Q}_1 \mathcal{Q}_3 \times \cdots \times \mathcal{Q}_N} \big[ \psi \big[  \textbf{E}_{\mathcal{Q}_2 } \big[ \mu \big( Q_1 , Q_2 ,  \cdots ,    Q_N \big)  \big]     \big] \big]   \big\}^{\prime}       =    \frac{\partial}{\partial N} \big\{       \textbf{E}_{\mathcal{Q}_1 \mathcal{Q}_3 \times \cdots \times \mathcal{Q}_N} \big[ \psi \big[  \textbf{E}_{\mathcal{Q}_2 } \big[ \mu \big( Q_1 , Q_2 , }  \\ \\ {\tiny \cdots ,    Q_N \big)  \big]     \big] \big]        \big\}    }    \\ \\ {\tiny <        \mathrm{sup}  \big\{            \big\{  \big\{ \textbf{E}_{\mathcal{Q}_1 \mathcal{Q}_3 \times \cdots \times \mathcal{Q}_N} \big[ \psi \big[  \textbf{E}_{\mathcal{Q}_2 } \big[ \mu \big( Q_1 , Q_2    ,  \cdots ,    Q_N \big)  \big]     \big] \big]   \big\}^{\prime}   , \cdots  ,      \big\{ \textbf{E}_{\mathcal{Q}_N} \big[ \psi^{\prime\cdots \prime}_{N-1} \big[  \textbf{E}_{\mathcal{Q}_1 \times \cdots \times \mathcal{Q}_{N-1}} \big[ \mu \big( Q_1  , Q_2  ,            } \\ \\ {\tiny   \cdots , Q_N \big)  \big]     \big] \big]          \big\}^{\prime}                                                                                                                }    , \\ \\  {\tiny  \big\{ \textbf{E}_{\mathcal{Q}_N} \big[ \psi^{\prime\cdots \prime}_{N-1} \big[  \textbf{E}_{\mathcal{Q}_1 \times \cdots \times \mathcal{Q}_{N-1}} \big[ \mu \big( Q_1  , Q_2  ,  \cdots , Q_N \big)  \big]     \big] \big]          \big\}^{\prime}         =    \frac{\partial}{\partial N} \big\{ \textbf{E}_{\mathcal{Q}_N} \big[ \psi^{\prime\cdots \prime}_{N-1} \big[  \textbf{E}_{\mathcal{Q}_1 \times \cdots \times \mathcal{Q}_{N-1}} \big[ \mu \big( Q_1  , Q_2  , } \\ \\ {\tiny      \cdots , Q_N \big)  \big]     \big] \big]          \big\}   } \\ \\ {\tiny <              \mathrm{sup}  \big\{            \big\{  \big\{ \textbf{E}_{\mathcal{Q}_1 \mathcal{Q}_3 \times \cdots \times \mathcal{Q}_N} \big[ \psi \big[  \textbf{E}_{\mathcal{Q}_2 } \big[ \mu \big( Q_1 , Q_2    ,  \cdots ,    Q_N \big)  \big]     \big] \big]   \big\}^{\prime}   , \cdots  ,      \big\{ \textbf{E}_{\mathcal{Q}_N} \big[ \psi^{\prime\cdots \prime}_{N-1} \big[  \textbf{E}_{\mathcal{Q}_1 \times \cdots \times \mathcal{Q}_{N-1}} \big[ \mu \big( Q_1  , Q_2  ,           \cdots } \\ \\ {\tiny , Q_N \big)  \big]     \big] \big]          \big\}^{\prime}                                                                                   }                   ,  \\ 
\end{align*}
}

\noindent together imply the desired claim that the limit can be made arbitrarily close to $0$ and as a result can be upper bounded by $1$ up to a constant. Therefore, to reiterate,

{\small \begin{align*}
{\tiny \frac{1}{C \big( N , n \big)} }  \end{align*}

\begin{align*} {\tiny \times      \frac{ \Psi^{-1} \bigg[   2^{-1}   \mathrm{sup} \big\{  \textbf{E}_{\mathcal{Q}_1 \mathcal{Q}_3 \times \cdots \times \mathcal{Q}_N} \big[ \psi \big[  \textbf{E}_{\mathcal{Q}_2 } \big[ \mu \big( Q_1 , Q_2 ,   \cdots , Q_N \big)  \big]     \big] \big]        , \textbf{E}_{\mathcal{Q}_2\mathcal{Q}_3 \times \cdots \times \mathcal{Q}_N } \big[ \psi \big[  \textbf{E}_{\mathcal{Q}_1} \big[ \mu \big( Q_1 ,   Q_2 ,    \cdots  , Q_N \big)  \big]     \big] \big]       \big\}    \bigg]^{2} }{ \Psi^{-1} \bigg[    2^{-N}       \mathrm{sup} \big\{  \textbf{E}_{\mathcal{Q}_1 \mathcal{Q}_3 \times \cdots \times \mathcal{Q}_N} \big[ \psi \big[  \textbf{E}_{\mathcal{Q}_2 } \big[ \mu \big( Q_1 , Q_2 ,   \cdots , Q_N \big)  \big]     \big] \big]      , \cdots ,          \textbf{E}_{\mathcal{Q}_N} \big[ \psi^{\prime\cdots \prime}_{N-1} \big[  \textbf{E}_{\mathcal{Q}_1 \times \cdots \times \mathcal{Q}_{N-1}} \big[ \mu \big( Q_1  , Q_2  ,   \cdots , Q_N \big)  \big]     \big] \big]                 \big\}                 \bigg]^{2^N}  }    }   \end{align*}

\begin{align*}       \overset{(\mathrm{LR})}{\lesssim} 1  . \\ 
\end{align*}}

\noindent from the fact that,

{\small  \begin{align*}
   \Psi^{-1} \bigg[   2^{-1}   \mathrm{sup} \big\{  \textbf{E}_{\mathcal{Q}_1 \mathcal{Q}_3 \times \cdots \times \mathcal{Q}_N} \big[ \psi \big[  \textbf{E}_{\mathcal{Q}_2 } \big[ \mu \big( Q_1 , Q_2 ,   \cdots , Q_N \big)  \big]     \big] \big]        , \textbf{E}_{\mathcal{Q}_2\mathcal{Q}_3 \times \cdots \times \mathcal{Q}_N } \big[ \psi \big[  \textbf{E}_{\mathcal{Q}_1} \big[ \mu \big( Q_1 ,   Q_2 ,    \cdots  , Q_N \big)  \big]     \big] \big]       \big\}    \bigg]^{2} \\ \lesssim  \Psi^{-1} \bigg[    2^{-N}       \mathrm{sup} \big\{  \textbf{E}_{\mathcal{Q}_1 \mathcal{Q}_3 \times \cdots \times \mathcal{Q}_N} \big[ \psi \big[  \textbf{E}_{\mathcal{Q}_2 } \big[ \mu \big( Q_1 , Q_2 ,   \cdots , Q_N \big)  \big]     \big] \big]      , \cdots ,          \textbf{E}_{\mathcal{Q}_N} \big[ \psi^{\prime\cdots \prime}_{N-1} \big[  \textbf{E}_{\mathcal{Q}_1 \times \cdots \times \mathcal{Q}_{N-1}} \big[ \mu \big( Q_1  , Q_2  , \\   \cdots , Q_N \big)  \big]     \big] \big]                 \big\}                 \bigg]^{2^N}   , \\ 
\end{align*} }

\noindent implies,

{\small \begin{align*}
    C \big( N , n \big)                \Psi^{-1} \bigg[  \frac{1}{2} \mathrm{sup} \big\{    \textbf{E}_{\mathcal{Q}_1 \mathcal{Q}_3 \times \cdots \times \mathcal{Q}_N} \big[ \psi \big[  \textbf{E}_{\mathcal{Q}_2 } \big[ \mu \big( Q_1 , Q_2 ,   \cdots , Q_N \big)  \big]     \big] \big]        , \textbf{E}_{\mathcal{Q}_2\mathcal{Q}_3 \times \cdots \times \mathcal{Q}_N } \big[ \psi \big[  \textbf{E}_{\mathcal{Q}_1} \big[ \mu \big( Q_1 ,   Q_2 ,     \cdots \\  , Q_N \big)  \big]     \big] \big]  \big\}    \bigg]^{2}   \\        \leq    C \big( N , n \big)  \Psi^{-1}  \bigg[ 2^{-N }   \mathrm{sup}   \big\{        \textbf{E}_{\mathcal{Q}_1} \big[ \psi \big[  \textbf{E}_{\mathcal{Q}_2} \big[ \mu \big( Q_1 , Q_2 ,   \cdots , Q_N \big)  \big]     \big] \big]   , \textbf{E}_{\mathcal{Q}_2} \big[ \psi^{\prime} \big[  \textbf{E}_{\mathcal{Q}_1} \big[ \mu \big( Q_1  , Q_2 ,  \cdots , Q_N \big)  \big]     \big] \big]  ,    \cdots \\ ,        \textbf{E}_{\mathcal{Q}_N} \big[ \psi \big[  \textbf{E}_{\mathcal{Q}_1} \big[ \mu \big( Q_1  , Q_2 ,  \cdots , Q_N \big)  \big]     \big] \big]  , \cdots  , \textbf{E}_{\mathcal{Q}_N} \big[ \psi^{\prime\cdots \prime}_{N-1} \big[  \textbf{E}_{\mathcal{Q}_1 \times \cdots \times \mathcal{Q}_{N-1}} \big[ \mu \big( Q_1  , Q_2  ,   \cdots  , Q_N \big)  \big]     \big] \big]                   \big\}               \bigg]^{2^N}                  , \\ \end{align*}   }

    \noindent from which we conclude the argument. \boxed{}

    \bigskip

    \noindent \noindent \textit{Proof of Theorem 3}. To demonstrate that the two desired inequalities hold, from the objects,

    {\small \begin{align*}
    \underset{i \neq i^{\prime} \neq \cdots \neq i^{\prime\cdots\prime}}{\underset{i^{\prime\cdots\prime} \in [n]}{\underset{\vdots}{\underset{i^{\prime} \in [n]}{\underset{i \in [ n ] }{\prod}}}}} \big[  \epsilon_i + \epsilon_{i^{\prime}} + \cdots + e_{i^{\prime\cdots\prime}} \big]  +  \underset{i \neq i^{\prime} \neq \cdots \neq i^{\prime\cdots\prime}}{\underset{i^{\prime\cdots\prime} \in [n]}{\underset{\vdots}{\underset{i^{\prime} \in [n]}{\underset{i \in [ n ] }{\sum}}}}} \big[ \delta_i + \delta_{i^{\prime}} + \cdots + \delta_{i^{\prime\cdots\prime}} \big]  , \\  \tag{\textbf{4}} \\ \end{align*}

{\small \begin{align*}
      \textbf{E}_{\mathcal{Q}_1 \times \cdots \times \mathcal{Q}_N} \big[ \mu \big( Q_1 , \cdots , Q_N \big)  \big]     , \\ \tag{\textbf{6}} \\    
\end{align*} }

    \noindent by direct computation observe,

      {\small \begin{align*}
{\tiny \mathrm{sup} \bigg\{       \underset{i \neq i^{\prime} \neq \cdots \neq i^{\prime\cdots\prime}}{\underset{i^{\prime\cdots\prime} \in [n]}{\underset{\vdots}{\underset{i^{\prime} \in [n]}{\underset{i \in [ n ] }{\prod}}}}} \bigg[  \epsilon_i + \epsilon_{i^{\prime}} + \cdots + e_{i^{\prime\cdots\prime}} \bigg]  ,   \underset{i \neq i^{\prime} \neq \cdots \neq i^{\prime\cdots\prime}}{\underset{i^{\prime\cdots\prime} \in [n]}{\underset{\vdots}{\underset{i^{\prime} \in [n]}{\underset{i \in [ n ] }{\sum}}}}} \bigg[ \delta_i + \delta_{i^{\prime}} + \cdots + \delta_{i^{\prime\cdots\prime}} \bigg]         \bigg\}    \leq   \underset{i \neq i^{\prime} \neq \cdots \neq i^{\prime\cdots\prime}}{\underset{i^{\prime\cdots\prime} \in [n]}{\underset{\vdots}{\underset{i^{\prime} \in [n]}{\underset{i \in [ n ] }{\prod}}}}} \bigg[  \epsilon_i + \epsilon_{i^{\prime}} + } \\ {\tiny \cdots + e_{i^{\prime\cdots\prime}} \bigg]  +  \underset{i \neq i^{\prime} \neq \cdots \neq i^{\prime\cdots\prime}}{\underset{i^{\prime\cdots\prime} \in [n]}{\underset{\vdots}{\underset{i^{\prime} \in [n]}{\underset{i \in [ n ] }{\sum}}}}} \bigg[ \delta_i + \delta_{i^{\prime}} + \cdots + \delta_{i^{\prime\cdots\prime}} \bigg] }  , \end{align*}

    \begin{align*}     {\tiny \mathrm{sup} \bigg\{   \underset{i \in [ n ] }{\prod}     \bigg[ \delta_i \bigg]         , \cdots ,    \underset{i \neq i^{\prime} \neq \cdots \neq i^{\prime\cdots\prime}}{\underset{i^{\prime\cdots\prime} \in [n]}{\underset{\vdots}{\underset{i^{\prime} \in [n]}{\underset{i \in [ n ] }{\prod}}}}}    \bigg[ \delta_i + \delta_{i^{\prime}} + \cdots + \delta_{i^{\prime\cdots\prime}} \bigg]  \bigg\}        \leq    \underset{i \neq i^{\prime} \neq \cdots \neq i^{\prime\cdots\prime}}{\underset{i^{\prime\cdots\prime} \in [n]}{\underset{\vdots}{\underset{i^{\prime} \in [n]}{\underset{i \in [ n ] }{\prod}}}}}    \bigg[ \delta_i + \delta_{i^{\prime}} + \cdots + \delta_{i^{\prime\cdots\prime}} \bigg] }   .  \\       
 \end{align*} }

 \noindent and hence, that,

{\small
\begin{align*}
        ( \textbf{6} )  =    \textbf{E}_{\mathcal{Q}_1 \times \cdots \times \mathcal{Q}_N} \big[ \mu \big( Q_1 , \cdots , Q_N \big)  \big]       \overset{(\textbf{Corollary} \text{ } \textit{1})}{\leq}        \mathrm{sup} \big\{ \mathcal{C}_1 , \mathcal{C}_2 , \cdots , \mathcal{C}_7 \big\}              <  \underset{i \neq i^{\prime} \neq \cdots \neq i^{\prime\cdots\prime}}{\underset{i^{\prime\cdots\prime} \in [n]}{\underset{\vdots}{\underset{i^{\prime} \in [n]}{\underset{i \in [ n ] }{\prod}}}}} \big[  \epsilon_i + \epsilon_{i^{\prime}} + \cdots \\ \\ + e_{i^{\prime\cdots\prime}} \big] \\ \\  +  \underset{i \neq i^{\prime} \neq \cdots \neq i^{\prime\cdots\prime}}{\underset{i^{\prime\cdots\prime} \in [n]}{\underset{\vdots}{\underset{i^{\prime} \in [n]}{\underset{i \in [ n ] }{\sum}}}}} \big[ \delta_i + \delta_{i^{\prime}} + \cdots + \delta_{i^{\prime\cdots\prime}} \big]    = ( \textbf{4} )                         , \\ 
\end{align*} }

\noindent from which we conclude the argument. \boxed{}

    \bigskip

    \noindent \textit{Proof of Corollary 2}. To demonstrate that the two desired inequalities hold, from the objects,

    \begin{align*}  \underset{i \neq i^{\prime} \neq \cdots \neq i^{\prime\cdots\prime}}{\underset{i^{\prime\cdots\prime} \in [n]}{\underset{\vdots}{\underset{i^{\prime} \in [n]}{\underset{i \in [ n ] }{\mathrm{sup}}}}}}    {\tiny \bigg[   \underset{k , k^{\prime} \in \{ i , i^{\prime} , \cdots , i^{\prime\cdots\prime} \} }{\prod} \delta_k \epsilon_{k^{\prime}}        \bigg]^N } ,     \\ \tag{\textbf{5}}      
 \end{align*} }

{\small \begin{align*}
      \textbf{E}_{\mathcal{Q}_1 \times \cdots \times \mathcal{Q}_N} \big[ \mu \big( Q_1 , \cdots , Q_N \big)  \big]     , \\ \tag{\textbf{6}} \\    
\end{align*} }

\noindent write,

{\small
\begin{align*}
        ( \textbf{6} )  =    \textbf{E}_{\mathcal{Q}_1 \times \cdots \times \mathcal{Q}_N} \big[ \mu \big( Q_1 , \cdots , Q_N \big)  \big]       \overset{(\textbf{Corollary} \text{ } \textit{1})}{\leq}        \mathrm{sup} \big\{ \mathcal{C}_1 , \mathcal{C}_2 , \cdots , \mathcal{C}_7 \big\}   \\  \\    {\tiny       \overset{(\textbf{Theorem} \text{ } \textit{3})}{<}         \underset{i \neq i^{\prime} \neq \cdots \neq i^{\prime\cdots\prime}}{\underset{i^{\prime\cdots\prime} \in [n]}{\underset{\vdots}{\underset{i^{\prime} \in [n]}{\underset{i \in [ n ] }{\prod}}}}} \big[  \epsilon_i + \epsilon_{i^{\prime}}  + \cdots  + e_{i^{\prime\cdots\prime}} \big]   +  \underset{i \neq i^{\prime} \neq \cdots \neq i^{\prime\cdots\prime}}{\underset{i^{\prime\cdots\prime} \in [n]}{\underset{\vdots}{\underset{i^{\prime} \in [n]}{\underset{i \in [ n ] }{\sum}}}}} \bigg[ \delta_i + \delta_{i^{\prime}}  + \cdots} \\ {\tiny + \delta_{i^{\prime\cdots\prime}} \bigg] }  \\ \\ <           \underset{i \neq i^{\prime} \neq \cdots \neq i^{\prime\cdots\prime}}{\underset{i^{\prime\cdots\prime} \in [n]}{\underset{\vdots}{\underset{i^{\prime} \in [n]}{\underset{i \in [ n ] }{\prod}}}}}   \big\{  \big[ \epsilon_i + \epsilon_{i^{\prime}} + \cdots + \epsilon_{i^{\prime\cdots\prime}} \big]     \big[ \delta_i + \delta_{i^{\prime}} + \cdots + \delta_{i^{\prime\cdots\prime}} \big]       \big\} \\ \\ <   \underset{i \neq i^{\prime} \neq \cdots \neq i^{\prime\cdots\prime}}{\underset{i^{\prime\cdots\prime} \in [n]}{\underset{\vdots}{\underset{i^{\prime} \in [n]}{\underset{i \in [ n ] }{\prod}}}}}     \big[ \epsilon_i + \epsilon_{i^{\prime}} + \cdots + \epsilon_{i^{\prime\cdots\prime}} \big]  \underset{i \neq i^{\prime} \neq \cdots \neq i^{\prime\cdots\prime}}{\underset{i^{\prime\cdots\prime} \in [n]}{\underset{\vdots}{\underset{i^{\prime} \in [n]}{\underset{i \in [ n ] }{\mathrm{sup}}}}}}    {\tiny \bigg[    \underset{k \in \{ i , i^{\prime} , \cdots , i^{\prime\cdots\prime} \} }{\sum} \delta_k     \bigg]^N }       \\ \\ <   \underset{i \neq i^{\prime} \neq \cdots \neq i^{\prime\cdots\prime}}{\underset{i^{\prime\cdots\prime} \in [n]}{\underset{\vdots}{\underset{i^{\prime} \in [n]}{\underset{i \in [ n ] }{\mathrm{sup}}}}}}    {\tiny \bigg[    \underset{k^{\prime} \in \{ i , i^{\prime} , \cdots , i^{\prime\cdots\prime} \} }{\sum} \epsilon_{k^{\prime}}     \bigg]^N    }         \underset{i \neq i^{\prime} \neq \cdots \neq i^{\prime\cdots\prime}}{\underset{i^{\prime\cdots\prime} \in [n]}{\underset{\vdots}{\underset{i^{\prime} \in [n]}{\underset{i \in [ n ] }{\mathrm{sup}}}}}}    {\tiny \bigg[    \underset{k \in \{ i , i^{\prime} , \cdots , i^{\prime\cdots\prime} \} }{\sum} \delta_k     \bigg]^N   }      \\ \\ =            \underset{i \neq i^{\prime} \neq \cdots \neq i^{\prime\cdots\prime}}{\underset{i^{\prime\cdots\prime} \in [n]}{\underset{\vdots}{\underset{i^{\prime} \in [n]}{\underset{i \in [ n ] }{\mathrm{sup}}}}}}    {\tiny \bigg[   \underset{k , k^{\prime} \in \{ i , i^{\prime} , \cdots , i^{\prime\cdots\prime} \} }{\sum} \delta_k \epsilon_{k^{\prime}}        \bigg]^N }  \\ \\        <       \underset{i \neq i^{\prime} \neq \cdots \neq i^{\prime\cdots\prime}}{\underset{i^{\prime\cdots\prime} \in [n]}{\underset{\vdots}{\underset{i^{\prime} \in [n]}{\underset{i \in [ n ] }{\mathrm{sup}}}}}}    {\tiny \bigg[    \underset{k^{\prime} \in \{ i , i^{\prime} , \cdots , i^{\prime\cdots\prime} \} }{\prod} \epsilon_{k^{\prime}}     \bigg]^N    }         \underset{i \neq i^{\prime} \neq \cdots \neq i^{\prime\cdots\prime}}{\underset{i^{\prime\cdots\prime} \in [n]}{\underset{\vdots}{\underset{i^{\prime} \in [n]}{\underset{i \in [ n ] }{\mathrm{sup}}}}}}    {\tiny \bigg[    \underset{k \in \{ i , i^{\prime} , \cdots , i^{\prime\cdots\prime} \} }{\prod} \delta_k     \bigg]^N   }        \\ \\    =         \underset{i \neq i^{\prime} \neq \cdots \neq i^{\prime\cdots\prime}}{\underset{i^{\prime\cdots\prime} \in [n]}{\underset{\vdots}{\underset{i^{\prime} \in [n]}{\underset{i \in [ n ] }{\mathrm{sup}}}}}}    {\tiny \bigg[   \underset{k , k^{\prime} \in \{ i , i^{\prime} , \cdots , i^{\prime\cdots\prime} \} }{\prod} \delta_k \epsilon_{k^{\prime}}        \bigg]^N }       = (\textbf{5} ) , \\  
\end{align*} }

\noindent for each $\delta_k$ and $\epsilon_k$ strictly positive over $\big[ 0 , 1 \big]$, from which we conclude the argument. \boxed{}

\bigskip

\noindent \textit{Proof of Corollary 3}. 
To demonstrate that the family of inequalities provided in (\textbf{1}) holds, it suffices to demonstrate that the last inequality,

{\small
\begin{align*}
    \textbf{E}_{\mathcal{Q}_1 \cdots \mathcal{Q}_{i-1} \mathcal{Q}_{i+1} \cdots \mathcal{Q}_{j-1} \mathcal{Q}_{j+1} \cdots \mathcal{Q}_{k-1} \mathcal{Q}_{k+1}  \cdots \mathcal{Q}_N} \big[ \psi^{\prime\cdots\prime}_{i} \big( \textbf{E}_{\mathcal{Q}_i} \big)   \psi^{\prime\cdots\prime}_{j} \big( \textbf{E}_{\mathcal{Q}_j} \big)   \psi^{\prime\cdots\prime}_{k} \big( \textbf{E}_{\mathcal{Q}_k} \big)   \times \cdots \times  \psi^{\prime\cdots\prime}_{N^{\prime}} \big( \textbf{E}_{\mathcal{Q}_{N^{\prime}}} \big)       \big] \\  \\ \leq  \frac{1}{\sqrt{N} } \bigg[  \epsilon_i   \psi^{\prime\cdots\prime}_{i} \big[  \underset{l \in \{ n \}  ,  l \neq i  }{\prod}  \delta_l   \big]                +  \epsilon_j \psi^{\prime\cdots\prime}_{j}  \big[  \underset{l \in \{ n \}  ,  l  \neq i \neq j }{\prod}   \delta_l \big]  +     \epsilon_j  \psi^{\prime\cdots\prime}_{k}  \big[  \underset{l \in \{ n \}  ,  l  \neq i \neq j \neq k }{\prod}   \delta_l \big]       + \cdots +    \epsilon_{N^{\prime}}  \psi^{\prime\cdots\prime}_{N^{\prime} }  \\ \\ \times  \big[  \underset{l \in \{ n \}  ,  l  \neq i \neq j \neq k \neq \cdots \neq N^{\prime} }{\prod}   \delta_l \big]                   \bigg]       ,  \\ 
\end{align*}
}

\noindent from (\textbf{1}) holds, from which the remaining inequalities in the system hold by a similar argument. Observe,

{\small
\begin{align*}
         \textbf{E}_{\mathcal{Q}_1 \cdots \mathcal{Q}_{i-1} \mathcal{Q}_{i+1} \cdots \mathcal{Q}_{j-1} \mathcal{Q}_{j+1} \cdots \mathcal{Q}_{k-1} \mathcal{Q}_{k+1}  \cdots \mathcal{Q}_N} \big[    \Psi  \big( \textbf{E}_{\mathcal{Q}_i} \big)    \big( \textbf{E}_{\mathcal{Q}_j} \big) \big( \textbf{E}_{\mathcal{Q}_k} \big)   \times \cdots \times  \big( \textbf{E}_{\mathcal{Q}_{N^{\prime}}} \big)       \big]                                  \\ \\ \leq     \Psi \bigg\{  \textbf{E}_{\mathcal{Q}_1 \cdots \mathcal{Q}_{i-1} \mathcal{Q}_{i+1} \cdots \mathcal{Q}_{j-1} \mathcal{Q}_{j+1} \cdots \mathcal{Q}_{k-1} \mathcal{Q}_{k+1}  \cdots \mathcal{Q}_N} \big[        \big( \textbf{E}_{\mathcal{Q}_i} \big)    \big( \textbf{E}_{\mathcal{Q}_j} \big) \big( \textbf{E}_{\mathcal{Q}_k} \big)   \times \cdots \\ \times  \big( \textbf{E}_{\mathcal{Q}_{N^{\prime}}} \big)       \big]  \bigg\}              \end{align*}

         \begin{align*} =   \Psi \big\{  \textbf{E}_{\mathcal{Q}_1 \cdots \mathcal{Q}_{i-1} \mathcal{Q}_{i+1} \cdots \mathcal{Q}_{j-1} \mathcal{Q}_{j+1} \cdots \mathcal{Q}_{k-1} \mathcal{Q}_{k+1}  \cdots \mathcal{Q}_N} \big[        \big( \textbf{E}_{\mathcal{Q}_i} \big)  \big]  \big\} \\ \\ \times   \Psi \big\{ \textbf{E}_{\mathcal{Q}_1 \cdots \mathcal{Q}_{i-1} \mathcal{Q}_{i+1} \cdots \mathcal{Q}_{j-1} \mathcal{Q}_{j+1} \cdots \mathcal{Q}_{k-1} \mathcal{Q}_{k+1}  \cdots \mathcal{Q}_N} \big[   \big( \textbf{E}_{\mathcal{Q}_j} \big)   \big\}  \times \cdots  \\ \times  \Psi \big\{ \textbf{E}_{\mathcal{Q}_1 \cdots \mathcal{Q}_{i-1} \mathcal{Q}_{i+1} \cdots \mathcal{Q}_{j-1} \mathcal{Q}_{j+1} \cdots \mathcal{Q}_{k-1} \mathcal{Q}_{k+1}  \cdots \mathcal{Q}_N} \big[   \big( \textbf{E}_{\mathcal{Q}_{N^{\prime}}} \big)       \big]    \big\}                       \\ \\  \\  =      \psi^{\prime\cdots\prime}_{i}  \big\{  \textbf{E}_{\mathcal{Q}_1 \cdots \mathcal{Q}_{i-1} \mathcal{Q}_{i+1} \cdots \mathcal{Q}_{j-1} \mathcal{Q}_{j+1} \cdots \mathcal{Q}_{k-1} \mathcal{Q}_{k+1}  \cdots \mathcal{Q}_N} \big[        \big( \textbf{E}_{\mathcal{Q}_i} \big)   \big\} \\ \\ \times  \psi^{\prime\cdots\prime}_{j} \big\{ \textbf{E}_{\mathcal{Q}_1 \cdots \mathcal{Q}_{i-1} \mathcal{Q}_{i+1} \cdots \mathcal{Q}_{j-1} \mathcal{Q}_{j+1} \cdots \mathcal{Q}_{k-1} \mathcal{Q}_{k+1}  \cdots \mathcal{Q}_N} \big[ \big(  \psi^{\prime\cdots\prime}_{j}            \big) \big]  \big\}       \\ \\  \times  \psi^{\prime\cdots\prime}_{N} \big\{ \textbf{E}_{\mathcal{Q}_1 \cdots \mathcal{Q}_{i-1} \mathcal{Q}_{i+1} \cdots \mathcal{Q}_{j-1} \mathcal{Q}_{j+1} \cdots \mathcal{Q}_{k-1} \mathcal{Q}_{k+1}  \cdots \mathcal{Q}_N} \big[ \big(  \psi^{\prime\cdots\prime}_{N}            \big) \big]   \big\}                                                      \\ \\ \\  < C^{\prime}       \psi^{\prime\cdots\prime}_{i}  \big\{  \textbf{E}_{\mathcal{Q}_1 \cdots \mathcal{Q}_{i-1} \mathcal{Q}_{i+1} \cdots \mathcal{Q}_{j-1} \mathcal{Q}_{j+1} \cdots \mathcal{Q}_{k-1} \mathcal{Q}_{k+1}  \cdots \mathcal{Q}_N} \big[        \big( \textbf{E}_{\mathcal{Q}_i} \big)   \big\} \\ \\ \times  \psi^{\prime\cdots\prime}_{j} \big\{ \textbf{E}_{\mathcal{Q}_1 \cdots \mathcal{Q}_{i-1} \mathcal{Q}_{i+1} \cdots \mathcal{Q}_{j-1} \mathcal{Q}_{j+1} \cdots \mathcal{Q}_{k-1} \mathcal{Q}_{k+1}  \cdots \mathcal{Q}_N} \big[ \big(  \psi^{\prime\cdots\prime}_{j}            \big) \big]  \big\}       \\ \\  \times  \psi^{\prime\cdots\prime}_{N} \big\{ \textbf{E}_{\mathcal{Q}_1 \cdots \mathcal{Q}_{i-1} \mathcal{Q}_{i+1} \cdots \mathcal{Q}_{j-1} \mathcal{Q}_{j+1} \cdots \mathcal{Q}_{k-1} \mathcal{Q}_{k+1}  \cdots \mathcal{Q}_N} \big[ \big(  \psi^{\prime\cdots\prime}_{N}            \big) \big]   \big\}                 \\ \\ = C^{\prime}      \bigg\{  \psi^{\prime\cdots\prime}_{i}  \big\{  \textbf{E}_{\mathcal{Q}_1 \cdots \mathcal{Q}_{i-1} \mathcal{Q}_{i+1} \cdots \mathcal{Q}_{j-1} \mathcal{Q}_{j+1} \cdots \mathcal{Q}_{k-1} \mathcal{Q}_{k+1}  \cdots \mathcal{Q}_N} \big[        \big( \textbf{E}_{\mathcal{Q}_i} \big)   \big\} \\ \\ \times  \psi^{\prime\cdots\prime}_{j} \big\{ \textbf{E}_{\mathcal{Q}_1 \cdots \mathcal{Q}_{i-1} \mathcal{Q}_{i+1} \cdots \mathcal{Q}_{j-1} \mathcal{Q}_{j+1} \cdots \mathcal{Q}_{k-1} \mathcal{Q}_{k+1}  \cdots \mathcal{Q}_N} \big[ \big(  \psi^{\prime\cdots\prime}_{j}            \big) \big]  \big\}       \\ \\  \times  \psi^{\prime\cdots\prime}_{N} \big\{ \textbf{E}_{\mathcal{Q}_1 \cdots \mathcal{Q}_{i-1} \mathcal{Q}_{i+1} \cdots \mathcal{Q}_{j-1} \mathcal{Q}_{j+1} \cdots \mathcal{Q}_{k-1} \mathcal{Q}_{k+1}  \cdots \mathcal{Q}_N} \big[ \big(  \psi^{\prime\cdots\prime}_{N}            \big) \big]   \big\}       \bigg\} \textbf{1} \\ \\ \\   = C^{\prime}     \psi^{\prime\cdots\prime}_{i}  \big\{  \textbf{E}_{\mathcal{Q}_1 \cdots \mathcal{Q}_{i-1} \mathcal{Q}_{i+1} \cdots \mathcal{Q}_{j-1} \mathcal{Q}_{j+1} \cdots \mathcal{Q}_{k-1} \mathcal{Q}_{k+1}  \cdots \mathcal{Q}_N} \big[        \big( \textbf{E}_{\mathcal{Q}_i} \big) \big[  \textbf{1}   \big]  \big\} \\ \\ \times  \psi^{\prime\cdots\prime}_{j} \big\{ \textbf{E}_{\mathcal{Q}_1 \cdots \mathcal{Q}_{i-1} \mathcal{Q}_{i+1} \cdots \mathcal{Q}_{j-1} \mathcal{Q}_{j+1} \cdots \mathcal{Q}_{k-1} \mathcal{Q}_{k+1}  \cdots \mathcal{Q}_N} \big[  \big( \textbf{E}_{\mathcal{Q}_j} \big) \big[  \textbf{1}   \big]   \big]  \big\}       \\ \\  \times  \psi^{\prime\cdots\prime}_{N} \big\{ \textbf{E}_{\mathcal{Q}_1 \cdots \mathcal{Q}_{i-1} \mathcal{Q}_{i+1} \cdots \mathcal{Q}_{j-1} \mathcal{Q}_{j+1} \cdots \mathcal{Q}_{k-1} \mathcal{Q}_{k+1}  \cdots \mathcal{Q}_N} \big[  \big( \textbf{E}_{\mathcal{Q}_N} \big) \big[  \textbf{1}   \big]   \big]   \big\}      \\     \\    \\ < \big[ \textit{last equation of system } (\textbf{1}) \big]  ,    \\  
\end{align*}
}

\noindent for some $C^{\prime} >0$ and $N$ taken sufficiently small from which we conclude the argument, as all of the remaining inequalities within the system ($\textbf{1}$) can be established with an adaptation of the above argument. \boxed{}

\section{Conclusion}

\noindent In this work we demonstrated how amplification in multiplayer game-theoretic settings can be obtained in the absence of dependency-breaking and anchoring variables. As first proposed in [2,3,4], such variables are of use for obtaining estimates on the decay of the optimal value (equivalently, the maximum winning probability of a game) under parallel repetition, which has previously been shown to decay exponential fast with respect to the number of repetitions. As demonstrated in the arguments for the main result of this work, if one instead makes use of assumptions surrounding multiplayer concave functions the behavior of the optimal value under parallel repetition is inversely proportional to $  \mathrm{exp} \big[ - N q_i \big] + N H_i $. To answer a question raised by  Lanzenberger and Maurer, we not only discussed how notions of asymmetry between Alice and Bob are formulated for an arbitrary number of participants, but also how combinatorial factors related to $2^N$ appearing in the preimage under $\Psi^{-1} \big[ \cdot \big]$. The combinatorial factor of $2^N$ to which the preimage of the multiplayer concave function is raised encodes $2^{N-1}$ more possible outcomes than the power $2$ to which the concave function previously obtained by Lanzenberger and Maurer is raised. In place of the dependency-breaking and anchoring variables we introduced a generalization of the monotonic, concave functions used for amplification in the two-player game-theoretic setting from [10] is obtained for functions $\Psi$. Along the lines of this work, it continues to remain of interest to determine the underlying assumptions from which decay on the optimal value for a game (whether two player or multiplayer) of interest can be obtained.

\section{References}


\bigskip


\noindent [1] Amr, A., Villanueva, I.: Quantum one way vs. classical two way communication
in xor games. Quantum Information Processing 20(79) (2021). https://doi.org/10.1007/s11128-021-03014-2.



\bigskip

\noindent [2] Bavarian, M. Parallel repetition of Multi-party and Quantum Games via Anchoring and Fortification. PhD Thesis, Massachussetts Institute of Technology (2017).

\bigskip

\noindent [3] Bavarian, M., Vidick, T., Yuen, H. Anchored Parallel Repetition for Nonlocal Games. \textit{SIAM Journal on Computing} \textbf{51}:2 (2022). https://doi.org/10.1137/21M1405927.

\bigskip

\noindent [4] Bavarian, M., Vidick, T., Yuen, H. Hardness amplification for entangled games via anchoring. \textit{STOC 2017: Proceedings of the 49th Annual ACM SIGACT Symposium on Theory of Computing}, 303-316. https://doi.org/10.1145/3055399.3055433.  



\bigskip

\noindent [5] Hur, T. and Kim, L. and Park, D.K. Quantum convolutional neural network for classical data classification. \textit{Quantum Machine Intelligence} 4: 3 (2022). https://doi.org/10.1007/s42484-021-00061-x. 



\bigskip

\noindent [6] Holmes, Z. and Coble, N.J. and Sornborger, A.T. and Subasi, Y. On nonlinear transformations in quantum computation. \textit{Phys. Rev. Research} 5: 013105 (2023). https://doi.org/10.1103/PhysRevResearch.5.013105. 



\bigskip

\noindent [7] Jing, H. and Wang, Y. and Li, Y. Data-Driven Quantum Approximate Optimization Algorithm for Cyber-Physical Power Systems. \textit{arXiv}: 2204.00738 (2022). https://doi.org/10.48550/arXiv.2204.00738.


\bigskip

\noindent [8] Junge, M., Palazuelos, C. On the power of quantum entanglement in multipartite quantum XOR games. \textit{Journal of the London Mathematical Society} \textbf{110}(5) (2024).

\bigskip

\noindent [9] Kubo, K. and Nakagawa, Y.O. and Endo, S. and Nagayama, S. Variational quantum simulations of stochastic differential equations. \textit{Physical Review A} 103: 052425 (2021). https://doi.org/10.1103/PhysRevA.

\noindent 103.052425.


\bigskip



\noindent [10] Lanzenberger, D., Maurer, U. Direct Product Hardness Amplification. In: Nissim, K., Waters, B. (eds) Theory of Cryptography. TCC 2021. Lecture Notes in Computer Science, vol 13043 (2021). Springer, Cham. https://doi.org/10.1007/978-3-030-90453-1 21.

\bigskip

\noindent [11] Ostrev, D. Composable, unconditionally secure message authentication without any secret key. IEEE International Symposium on Information Theory \textbf{10}(1109), 622-626 (2019). https://doi.org/10.1109/ISIT.2019.8849510.


\bigskip

\noindent [12] Ostrev, D.. QKD parameter estimation by two-universal hashing. Quantum \textbf{7}, 894 (2023) https://doi.org/10.22331/q-2023-01-13-894.

\bigskip


\noindent [13] Paine, A.E. and Elfving, V.E. and Kyriienko, O. Quantum Kernel Methods for Solving Differential Equations. \textit{Physical Review A} 107: 032428 (2023). https://doi.org/10.1103/PhysRevA.107.032428. 


\bigskip


\noindent [14] Paudel, H.P., Syamlal, M., Crawford, S.E., Lee, Y-L, Shugayev, R.A., Lu, P., Ohodnicki, P.R., Mollot, D., Duan, Y. \textit{Quantum Computing and Simulations for Energy Applications: Review and Perspective. ACS Eng. Au}: 3 151-196 (2022). $\mathrm{https://doi.org/10.1021/ac}$$\mathrm{sengineeringau.1c00033}$.


\bigskip


\noindent [15] Przhiyalkovskiy, Y.V. Quantum process in probability representation of quantum mechanics. \textit{Journal of Physics A: Mathematical and Theoretical} 55: 085301 (2022). https://doi.org/10.1088/1751-8121/ac4b15.


\bigskip

\noindent [16] Perc, M. Statistical physics of human cooperation. \textit{Physics Reports} \textbf{687}: 1-51 (2017). $\mathrm{https://papers.ssrn}$ $\mathrm{.com/sol3/papers.cfm?abstract id=2972841}$.

\bigskip

\noindent [17] Renner, R., Wolf, S. The Exact Price for Unconditionally Secure Asymmetric Cryptography. In: Cachin, C., Camenisch, J.L. (eds) Advances in Cryptology - EUROCRYPT 2004. EUROCRYPT 2004. Lecture Notes in Computer Science, \textbf{3027}. Springer, Berlin, Heidelberg. $https://doi.org/10.1007/978-3-540-24676-3 7$.

\bigskip

\noindent [18] Rigas, P. Optimal, and approximately optimal, quantum strategies for $\mathrm{XOR^{*}}$ and $\mathrm{FFL}$ games. \textit{arXiv: 2311.12887} (2023), submitted. https://doi.org/10.48550/arXiv.2311.12887.

\bigskip


\noindent [19] Rigas, P. Quantum strategies, error bounds, optimality, and duality gaps for multiplayer XOR, $\mathrm{XOR}^{*}$, compiled XOR, $\mathrm{XOR}^{*}$, and strong parallel repetiton of XOR, $\mathrm{XOR}^{*}$, and FFL games. 	arXiv:2505.06322, submitted (2025). $\mathrm{
https://doi.org/10.48550/arXiv.2505.06322
}$


\bigskip

\noindent [20] Rigas, P. Error correction, authentication, and false acceptance, probabilities for communication over noisy quantum channels: converse upper bounds on the bit transmission rate. arXiv:2507.03035, submitted (2025). 
https://doi.org/10.48550/arXiv.2507.03035.

\bigskip

\noindent [21] Rigas, P. Parallel repetition of expanded, and multiplayer, Quantum games: anchoring, optimal values, generalized error bounds, dependency-breaking as symmetry-breaking. arXiv: 2508.09380, submitted (2025). 
https://doi.org/10.48550/arXiv.2508.09380.
.

\bigskip

\noindent [22] Rigas, P. Probability distributions over CSS codes: two-universality, QKD hashing, collision bounds, security. arXiv:2510.02402, submitted (2025). https://doi.org/10.48550/arXiv.2510.02402.

\bigskip

\noindent [23] Rigas, P. Composable, unconditional security without a Quantum secret key: public broadcast channels and their conceptualizations, adaptive bit transmission rates, fidelity pruning under wiretaps. arXiv: 2512.19759, submitted (2025). 
https://doi.org/10.48550/arXiv.2512.19759.

\bigskip

\noindent [24] Rigas, P. Eve's forgery probability from her false acceptance probability:  interactive authentication, Holevo information and the min-entropy, submitted (2026). 
https://doi.org/10.48550/arXiv.2603.06645.

\end{document}